\documentclass[useAMS,usedcolumn,usenatbib,usegraphicx]{mn2e}
\usepackage{epsfig,graphicx,natbib}
\usepackage{./reference/mycite}
\usepackage{amssymb}
\usepackage{amsfonts}
\usepackage{amsmath}
\usepackage{color}
\usepackage{lscape,longtable,multirow,threeparttablex}
 
\begin{document}

\title[Redshift 1--2 Luminous Starbursts]{Luminous Starbursts in the Redshift Desert at z$\sim$1--2: Star-Formation Rates, Masses \& Evidence for Outflows} 
\author[M. Banerji et al.]{ \parbox{\textwidth}{
Manda Banerji$^{1}$\thanks{E-mail: mbanerji@ast.cam.ac.uk},
S. C. Chapman$^{1}$,  
Ian Smail$^{2}$, S. Alaghband-Zadeh$^{1}$,  A. M. Swinbank$^{2}$, J. S. Dunlop$^{3}$, R. J. Ivison$^{4,3}$, A. W. Blain$^{5}$.
}
  \vspace*{6pt} \\
$^{1}$Institute of Astronomy, University of Cambridge, Madingley Road, Cambridge, CB3 0HA, UK.\\
$^{2}$Institute for Computational Cosmology, Durham University, South Road, Durham DH1 3LE \\
$^{3}$Institute for Astronomy, University of Edinburgh, Royal Observatory, Edinburgh, EH9 3HJ \\
$^{4}$UK Astronomy Technology Centre, Royal Observatory, Blackford Hill, Edinburgh, EH9 3HJ \\
$^{5}$Department of Physics \& Astronomy, University of Leicester, University Road, Leicester LE1 7RH, UK \\
} 

\maketitle

\begin{abstract}

We present a spectroscopic catalogue of 40 luminous starburst galaxies at z=0.7--1.7 (median z=1.3). 19 of these are submillimetre galaxies (SMGs) and 21 are submillimetre-faint radio galaxies (SFRGs). This sample helps to fill in the redshift desert at z=1.2--1.7 in previous studies as well as probing a lower luminosity population of galaxies. Radio fluxes are used to determine star-formation rates for our sample which range from around 50 to 500 M$_\odot$ yr$^{-1}$ and are generally lower than those in z$\sim$2 SMGs. We identify nebular [OII] 3727 emission in the rest-UV spectra and use the linewidths to show that SMGs and SFRGs in our sample have larger linewidths and therefore dynamical masses than optically selected star-forming galaxies at similar redshifts. The linewidths are indistinguishable from those measured in the z$\sim$2 SMG populations suggesting little evolution in the dynamical masses of the galaxies between redshift 1--2. [NeV] and [NeIII] emission lines are identified in a subset of the spectra indicating the presence of an active galactic nucleus (AGN). In addition, a host of interstellar absorption lines corresponding to transitions of MgII and FeII ions are also detected. These features show up prominently in composite spectra and we use these composites to demonstrate that the absorption lines are present at an average blueshift of $-$240$\pm$50 kms$^{-1}$ relative to the systemic velocities of the galaxies derived from [OII]. This indicates the presence of large-scale outflowing interstellar gas in these systems. We do not find any evidence for differences in outflow velocities between SMGs and SFRGs of similar infra-red luminosities.  We find that the outflow velocities of z$\sim$1.3 SMGs and SFRGs are consistent with the V $\propto$ SFR$^{0.3}$ local envelope seen in lower redshift Ultra-luminous Infra-red Galaxies (ULIRGs). These observations are well explained by a momentum-driven wind model.  

\end{abstract}


\section{INTRODUCTION}

Submillimetre (submm) galaxies (SMGs) selected based on their 850$\mu$m fluxes \citep{Blain:99, Chapman:05, Coppin:06, Weiss:09} play an important role in the overall star-formation history of the Universe. They are thought to be the high-redshift counterparts of the most infra-red luminous systems in the local Universe such as the Luminous Infra-Red Galaxies (LIRGs) and Ultra-luminous Infra-Red Galaxies (ULIRGs) and progenitors of the most massive ellipticals seen today \citep{Swinbank:06, Swinbank:08}. However, the sensitivities of current generation submm instruments such as LABOCA biases the detection of SMGs to those with cooler dust temperatures therefore missing a population of luminous but hotter starbursts at high redshifts. Submm Faint Radio Galaxies (SFRGs) have been proposed as an extension of the SMG population to hotter dust temperatures \citep{Chapman:04, Casey:09a, Magnelli:10, Chapman:10}. Radio observations of SFRGs indicate that they are very similar to SMGs in terms of radio morphology and size \citep{Casey:09b}. Together, these two populations have a SFR density at redshifts $\sim$1-3 that is roughly comparable to the rest-frame UV selected population. These galaxies therefore represent a reasonably complete sample of infrared luminous galaxies at the main epoch of galaxy formation.

The most extensive spectroscopic sample of SMGs and SFRGs to date has been presented in \citet{Chapman:04,Chapman:05} (C04 and C05 hereafter). This sample has a median redshift of 2--2.5 with a dip in the numbers below $z \sim 1.5$ where no strong spectral features fall into the Keck-LRIS window used for redshift identification. Redshifts obtained from the rest-UV analysis of this prodigiously star-forming massive galaxy population, have motivated further observations of these galaxies at multiple wavelengths all the way from the X-ray \citep{Alexander:05} through the infra-red \citep{Swinbank:04,Swinbank:10} to the millimeter \citep{Tacconi:08}. Such observations have been succesful in constraining a wide range of physical parameters such as the AGN content, masses, sizes, morphologies and merging histories of these galaxies. 

The rest-UV spectra themselves are rich in spectral features \citep{Leitherer:11} that can be used to trace cool interstellar gas in the galaxies in order to look for signatures of outflows. Provided that one can obtain a reasonable estimate of the systemic redshift of the galaxies, numerous absorption lines in the rest-UV can be used to probe the state of the cool atomic gas and characterise \textit{feedback} processes in the galaxies. A wealth of observational evidence now suggests that galactic winds are in fact ubiquitous in galaxies with star-formation surface density, $\Sigma_{SF} \gtrsim 10^{-1} M_\odot$yr$^{-1}$kpc$^2$ \citep{Heckman:90} and the most powerful starbursts can power winds that are approaching the escape velocity of the galaxy. Winds that can escape the galaxy are likely to be responsible for the metal enrichment of the IGM and the observed mass-metallicity relation of galaxies \citep{Tremonti:04, Erb:06, Maiolino:08}. 

There have been extensive studies of galactic scale outflows in local starbursts ranging from dwarfs \citep{Schwartz:06} to LIRGs and ULIRGs \citep{Heckman:00, Rupke:02, Martin:05}. Recently, \citet{Weiner:09} and \citet{Rubin:09} extended this analysis to optically-selected star-forming galaxies at z$\sim$1.5 using stacked spectra from the DEEP2 survey to demonstrate that outflows are also ubiquitous at high redshifts. At even higher redshifts, studies of the most luminous Lyman break galaxies also show signs of outflowing gas in these systems \citep{Pettini:01,Pettini:02,Shapley:03,Steidel:10} and outflow velocities of several hundred kms$^{-1}$ have been measured in lensed Lyman-$\alpha$ emitting systems \citep{Frye:02}. 

The Keck-LRIS spectra in C04 and C05 did not allow for a study of galactic scale outflows due to the lack of suitable spectral features for systemic redshift determination at the observed wavelengths. However, numerous individual submm detected galaxies have been studied in detail and shown to drive powerful outflows. These include N2 850.4 \citep{Smail:03, Swinbank:05} which shows blueshifted SiII and CaII interstellar absorption features with a velocity of 700 kms$^{-1}$. \citet{Alexander:10} use integral field spectroscopy to identify a highly energetic outflow from the submm galaxy SMMJ1237+6203 at z$\sim$2 being radiatively powered by the AGN and/or supernovae. Similary, \citet{Nesvadba:07} also show that the lensed starburst SMM J14011+0252 at z$\sim$2.6 is powering a superwind with velocity 350-500 kms$^{-1}$.

Motivated by these previous studies, our objective in the current paper is therefore two-fold. Firstly, we present a spectroscopic catalogue of SMGs and SFRGs observed using the Keck-II DEIMOS spectrograph with a view to filling in the \textit{redshift desert} at z$<$1.5 in the C04 and C05 samples. Our galaxies have lower redshifts and lower characteristic luminosities compared to these previous studies thus helping us to characterise both the redshift and luminosity evolution of SMG/SFRG properties. The wavelength window of the DEIMOS spectrograph allows us to identify nebular [OII] 3727 emission, which can be used to determine the systemic redshifts of the galaxies as the warm ionized gas from which this emission originates is associated with young stars. The rest-UV spectra therefore also allow us to search for signatures of outflowing gas in the galaxies in order to compare to both the local starbursts as well as optically-selected star-forming galaxies at similar redshifts to our sample. 

The paper is structured as follows. $\S$ \ref{sec:obs} describes our sample selection, spectroscopic observations, galaxy properties derived from broadband photometry and creation of composite spectra. In $\S$ \ref{sec:results} we present the main results of our spectroscopic study including characterisation of the [OII] line profiles in the galaxies and derivations of outflow velocities and their dependence on galaxy properties. We discuss these trends in $\S$ \ref{sec:disc}. Finally in $\S$ \ref{sec:conclusion} we give our conclusions. Throughout this paper we assume a flat $\Lambda$CDM cosmology with $\Omega_m$=0.3, $\Omega_\Lambda$=0.7 and $h$=0.7. All magnitudes quoted are on the AB system. 

\section{OBSERVATIONS}

\label{sec:obs}

We begin with a description of our sample selection and the spectroscopic observations. 

\subsection{Sample Selection}

\label{sec:data}

C05 have presented the most extensive catalogue of SMG redshifts to date with the selection criteria aimed at identifying the bulk of the SMG population brighter than S$_{\rm{850}}$=5mJy. The redshift distribution for these galaxies peaks around $z \sim 2-2.5$ with a dip in the numbers below $z \sim 1.5$ due to Keck-LRIS's spectral coverage. Hotter dust counterparts to these galaxies have also been presented in C04 spanning similar redshifts. With the aim of filling in the \textit{redshift desert} in these samples, a redshift survey was carried out using the red-sensitive KeckII DEep Imaging Multi-Object Spectrograph (DEIMOS; \citet{Faber:03}) in order to target emission line galaxies at $z \sim 1.3$ predominantly through their [OII] 3727 emission. The survey was carried out in 4 fields - HDF-N, Lockman Hole, SXDF and CFRS03. All the targetted galaxies have robust detections in the radio, 850 $\mu$m fluxes from SCUBA/JCMT (see Table \ref{tab:deimos1}) or are detected in \textit{Herschel} SPIRE catalogues at 250-500$\mu$m \citep{Chapman:10}. Radio data in the Lockman Hole is taken from \citet{Ivison:02} and \citet{Ivison:07} who identified the counterparts to the submm galaxies in these regions from \citet{Coppin:06}. The radio data for SXDF is from a deeper map from Arumugam et al. (in preparation) and the submm counterparts are again identified from \citet{Coppin:06}. The submm and radio data in the CFRS03 field is taken from \citet{Webb:03}. Radio data in the HDF has been presented in \citet{Biggs:06} and \citet{Morrison:10} and submm detections in this field are taken from \citet{Borys:02}.

\subsection{Spectroscopic Observations}

\label{sec:deimos}

Spectroscopic observations were made with the Keck II DEIMOS spectrograph on December 23, 2003, October 8-9, 2005, and February 22, 2005 (with seeing ranging from 0.6-1.0$^{\prime \prime}$) and the full sample of z$\sim$1.3 galaxies with spectroscopic redshifts is presented in Table \ref{tab:deimos1}. The lower resolution mode was used in all cases to gain the widest wavelength coverage, with a grating of 600 lines/mm yielding a resolution of 3\AA\@ or $\sim$100 kms$^{-1}$ (measured from the sky lines) and a scale of 0.65\AA/pixel. For a slit in the middle of the mask the observed wavelength range was 5600 to 9800 \AA. The [OII] doublet is only moderately resolved in a handful of spectra in this setting and this and other red spectral features studied can be affected by proximity to the strong night sky OH recombination lines. All data was reduced using the DEIMOS-DEEP2 pipeline \citep{Faber:03}.

The candidate redshifts in all cases are based on identification of [OII] 3727 emission in the 2D spectra. In most cases, the redshift is then corroborated through identification of multiple emission and absorption lines, including subsets of FeII(2344, 2374, 2383, 2586, 2600\AA\@), MgII(2803, 2853\AA\@), H$\delta$(4103\AA\@), H$\gamma$(4340\AA\@) and Ca H\&K(3933,3969\AA\@) in absorption, and [NeIII](3869\AA\@), [NeV](3426\AA\@) and [OIII](4959,5007\AA\@) in emission. The DEIMOS coverage allows us to target emission lines in galaxies between z=0.7--1.7. In this redshift range, the spectroscopic completeness is estimated to be $\sim$70\% from the fraction of radio IDs in the different masks with well-detected spectral features. 

Redshifts are measured by fitting two Gaussians of equal width and intensity centred at 3726.1 and 3728.8 \AA\@ to the blended [OII] line observed in most of the spectra. The [OII] redshift errors estimated from adjusting the Gaussian fits to give $\Delta$ $\chi^2$=1 are of the order of 60kms$^{-1}$.

As much of the paper is concerned with characterising velocity offsets of interstellar absorption lines from [OII], it is important to devise a sensible method of measuring the centroids and widths of the absorption lines. The absorption line profiles are often highly asymmetric with large velocity tails and it is therefore not appropriate to fit Gaussians to the line profiles. The velocities and equivalent widths are instead calculated by measuring the line centroids according to the apparent optical depth (AOD) formalism of \citet{Savage:91}.
 
Errors in the [OII] redshift estimates, poor wavelength calibration at the blue-end of the spectrum and poor S/N could all lead to systematic differences between the [OII] and interstellar absorption line redshifts. Redshift errors arising from the Gaussian fits to the [OII] line have already been discussed. There are five examples of spectra where, in addition to [OII], several other emission lines such as [OIII], [NeIII], [NeV] and various Balmer absorption features are also detected. We find these features to be shifted by +10$\pm$80 kms$^{-1}$ relative to [OII]. This shift is within the typical redshift error associated with the Gaussian fits.

\subsection{Creating Composite Spectra}

\label{sec:composite}

Most of the individual 1D spectra in our sample do not have sufficient signal-to-noise (S/N) to be able to conduct a detailed spectroscopic characterisation. We therefore created composite spectra of various subsamples of galaxies for detailed analysis.  We masked out sky lines in the individual observed frame spectra located at 5577, 5592 and 6302 \AA\@. The spectra were then shifted into the rest-frame using the systemic redshifts derived from a double Gaussian fit to the [OII] line as described in $\S$ \ref{sec:deimos}. We defined a continuum using a 9th order polynomial fit in seven wavelength windows in which we do not expect any emission or absorption line features and divided each spectrum by this continuum. All spectra were then normalised to a common median of 1 in the wavelength range 2650-2750 \AA\@ (a featureless window near the ISM absorption line features that we are interested in studying). The spectra are then smoothed by a 5 pixel boxcar (1.5\AA\@) in order to reduce noise in the spectra, before co-adding.  

The composite spectra are clipped averages of all the individual spectra making up the stack where the clipping procedure is such that 10\% of the lowest and highest points contributing at each pixel are rejected from the stack leading to a total of 20\% of the pixels being rejected. As can be seen later, this procedure can also remove some signal from the composite spectra but due to the presence of significant noise in many of the composites, such a clipping is essential in order to ensure reliable line detections. Errors on these composites are determined from jackknife sampling the individual spectra making up the stack.

\subsection{Galaxy Properties}

\label{sec:properties}

Our spectroscopic survey was conducted in well-studied fields meaning that our galaxies have ancillary photometric data from optical to radio wavelengths. These data can be used to derive a range of physical properties such as the star-formation rates, bolometric luminosities, dust temperatures and crude stellar masses of the galaxies. 

\subsubsection{Star-Formation Rates \& Dust Temperatures}

The star-formation rates are calculated from the $S_{1.4\rm{GHz}}$ radio fluxes as the [OII] luminosity is highly dependent on both the metallicity and extinction and therefore not a good proxy for unobscured star-formation. We use the latest \textit{Herschel} radio-FIR correlation of $q=2.4 \pm 0.24$ \citep{Ivison:10} to calculate the 8-1000$\mu$m infra-red luminosities from the radio fluxes after k-correcting the radio flux assuming a spectrum with S$_\nu \propto \nu^{-0.8}$. The star-formation rate is then calculated using the relation from \citet{Kennicutt:98}: 

\begin{equation}
\rm{SFR} (M_\odot yr^{-1}) = 4.55\times10^{-44}L_{\rm{IR}}(\rm{erg} \rm{s}^{-1} \rm{cm}^{-2} \rm{Hz}^{-1})
\label{eq:SFR} 
\end{equation}

\noindent In galaxies with strong AGN signatures, there will also be some contribution to the radio flux from the AGN so the star-formation rates should be taken as upper limits. Both the total infra-red luminosity and the star-formation rates derived for all galaxies are presented in Table \ref{tab:deimos2}.

We also calculate the dust temperatures for our SMGs using the empirical relation derived by C05:

\begin{equation}
T_d=\frac{6.25(1+z)}{(S_{850}/S_{1.4})^{0.26}}
\label{eq:Td}
\end{equation}

\begin{figure}
\begin{center}
\includegraphics[width=8.5cm,height=6.0cm,angle=0]{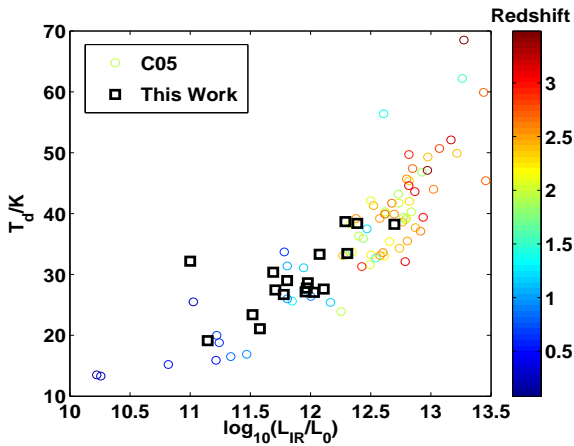}
\caption{Far infra-red luminosity versus dust temperature for our sample of SMGs compared to the higher redshift SMGs from C05. The C05 sample has been colour-coded according to redshift. It can be seen that the 850$\mu$-m selection of these sources results in an apparent trend of dust temperature with infra-red luminosity where lower luminosity galaxies are seen to have cooler dust temperatures.}
\label{fig:Td}
\end{center}
\end{figure}

These are plotted as a function of the infra-red luminosity in Figure \ref{fig:Td} and compared to the higher redshift SMGs from C05. As can be seen, the lower luminosity SMGs also have cooler dust temperatures on account of their 850$\mu$m selection which precludes detection of galaxies with hotter dust temperatures \citep{Chapman:10, Magnelli:10}. 

\subsubsection{Stellar Masses}

\label{sec:stellarmass}

We derive stellar masses and corresponding errors for all our SMGs and SFRGs using the \textit{Hyperzmass} code (Micol Bolzonella: private communication) which is based on the SED fitting code for photometric redshift determination, \textit{HyperZ} \citep{Bolzonella:Hyperz}. The code is used to fit an SED to the available optical and infra-red galaxy photometry in each field including subsets of MUSYC U, Subaru B, R, i, z, \textit{HST}-ACS B, V, i, z, MUSYC J, H, K and \textit{Spitzer} IRAC 3.6, 4.5, 5.8 and 8.0$\mu$m. The \textit{Spitzer} IRAC photometry for the Lockman and CFRS03 fields came from our own General Observer (GO) program as well as from publicly available data from various Legacy and Guaranteed Time Observation (GTO) programs obtained through the \textit{Spitzer Science Center (SSC)} archive, and was reduced and combined using standard SSC pipeline products. Optical and infra-red photometry for the Lockman SMGs was taken from the catalogues of \citet{Dye:08}. The photometry  for SXDF SMGs was obtained from the catalogues in \citet{Clements:08}, while that of the SFRGs came from \citet{Cirasuolo:10}. The HDF IRAC and {\it HST}-ACS photometry came from \citet{Pope:06}  and \citet{Giavalisco:04}. We use a set of 9 \citet{BC:03} templates with exponentially declining star-formation histories with timescales, $\tau$ between 0.1 and 30 Gyr as well as a constant star-formation history template. All the templates assume a \citet{Chabrier:03} IMF and are at solar metallicity and we fit extinction values between $A_v$=0 and 6.0. 

Mass-to-light ratios in the H-band as well as dust extinction are derived from the best SED fits and the errors quoted represent the 68\% lower and upper confidence limits obtained from the full probability distribution associated with the best chi-squared fit. These errors are very large (of the order of 1 dex) in some cases where a poor fit to the photometry was obtained. The derived stellar masses and dust extinctions for our sample can be found in Table \ref{tab:deimos2}. We caution however that most of our galaxies are highly obscured systems with large degeneracies between age, reddening and star-formation histories. The actual errors in the stellar masses due to uncertainties in template-fitting and dust extinction are probably therefore considerable. \citet{Wardlow:10} for example find stellar masses that are 5$\times$ lower than those estimated by \citet{Dye:08} for SMGs due to difficulties in disentangling the underlying old stellar population in SMGs from the recent burst even when correcting to the same IMF. 

The median stellar mass for SMGs in our sample is (3$\pm$0.2)$\times$10$^{10}$M$_\odot$ and for SFRGs, (6$\pm$0.8)$\times$10$^{10}$M$_\odot$. If we had instead adopted a constant mass-to-light ratio in the H-band of 5.6 \citep{Hainline:10} this would lead to median stellar masses of (4$\pm$0.2)$\times$10$^{10}$M$_\odot$ and (2$\pm$0.2)$\times$10$^{10}$M$_\odot$ for SMGs and SFRGs respectively. The quoted uncertainties reflect the variance in the sample average. For comparison to these values, \citet{Wardlow:10} measure a median stellar mass of (9$\pm$1)$\times$10$^{10}$M$_\odot$ for SMGs with photometric redshift $\sim$2.2 while \citet{Hainline:10} find a stellar mass of (7$\pm$1)$\times$10$^{10}$M$_\odot$ for a spectroscopic sample at similar redshifts. Unlike in \citet{Hainline:10} however, we have not subtracted any AGN contribution to the continuum at 8$\mu$m for the galaxies in our sample so our stellar masses for AGN dominated systems should be interpreted as upper limits. Given the large uncertainties in the mass estimates, we conclude that the stellar masses at z$\sim$1.3 and z$\sim$2 are not significantly different. In all subsequent analysis, we use the stellar masses derived from the full SED fits as presented in Table \ref{tab:deimos2}. 

\subsubsection{Metallicity and AGN Content}

\label{sec:agn}

A subset of the galaxies in our sample have been classified as AGN versus starburst dominated (Alaghband-Zadeh et al., (in preparation)) based on various factors such as X-ray properties, [NeV] detection and the [NeIII]/[OII] ratio. These authors also derive metallicities by considering the [NeIII]/[OII] ratio as a metallicity indicator \citep{Shi:07}. The metallicities and classifications derived by these authors are presented in Table \ref{tab:deimos1} and will be used to divide galaxies into different subsamples later in the paper. We note here however, that some of our galaxies appear to have very low metallicities derived from the high [NeIII]/[OII] ratio. This high flux ratio could arise from the presence of a strong AGN rather than an intrinsically low metallicity galaxy \citep{Nagao:01} and the former is more likely as SMGs and SFRGs are seen to be very massive systems ($\S$\ref{sec:stellarmass}) which are likely to already have been enriched to solar metallicities \citep{Swinbank:04}. 

\section{ANALYSIS \& RESULTS}

\label{sec:results}

\subsection{Redshift \& Luminosity Distribution}

\begin{figure*}
\begin{center}
\begin{minipage}[c]{1.00\textwidth}
\centering
\includegraphics[width=8.5cm,height=6.0cm,angle=0] {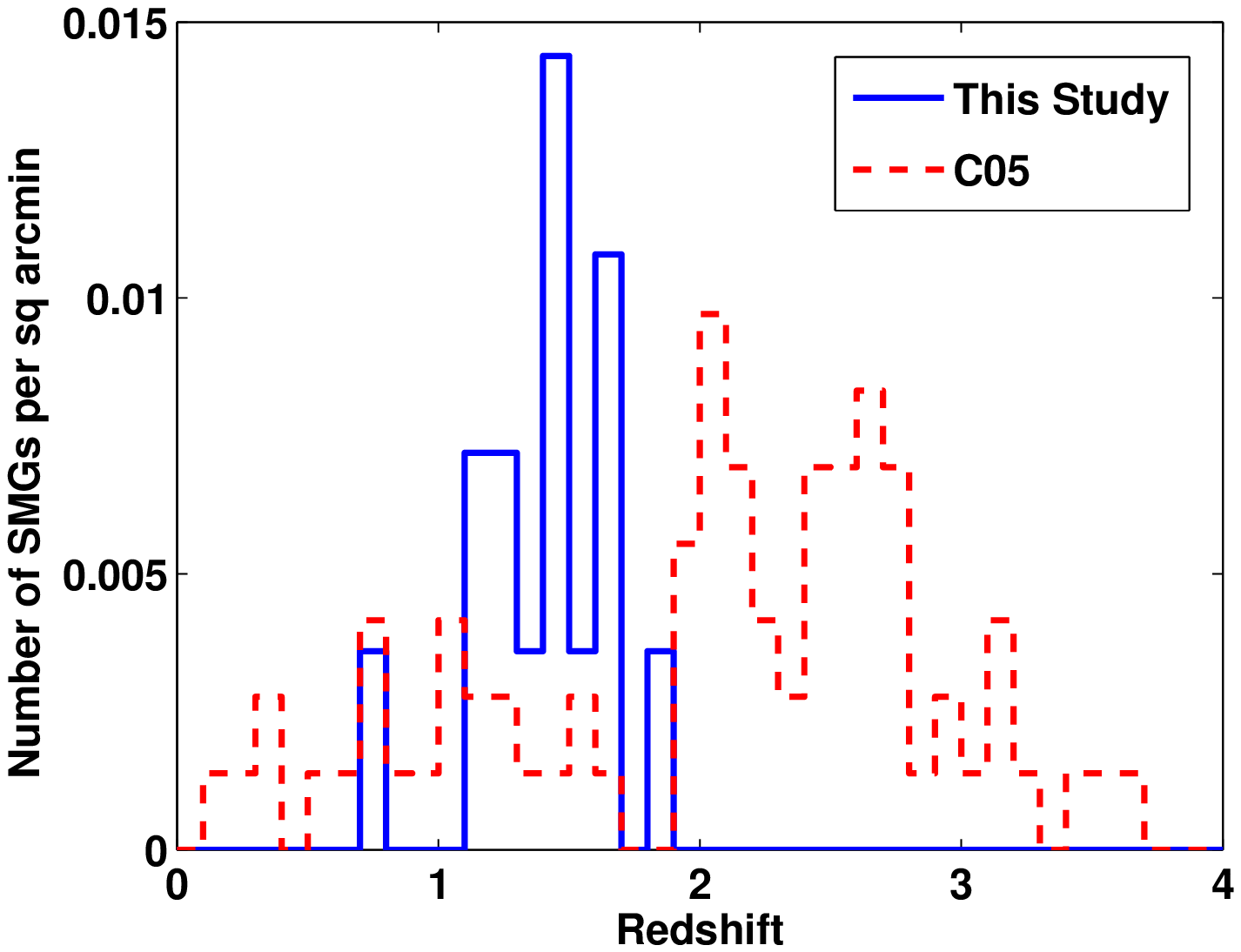}
\includegraphics[width=8.5cm,height=6.0cm,angle=0] {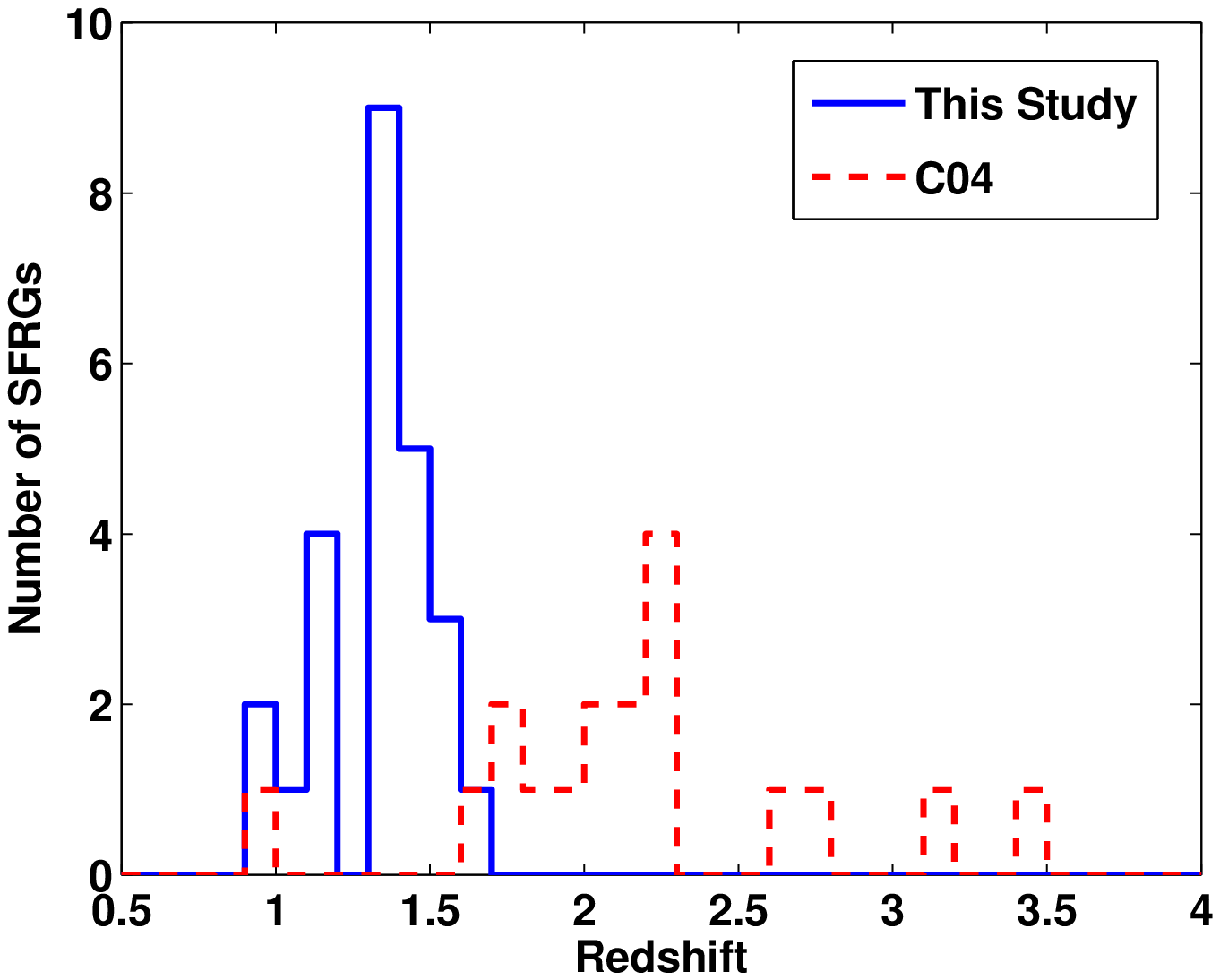}
\end{minipage}
\caption{Redshift distribution of SMGs per unit area (left) and SFRGs (right) for our sample compared to the SMGs and SFRGs studied by C05 and C04 respectively. The SFRG distribution is not scaled by total area as the SFRGs were selected as mask fillers making it difficult to assess their total survey area and completeness. It can be seen that our sample helps fill in the redshift desert between redshifts of 1.2 and 1.7 in previous studies. }
\label{fig:compare}
\end{center}
\end{figure*}

The sample of SMGs and SFRGs with secure redshifts that are analysed in this study has been summarised in Tables \ref{tab:deimos1} and \ref{tab:deimos2}. In Figure \ref{fig:compare} we show the redshift distribution for SMGs and SFRGs in our sample compared to the higher redshift SMGs (C05) and SFRGs (C04) observed with LRIS. The SMG distribution has been scaled by the total area of each survey - 721 sq arcmins in C05 and 278 sq arcmins in this study. The SFRG distribution was not scaled as the SFRGs were chosen as mask fillers making it difficult to assess their total survey area and completeness. As such, they are over-represented in this study while being under-represented in C04 relative to C05. It can clearly be seen that this spectroscopic sample of SMGs and SFRGs helps to effectively fill in the \textit{redshift desert} at z$\sim$1.5.  

\begin{figure}
\begin{center}
\includegraphics[width=8.5cm,height=6.0cm,angle=0]{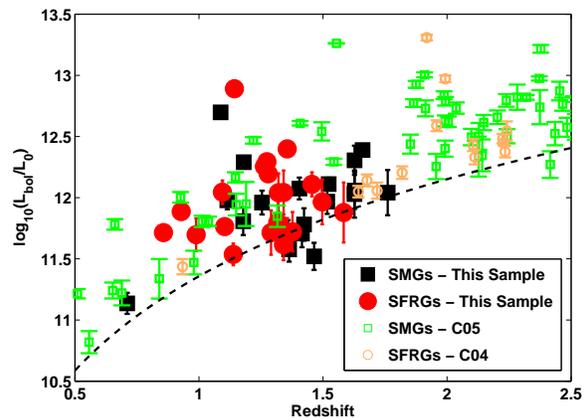}
\caption{Redshift versus total infra-red luminosity for our sample of SMGs and SFRGs compared to the SMGs and SFRGs studied by C04 and C05 (light symbols). The black dashed line shows the variation in infra-red luminosity with redshift for a radio source with S$_{1.4\rm{GHz}}$=30$\mu$Jy. It can be seen that selecting sources at a constant radio flux allows us to probe lower luminosity populations at lower redshifts.}
\label{fig:compare2}
\end{center}
\end{figure}

In Figure \ref{fig:compare2}, we also show the redshift versus bolometric luminosity distribution for our galaxies compared to those of C04 and C05. Both this sample and the C04,C05 samples have been selected in the same manner using their 850$\mu$m fluxes and radio data. As can be seen in Figure \ref{fig:compare2}, selecting sources at a constant radio flux allows us to probe lower luminosity populations at lower redshifts. However, the 850$\mu$m fluxes of SMGs at z$\sim$1.5 and z$\sim$2 should be reasonably similar due to the effects of the negative k-correction at submm wavelengths. As the SMGs in both this work and C05 were selected based on both their radio and 850$\mu$m fluxes, the differences in luminosity between the two populations likely arises due to a combination of the radio k-correction and some intrinsic luminosity evolution.  

\subsection{[OII] Linewidths and Dynamical Masses}

\label{sec:OII}

While the S/N in the majority of our spectra is not good enough to be able to detect many interstellar absorption features, the [OII] emission line is detected at high significance in almost all the spectra and is the main line used for redshift measurement. In this section, we characterise the [OII] linewidth in the galaxies in our sample and use these to derive circular velocities and dynamical masses. The $\sigma$-values obtained from fitting two Gaussians of equal width to the [OII] doublet are tabulated in Table \ref{tab:deimos1}. There were a few sources for which the [OII] fit was poor as a result of the line being close to the edge of the spectrum or close to sky line residuals. While crude redshifts were still derived for these galaxies, the $\sigma$-measurement was highly uncertain and therefore not quoted. These galaxies have been flagged in Table \ref{tab:deimos1}. In addition, a few galaxies show very broad [OII] emission. In most cases, this is due to the proximity to night sky lines although one galaxy - lock492 - does show some evidence for multiple components in the [OII] emission seen in the 2D spectrum. Again, these galaxies have been flagged in Table \ref{tab:deimos1}.  

In order to derive circular velocities and dynamical masses for the galaxies, the line is corrected for an instrumental broadening of 120 kms$^{-1}$. We calculate the dynamical mass:  

\begin{equation}
M_{dyn}=\frac{C v_{c}^2R}{G}
\label{eq:Mdyn}
\end{equation}

\noindent where $R$ is the radius, $G$ is the gravitational constant and $C$ is a constant that depends on the galaxy's mass distribution, velocity field as well as the inclination. $C$ is assumed to be 2.1 for a rotating disk \citep{Neri:03} assuming an inclination correction of sin(i)=0.5 \citep{Carilli:06}, and $C=$5 for an isotropic virialised system. We compute dynamical masses for both the isotropic and rotating disk estimators and take the true dynamical mass of the galaxy to be the average of these two estimators. For the radius, we assume $R=6$kpc consistent with that found from H$\alpha$ observations of SMGs \citep{Swinbank:04} and with the typical size of the [OII] emission region in our 2D spectra. The dynamical masses are presented in Table \ref{tab:deimos2}. The mean dynamical mass is (3.0$\pm$0.3)$\times$10$^{10}$M$_\odot$ for SMGs and (5.9$\pm$0.3)$\times$10$^{10}$M$_\odot$ for SFRGs which is consistent with the higher redshift sample of SMGs studied by \citet{Swinbank:04} given differences in assumptions about the constant, $C$. The error bars on these mean estimates represent the sample variance but there are also significant systematic uncertainties in the computation of these dynamical masses in particular arising from assumptions about the mass distribution in the galaxies as well as their inclination on the sky. The virial mass estimate detailed above necessarily assumes that [OII] emission follows the total mass distribution of the galaxies. Nebular regions in galaxies are often centrally concentrated so this assumption will almost certainly lead to an under-estimate of the total dynamical mass. In addition, tidal tails from mergers may result in an additional contribution to the dynamical mass from the moment of inertia of the system although the departure from virial is likely to be small in early merger stages. \citet{Genzel:03} find that mergers may lead to a factor of 2 increase in the dynamical mass estimates, likely driven by differences in assumptions regarding the value of $R$ for mergers.   

In order to aid our comparison of the dynamical and stellar masses, we consider the typical sizes of the regions from which the nebular and stellar emission originates. The [OII] region has already been seen to extend to $\sim$6kpc. HST studies of z$\sim$2 SMGs suggest typical half-light radii of 2.8$\pm$0.4 kpc in the H-band \citep{Swinbank:10}. \citet{Targett:11} measure a slightly larger value of 3.4$\pm$0.3 kpc in the K-band from ground-based imaging. These near infra-red bands are typically used to calculate the mass-to-light ratio for the stellar mass estimates so we can see that the effective stellar mass aperture is around a factor of 2 smaller than the kinematic aperture assumed in the dynamical mass estimates.  

\begin{figure}
\begin{center}
\includegraphics[width=8.5cm,height=6.0cm,angle=0]{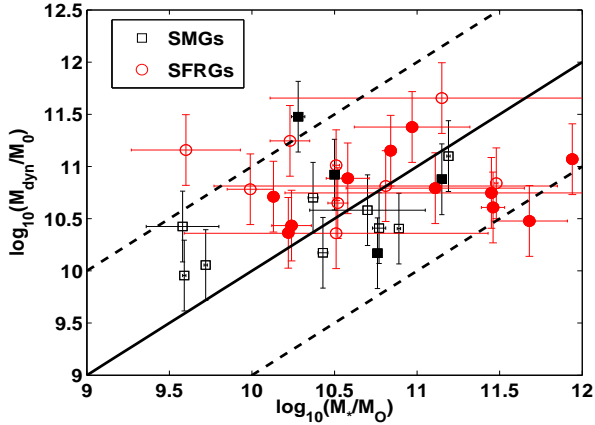}
\caption{Stellar versus dynamical mass for our sample of SMGs and SFRGs. Galaxies with AGN signatures are shown using filled symbols. The solid line indicates equal masses while the dashed lines represent systematic uncertainties of 1.0 dex due to uncertainties in the mass model in the case of the dynamical masses and SED fits in the case of the stellar masses.}
\label{fig:masses}
\end{center}
\end{figure}

In Figure \ref{fig:masses} we plot the stellar mass versus the dynamical mass for all our galaxies. The average dynamical masses are similar to the average stellar masses of SMGs and SFRGs computed in $\S$\ref{sec:stellarmass} and mostly consistent within the large systematic uncertainties in both mass estimates which we estimate to be of the order of 1 dex. Several of the AGN have their stellar masses over-estimated due to excess emission at 8$\mu$m.

\begin{figure}
\begin{center}
\includegraphics[width=8.5cm,height=6.0cm,angle=0]{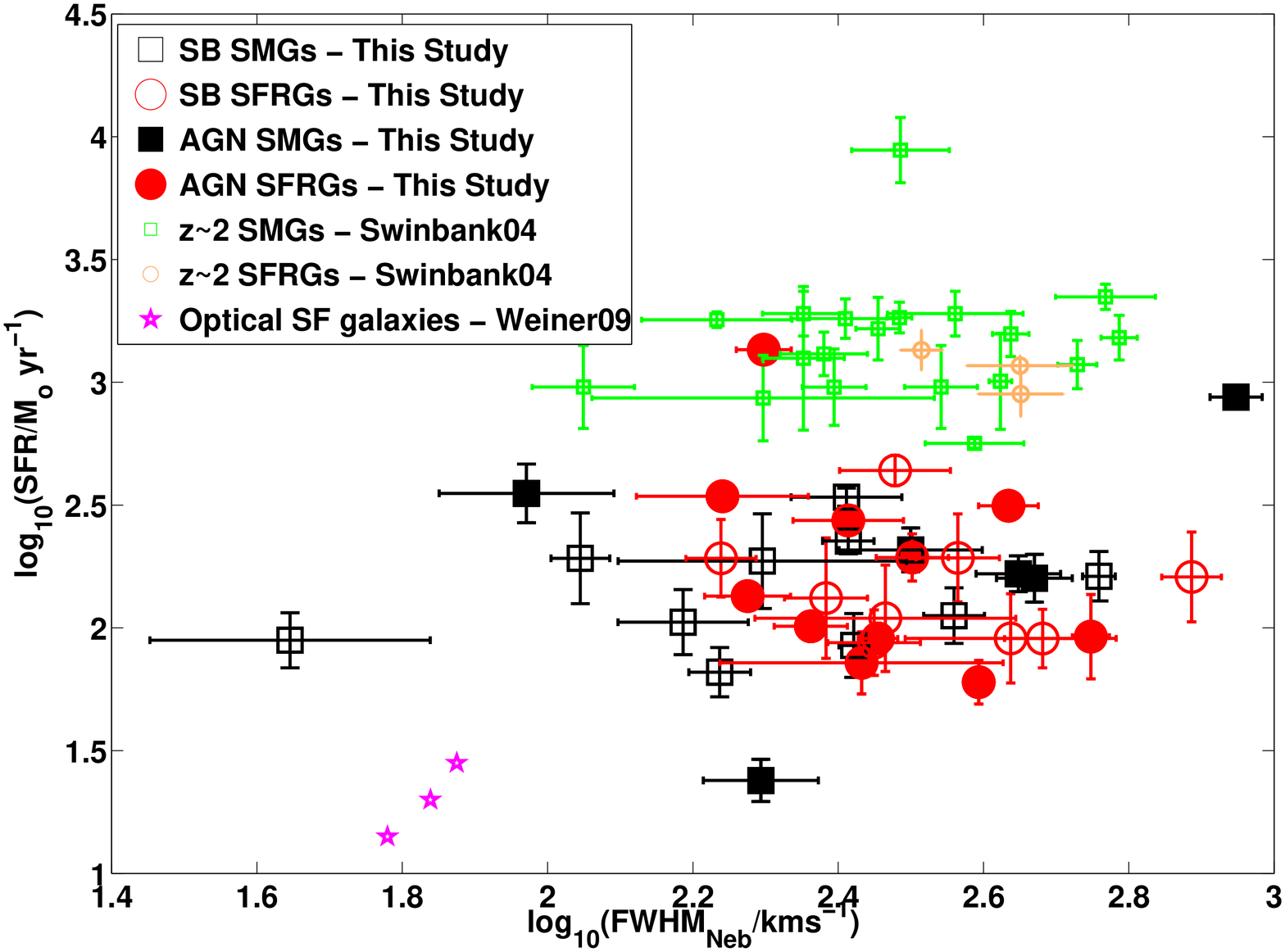}
\caption{Nebular linewidths for our sample of SMGs and SFRGs versus their star-formation rate derived from the infra-red luminosity compared with z$\sim$2 SMGs and SFRGs from \citet{Swinbank:04} and optically-selected star-forming galaxies from \citet{Weiner:09}. The filled symbols denote AGN. We see that our sample of galaxies has comparable nebular linewidths and therefore dynamical masses to z$\sim$2 SMGs and SFRGs but their star-formation rates are considerably smaller than those in the z$\sim$2 SMGs and SFRGs. By contrast, both the star-formation rates and nebular line widths are significantly larger than those in optically-selected star-forming galaxies at similar redshifts \citep{Weiner:09}.}
\label{fig:sigma}
\end{center}
\end{figure}

In Figure \ref{fig:sigma}, we plot the [OII] full-width-half-maximum as a function of the star-formation rate derived from the infra-red luminosities. For reference, we also plot the H$\alpha$ linewidths for the higher redshift sample of SMGs from \citet{Swinbank:04} as well as the [OII] linewidths and star-formation rates for optically-selected star-forming galaxies at z$\sim$1.5 \citep{Weiner:09}. There is a large scatter in the nebular linewidths for both samples of SMGs/SFRGs but they overlap, suggesting that many SMGs and SFRGs at redshifts 1.3 have similar dynamical masses to those at redshift 2. We conclude that while SMG bursts are occurring in halos of similar masses at z$\sim$1.3 and z$\sim$2.2, the lower redshift SMGs are intrinsically less luminous. In contrast to the SMG/SFRG population, the linewidths for the optically-selected sample are considerably smaller suggesting that dust-enshrouded submm galaxies at z$\sim$1.5 are more massive than optically-selected galaxies at the same redshift. 

\subsection{Composite Spectra}

\label{sec:class}

Having considered the [OII] line properties of individual galaxies in our sample, we now study the composite spectra in detail in order to characterise various lower S/N features that are difficult to detect in the individual spectra. The composite spectra are created as detailed in $\S$\ref{sec:composite}. The total composite has a continuum S/N of $\sim$7 at 2700\AA\@. The mean S/N of the individual spectra is $\sim$1.7 by comparison. 

\begin{table}
  \begin{center}
    \caption{Summary of various composite spectra analysed in this study including the number of galaxies comprising them and their continuum S/N at 2700\AA\@. \label{tab:sn}}
    \begin{tabular}{ccc}
      & Ngal & S/N at 2700\AA \\
      \hline
      \hline
      All (no e(a)) & 26 & 6.6 \\ \smallskip
      All (incl. e(a)) & 31 & 7.7 \\
      SMG & 13 & 6.0 \\ \smallskip
      SFRG & 13 & 3.9  \\
      AGN & 15 & 4.7  \\ \smallskip
      SF & 16 & 4.9 \\
      L$_{\rm{IR}}>$1.09$\times10^{12}$L$_\odot$ & 12 & 4.8 \\ \smallskip
      L$_{\rm{IR}}<$0.65$\times10^{12}$L$_\odot$ & 10 & 3.7  \\
      12+log(O/H)$>$8.4 (Low [NeIII]/[OII]) & 15 & 4.7  \\ \smallskip
      12+log(O/H)$<$8.2 (High [NeIII]/[OII]) & 10 & 5.2 \\
      $\rm{log}_{10}(M_*/M_\odot)>10.80$ & 11 & 4.7  \\ \smallskip
      $\rm{log}_{10}(M_*/M_\odot)<10.50$ & 11 & 5.5  \\
      $\rm{log}_{10}(M_{\rm{dyn}}/M_\odot)>10.50$ & 13 & 3.5 \\ \smallskip
      $\rm{log}_{10}(M_{\rm{dyn}}/M_\odot)<10.30$ & 13 & 6.3 \\
\hline
\end{tabular}	\vspace{2mm}
  \end{center}
\end{table} 

\begin{figure*}
\begin{center}
\centering
\begin{tabular}{c}
\includegraphics[width=18.0cm,height=7.0cm,angle=0]{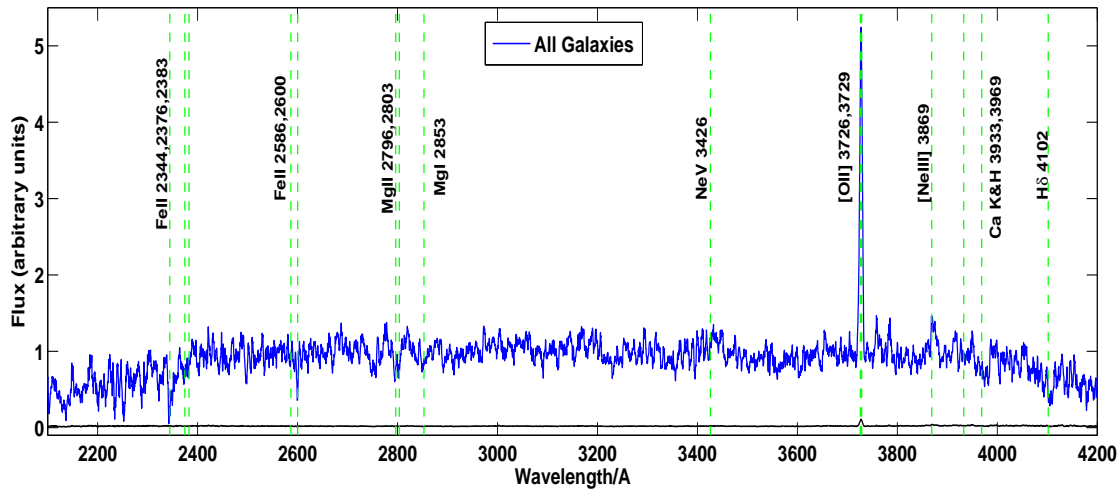} \\
\end{tabular}
\caption{Composite spectrum of all SMGs and SFRGs with robust redshift estimates in our sample in units of F$_\nu$. The line at the bottom of the panel denotes the error spectrum obtained from jackknife sampling the individual spectra making up the composite. As can be seen from the error spectrum, this clipping procedure also removes some signal from the emission lines but is necessary in order to get reliable absorption line identifications.}
\label{fig:all}
\end{center}
\end{figure*}

We begin by exploring different spectral features in the average spectra of our SMGs and SFRGs. In Figure \ref{fig:all} we show the composite spectra of all galaxies. In addition to [OII], there are a number of absorption features associated with MgII, FeII and MgI ions marked at the blue-end of the spectrum. These are discussed in more detail in $\S$\ref{sec:outflows}.

\begin{figure*}
\begin{center}
\centering
\begin{tabular}{c}
\includegraphics[width=18.0cm,height=7.0cm,angle=0]{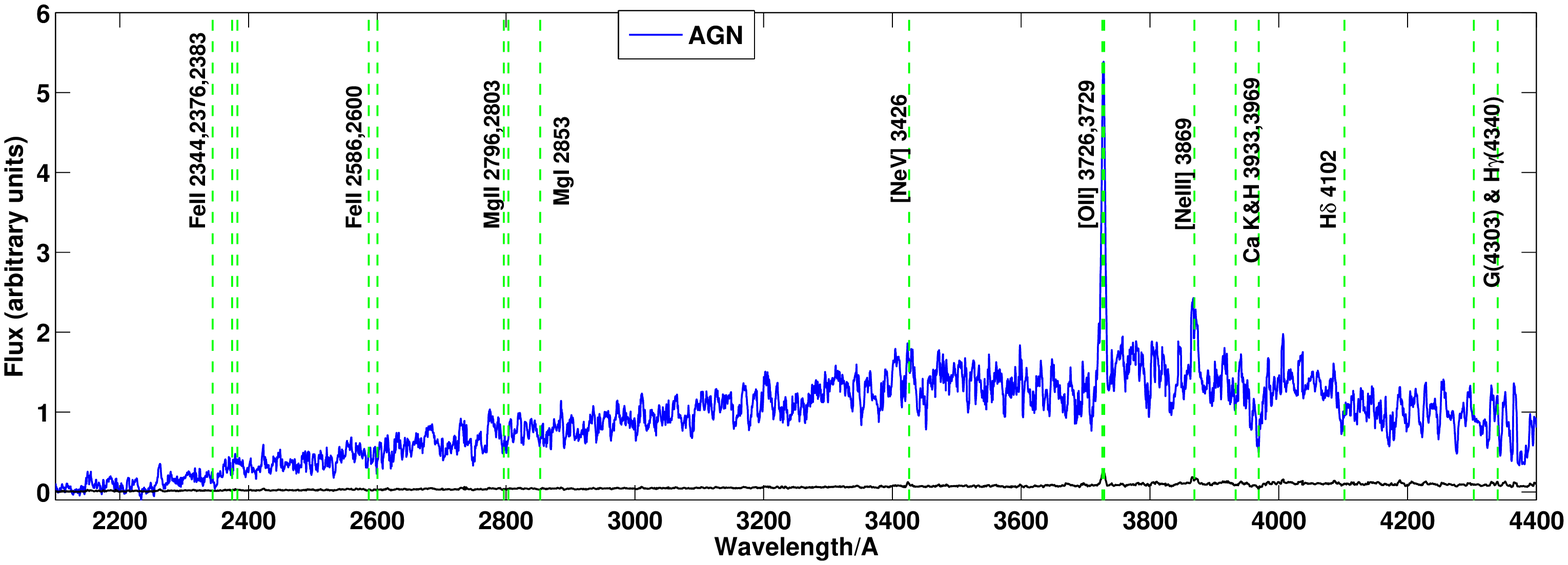} \\
\includegraphics[width=18.0cm,height=7.0cm,angle=0]{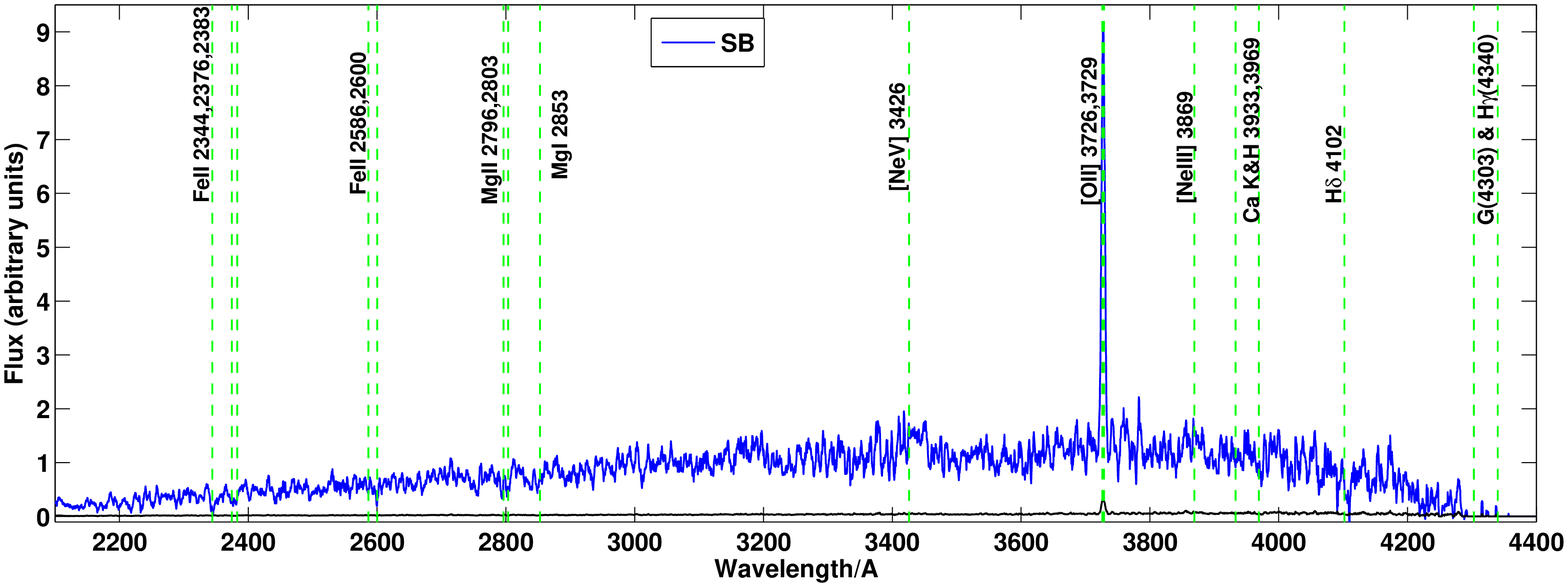} \\
\includegraphics[width=18.0cm,height=7.0cm,angle=0]{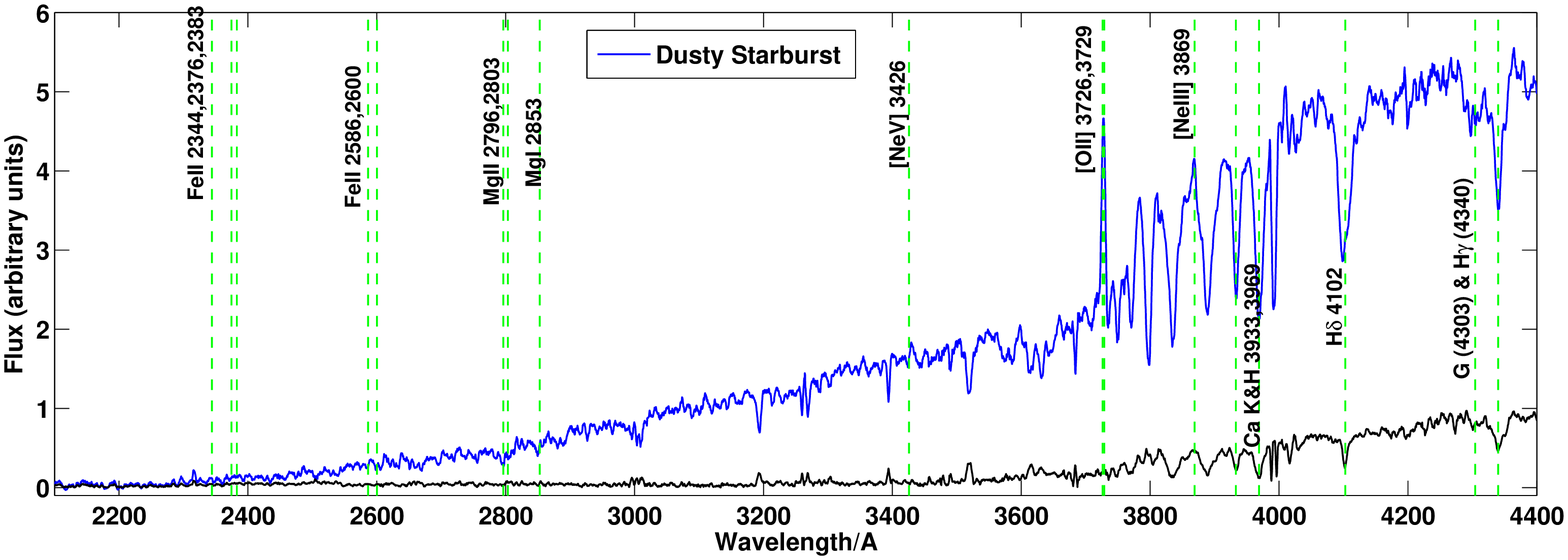} \\
\end{tabular}
\caption{Composite spectra of galaxies showing evidence for AGN activity (top), galaxies dominated by the starburst (middle) and dusty e(a) galaxies (bottom) in units of F$_\nu$, illustrating the spectral diversity observed in our SMG/SFRG sample at z$\sim$1.3. These spectra are shown with the continuum in order to illustrate differences in the continua between the star-forming galaxies, AGN and the dusty starbursts. The roll-off at the red-end in the AGN and starburst composites is due to the presence of multiple negative sky line residuals at these wavelengths. In each panel, the line at the bottom of the panel denotes the error spectrum obtained from jackknife sampling the individual spectra making up the composite. As can be seen in the error spectra, this clipping procedure also removes some signal - e.g. from weak emission lines such as [NeV] in the AGN composite.}
\label{fig:spectraltype}
\end{center}
\end{figure*}

As mentioned previously, just under half of the galaxies have been classified as AGN based on their X-ray properties, [NeIII] and [NeV] emission (Alaghband-Zadeh et al., (in preparation)). The composite spectrum for these sources is shown in the top panel of Figure \ref{fig:spectraltype}. We note, as expected, the presence of [NeIII] emission at 3869\AA\@ as well as weak [NeV] emission at 3426\AA\@. The middle panel of Figure \ref{fig:spectraltype} shows a composite spectrum of the starburst dominated galaxies, which, by contrast, has no detectable  [NeIII] emission despite its similar continuum S/N to the AGN composite.  

Four galaxies in the sample also show strong H$\delta$ absorption in addition to [OII] emission as well as a 4000\AA\@ break and Ca K \& H absorption features. These are lock76, c03.14, lock517 and sxdf1219. These sources are interpreted as extremely dusty starbursts or e(a) galaxies such as those studied by \citet{Poggianti:00} and the composite spectrum for these is shown in the bottom panel of Figure \ref{fig:spectraltype}. Once again [NeIII] emission is seen in this composite as all these galaxies also show evidence for the presence of an AGN. While the presence of [OII] emission in these galaxies suggests that there is significant ongoing star-formation as is the case for the rest of the sample, the large H$\delta$ equivalent widths could arise from significant dust extinction or a large population of older stars created in a massive burst around 500 Myr or so ago. The former explanation is more likely given that these galaxies are currently observed in the ULIRG phase. 

We find that the e(a) galaxies are generally brighter and bluer than the rest of the sample. A plausible interpretation is that these systems are seen at a slightly later evolutionary stage on their path to becoming post-starbursts where the stars have had time to diffuse out and the dust has cleared revealing a bluer burst population. We also find that degrading the spectra of the e(a) galaxies to a S/N of 1.7 (typical for the individual spectra in our sample), still allows for detection of the H$\delta$ absorption feature confirming that this feature is only present in a subset of our sample. The numbers in our sample are too small to be able to draw any definite conclusions but likely reveal an interesting sub-population in SMG/SFRG selection that merits further study.

\begin{figure}
\begin{center}
\centering
\begin{tabular}{c}
\includegraphics[width=8.5cm,height=6.0cm,angle=0]{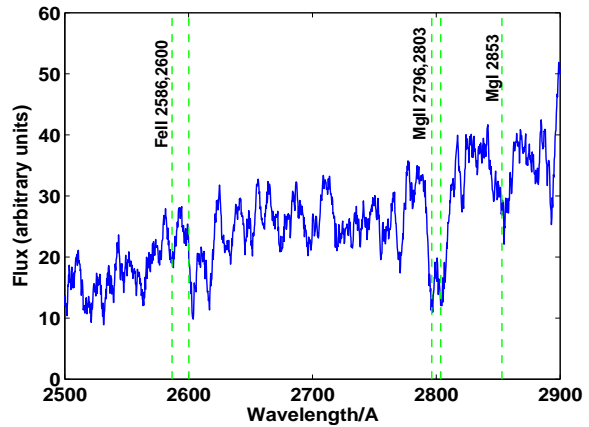} \\
\end{tabular}
\caption{ISM region in the highest S/N spectrum in our sample - c03.14 where the MgII doublet appears to be at the systemic redshift of the galaxy indicating that it is not associated with outflowing gas in this case. }
\label{fig:c14}
\end{center}
\end{figure}

The galaxy c03.14, one of the four e(a) dusty starbursts, is also the highest S/N spectrum in our sample and the MgII 2800 doublet is robustly detected in the spectrum as seen in Figure \ref{fig:c14}. The average MgII velocity relative to [OII] is found to be $-$20$\pm$60 kms$^{-1}$ consistent with these features originating at the systemic redshift of the galaxy which has a redshift error of 23kms$^{-1}$. We conclude that MgII here is not associated with any outflowing gas. We have already noted that the rest-UV spectra of e(a) galaxies such as c03.14 are likely to show signatures of both actively star-forming regions which suffer from heavy dust extinction, as well as older more diffuse stellar populations that are less obscured. The latter are predominantly composed of A, F and G stars. The stellar photospheres of these stars typically also show MgII absorption so the absorption features we are detecting in this galaxy may have a circumstellar origin. However, the equivalent width of MgII absorption is large (typically $\sim$1.5\AA\@) hinting at an interstellar rather than circumstellar origin. We discuss this point in more detail in the following section. 

\subsection{Outflows in Composite Spectra}

\label{sec:outflows}

The composite spectrum of SMGs and SFRGs at z$\sim$1.3 (Figure \ref{fig:all}) has numerous absorption features associated with the ISM in these galaxies. These features are now used to look for signatures of outflowing gas in these systems. However, before the absorption lines can be used to measure outflows, we need to convince ourselves that these spectral features are indeed associated with the ISM in the galaxies rather than the stellar photospheres of an old stellar population that is less obscured by dust in the optical/UV than the more recent burst. Interstellar lines can be distinguished from circumstellar lines as they are often seen to be asymmetric and highly saturated. As a consequence, they have large equivalent widths dominated by the velocity dispersion of the gas rather than its column density as is the case for circumstellar absorption \citep{Gonzalez:98}. Furthermore, MgII absorption originating in A and F stars is rarely blueshifted relative to systemic. In the rarest and brightest such stars, the maximum blueshift seen is $\sim$$-$100 kms$^{-1}$ \citep{Snow:94}. In order to remove any contaminating circumstellar absorption, we therefore only measure outflow velocities from pixels at $<-$100 kms$^{-1}$ relative to systemic.  

We start by excluding the four e(a) galaxies with obvious spectral signatures of $>$100Myr old stellar populations, from our composite spectrum in order to quantify the effect they have on the average velocities of the absorption lines. The galaxies for which the [OII] line fit was noisy making the systemic redshift estimates more uncertain as well as those where the [OII] line was very broad due to proximity to night-sky residuals or the presence of multiple components, are also excluded from the stack and have been flagged in Table \ref{tab:deimos1}. This leaves us with a total of 26 SMGs and SFRGs for our composite. 

\begin{figure*}
\begin{center}
\centering
\begin{tabular}{cc}
\includegraphics[width=8.5cm,height=6.0cm,angle=0]{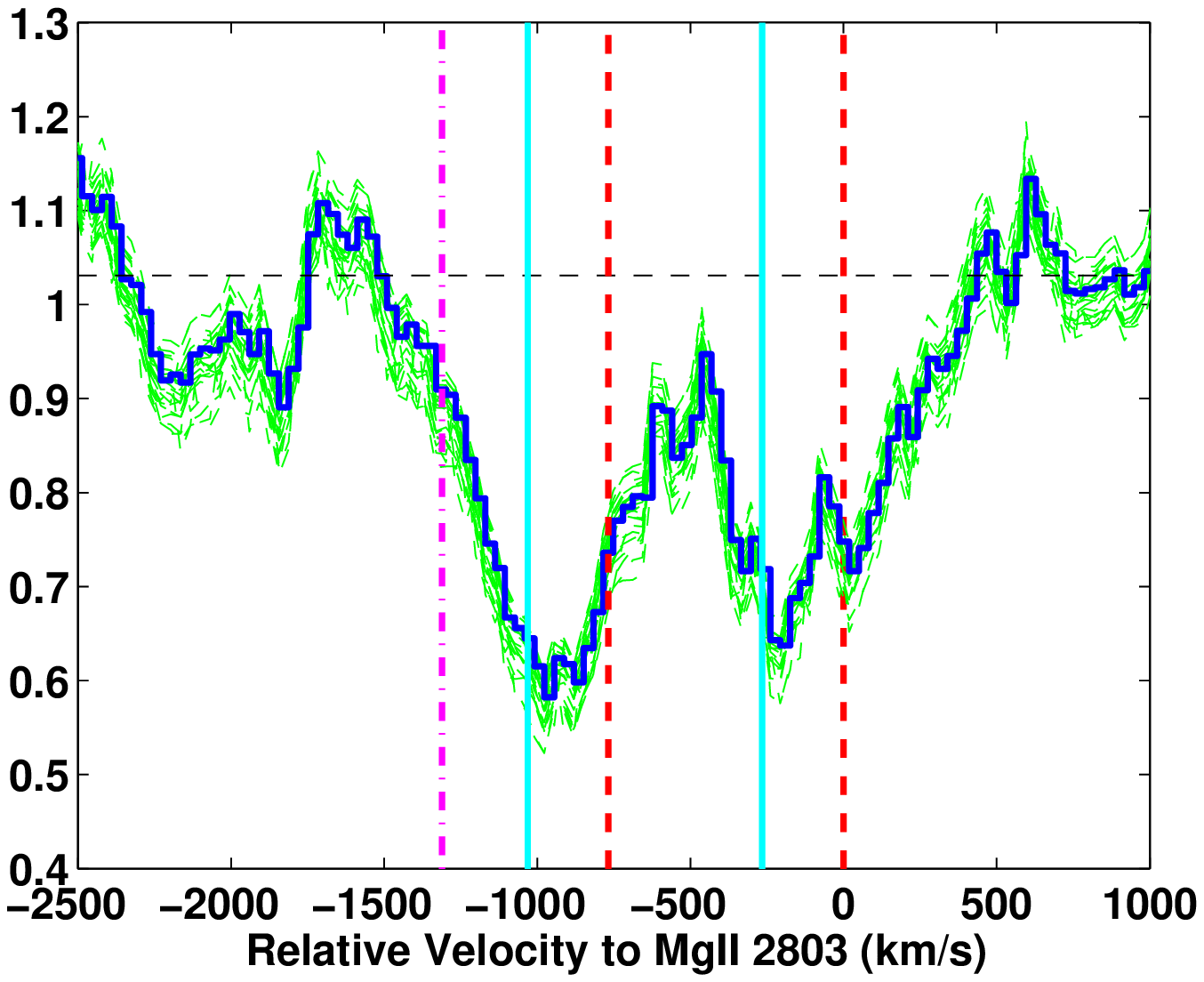} & \includegraphics[width=8.5cm,height=6.0cm,angle=0]{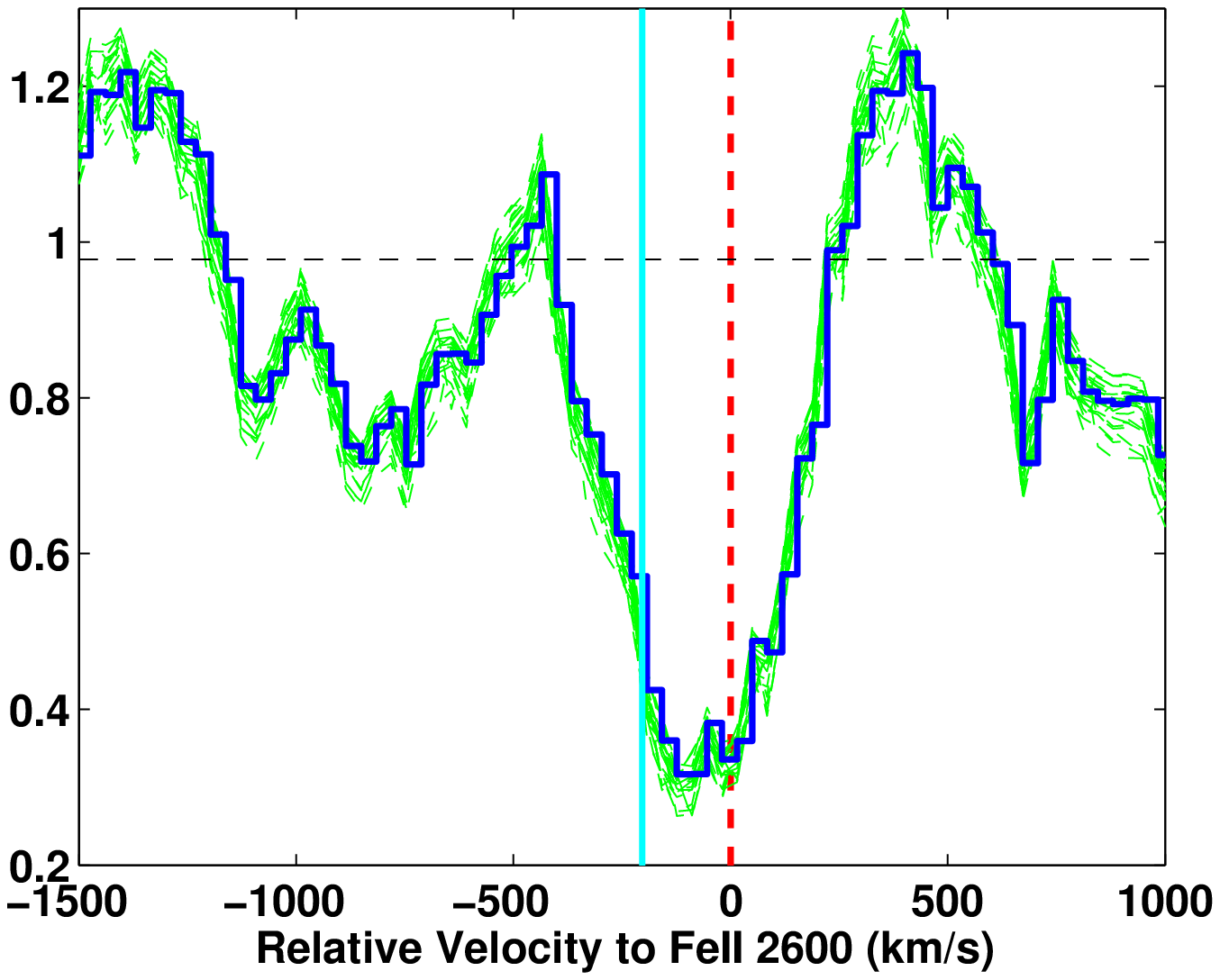} \\
\end{tabular}
\caption{Zoom-in around the MgII 2800 doublet (left) as well as the FeII 2600 absorption feature (right) for our composite spectrum of 26 SMGs and SFRGs with reliable [OII] redshifts and excluding the four dusty starbursts. The absorption line regions have been additionally smoothed for presentation purposes. The thick dark line is the composite spectrum while the lighter lines show the errors in the absorption line profile from the different jackknife samples. The dashed vertical lines mark the expected velocities of the various absorption features were they at the systemic redshifts of the galaxies. These features include MgII 2796.35 \AA\@ ($-$770 kms$^{-1}$ relative to MgII 2803), MgII 2803.53 \AA\@ (0 kms$^{-1}$ in the left plot) and FeII 2600.17 \AA\@ (0 kms$^{-1}$ in the right plot). The solid vertical lines indicate the actual centroid velocities of the lines measured between $-$500 and $-$100 kms$^{-1}$ using the AOD formalism. The dot-dashed vertical line in the left panel shows the terminal velocity of MgII 2796 measured by applying the AOD formalism to pixels in the absorption line profile that are $>$90\% of the continuum value. No terminal velocity was measured for FeII 2600 as the blue-wing of this line runs into a weaker MnII absorption feature at 2594.5\AA\@ making it difficult to determine where the line meets the continuum. A definite blueshift is seen in all absorption line velocities relative to systemic indicating the presence of outflowing gas in these galaxies.}

\label{fig:outflow}
\end{center}
\end{figure*}

In Figure \ref{fig:outflow}, we show a zoom-in around the absorption features for the composite spectrum of this sample of 26 galaxies in two windows around 2600 and 2800 \AA\@. The interstellar absorption lines are seen to have asymmetric tails that are blueshifted in velocity relative to the systemic [OII]. We interpret this average blueshift as indicating the presence of outflowing cold interstellar gas in these systems. 

Outflow velocities and equivalent widths are measured using the AOD formalism as described in $\S$ \ref{sec:deimos}, over a velocity window between $-$500 and $-$100 kms$^{-1}$ for all MgII and FeII lines seen in the composite spectrum. While there could in principle still be a contribution to these pixels from the blue wing of a systemic component, this contribution is expected to be small at v $<-100$ kms$^{-1}$. The S/N in our composite spectrum is not sufficient to be able to subtract this systemic component from the line profiles. Any pixels that lie above the continuum are discarded in order to exclude any strong emission from the line profiles. The limiting S/N however, does not allow us to effectively remove any weak emission that may be superimposed on an absorption line wing. The velocities and equivalent widths for MgII and FeII derived from our composite spectrum of 26 SMGs and SFRGs are presented in Table \ref{tab:velocities}. The velocities are the mean and RMS values derived by looking at the distribution in velocities of all detected lines of a particular ion. The errors on the equivalent widths are obtained from jackknife sampling the spectra. 

We have run a number of tests to check for systematic errors in our velocity and equivalent width determinations due to noise in the spectra. We find that applying the AOD method to randomly chosen windows in the continuum around the various absorption features, results in apparent absorption features being detected with equivalent widths of between 0.1 and 0.5 \AA\@ in the range $-$500 to $-$100 kms$^{-1}$. We therefore only consider lines with  $W_o > 0.7$\AA\@  as detections and flag lines with $W_o < 1.0$\AA\@ as marginal detections in Table \ref{tab:velocities}. The velocities and equivalent widths in Table \ref{tab:velocities} were derived by considering all pixels in the absorption lines in a fixed velocity window. We check that if we instead let the velocity window vary such that all pixels that are between 10\% and 90\% of the continuum value are used to characterise the line, the velocities and widths remain unchanged within the errors. This gives us confidence in the outflow velocity measurements presented despite the significant noise in the continua of the stacked spectra. We also measure terminal velocities from the MgII 2796 absorption feature by applying the AOD formalism to all pixels in the absorption line profile that are within 90\% of the continuum value. These terminal velocities are also presented in Table \ref{tab:velocities}. The errors are very large in some cases where the overall S/N of the composite spectrum is low making it difficult to determine where the line meets the continuum.   

For our 26 SMGs and SFRGs excluding the e(a) galaxies, we find a mean FeII velocity of $-$220$\pm$40 kms$^{-1}$ and a mean MgII velocity of $-$260$\pm$50 kms$^{-1}$. Both the MgII and FeII ions correspond to the same ionisation state and should therefore trace gas with similar physical conditions and therefore velocities. The velocities are consistent with each other as expected and we derive a mean ISM velocity of $-$240$\pm$50 kms$^{-1}$. The terminal velocity is around $-$500$\pm$100 kms$^{-1}$. Adding the four e(a) galaxies to the composite leaves the velocities and equivalent widths unchanged within the errors.


In Table \ref{tab:velocities}, we also present the equivalent width ratio between MgII 2796 and 2803\AA\@. In the case of optically thin absorption, we expect this ratio to be 2. The observed ratio is found to be 1.04$\pm^{0.11}_{0.16}$ suggesting that the lines are saturated and the absorption is optically thick. We note that the AOD formalism used to derive line centroids and widths implicitly assumes that the absorption material covers the source completely and homogenously. For optically thick absorption, the equivalent widths derived from a single parameter fit to the line profiles, do not allow us to disentangle the effects of optical depth and covering fraction. This requires multi-parameter fits to the absorption line profiles which, due to the limiting S/N of the spectra, is beyond the scope of this work. 

\begin{figure}
\begin{center}
\includegraphics[width=8.5cm,height=6.0cm,angle=0]{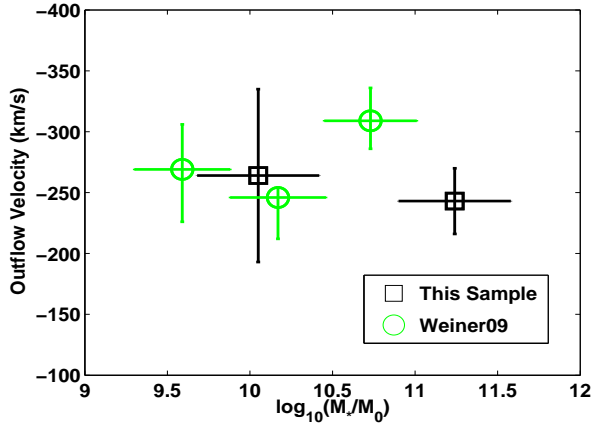}
\caption{Outflow velocities as a function of stellar mass for our SMGs and SFRGs compared to optically-selected star-forming galaxies from \citet{Weiner:09}. The luminous radio galaxy velocities are consistent with the optically-selected sample.}
\label{fig:massv}
\end{center}
\end{figure}

Having quantified the average outflow velocity in all SMGs and SFRGs, we now split the galaxies into various sub-samples in order to look at the dependence of outflow velocity on galaxy properties. These sub-samples are summarised in Table \ref{tab:sn} and the velocities and equivalent widths derived from them are given in Table \ref{tab:velocities}. As can be seen in Table \ref{tab:velocities}, there is no significant difference in outflow velocities between SMGs and SFRGs and we therefore treat the two populations as a single uniform sample in the remaining analysis. Similarly, there is no evidence for galaxies classified as AGN driving higher velocity winds. There is also no trend seen between the outflow velocities and the galaxy mass. Figure \ref{fig:massv} illustrates that the MgII velocities of both our stellar mass sub-samples are consistent with those derived by \citet{Weiner:09} for optically-selected galaxies at similar redshifts. Given the large systematic uncertainties in both the stellar and dynamical mass estimates of these galaxies ($\S$ \ref{sec:stellarmass} \& $\S$ \ref{sec:OII}), this lack of correlation is perhaps not surprising. 

We do however see some evidence for galaxies with high far infra-red luminosities (derived from the radio fluxes) as well as those with high [NeIII]/[OII] line ratios powering higher velocity winds. These trends are now explored in more detail. 

\subsubsection{Dependence on Infra-red Luminosity}

In order to study the dependence of outflow velocity on infra-red luminosity, we split our SMGs and SFRGs into two subsamples with L$_{\rm{IR}}<$0.68$\times$10$^{12}$L$_\odot$ and L$_{\rm{IR}}>$1.09$\times$10$^{12}$L$_\odot$ so there are roughly equal numbers of galaxies in each sample. The composite spectra are then generated as discussed in $\S$ \ref{sec:composite}. These composites are shown in Appendix A and the derived velocities and equivalent widths are summarised in Table \ref{tab:velocities}. 

Various absorption features are detected in both stacks corresponding to transitions of FeII and MgII. The average MgII velocity in the high-L$_{\rm{IR}}$ stack is found to be $-$300$\pm$60 kms$^{-1}$. The lower L$_{\rm{IR}}$ composite on the other hand has an average MgII velocity of $-$220$\pm$50 kms$^{-1}$. We use the MgII 2796 feature to measure the terminal velocities in both these subsamples by only considering pixels between 90\% and 100\% of the continuum value in the AOD formalism. This is measured to be $-$1000$\pm$200 kms$^{-1}$ in the high L$_{\rm{IR}}$ composite compared to $-$600$\pm$200 kms$^{-1}$ in the low-L$_{\rm{IR}}$ composite. 

A high far-infrared luminosity implies high star-formation rates assuming the far-IR radio correlation holds. However, this relation also has considerable scatter for example due to radio-loud AGN which can contribute significantly to the total radio flux even in the absence of significant star-formation. The slight increase in outflow velocity with total far infra-red luminosity seen in our sample, therefore has two possible interpretations. It could signal that galaxies with higher star-formation rates power higher velocity winds as already found in local samples \citep{Martin:05} as well as optically-selected galaxies at z$\sim$1.5 \citep{Weiner:09}. Alternatively, it could provide evidence for radio-loud AGN powering higher velocity winds as seen in \citet{Morganti:05, Alatalo:11} and \citet{Lehnert:11}. 

Two of our high-L$_{\rm{IR}}$ galaxies, lock332 with a radio flux of 720$\mu$Jy and sxdf05 with a radio flux of 526$\mu$Jy clearly show evidence for the presence of powerful AGN. The submm counterpart of lock332 is coincident with one lobe of a compact FRII galaxy and both galaxies show strong X-ray emission. The remainder of the high-L$_{\rm{IR}}$ galaxies however do not show obvious AGN signatures in the radio. We remove the two radio-loud AGN from our high-L$_{\rm{IR}}$ composite but find the outflow velocities to remain unchanged within the errors. We therefore conclude that these galaxies do not contribute significantly to the higher outflow velocities in the high-L$_{\rm{IR}}$ composite and that this signature is therefore primarily dominated by differences in star-formation rates between the two sub-samples. 

The centroid velocities of MgII for both sub-samples are consistent with the MgII velocities derived by \citet{Weiner:09} for galaxies at the same redshifts with considerably lower star-formation rates. We discuss this trend further and compare to other samples of galaxies in $\S$ \ref{sec:discSFR}. 

\begin{figure}
\begin{center}
\centering
\begin{tabular}{c}
\includegraphics[width=8.5cm,height=6.0cm,angle=0]{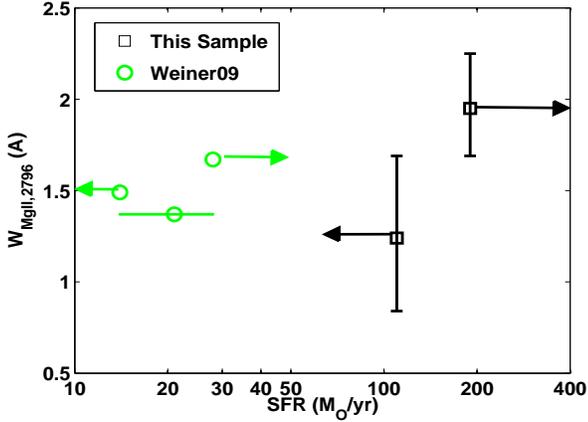} \\
\end{tabular}
\caption{star-formation rate versus equivalent width of MgII 2796\AA\@ for our sample of luminous radio galaxies as well as optically-selected galaxies at similar redshifts from \citet{Weiner:09}. The equivalent widths for the two samples are consistent with one another within the errors but show a slight increase at high star-formation rates.}
\label{fig:ew}
\end{center}
\end{figure}

\citet{Weiner:09} also find that the equivalent width of MgII absorption increases with the star-formation rate with higher SFR galaxies having equivalent widths that are $\sim$30\% higher than lower SFR galaxies. Having interpreted the high-L$_{\rm{IR}}$ galaxies as having high star-formation rates, in Figure \ref{fig:ew} we plot the equivalent widths versus star-formation rate for both our sample and that of \citet{Weiner:09}. Again, our results are consistent with the optically-selected galaxies.     

\subsubsection{Dependence on [NeIII]/[OII] Line Ratio}

Alaghband-Zadeh et al. (in preparation) study the mass-metallicity relation for our sample of galaxies by considering the [NeIII]/[OII] line ratio as a metallicity indicator. However, we have already discussed in $\S$ \ref{sec:agn} how a high flux ratio in these lines could also arise due to the presence of a strong AGN. In this section we divide our sample into two subsets with  12+log(O/H)$>$8.4 and 12+log(O/H)$<$8.2. Due to the ambiguity regarding whether a [NeIII]/[OII] selection is in fact a metallicity cut or simply a different AGN selection to that in $\S$ \ref{sec:agn}, we refer to the two sub-samples as the \textit{Low [NeIII]/[OII]} and \textit{High [NeIII]/[OII]} subsamples

\begin{figure}
\begin{center}
\centering
\begin{tabular}{c}
\includegraphics[width=8.5cm,height=6.0cm,angle=0]{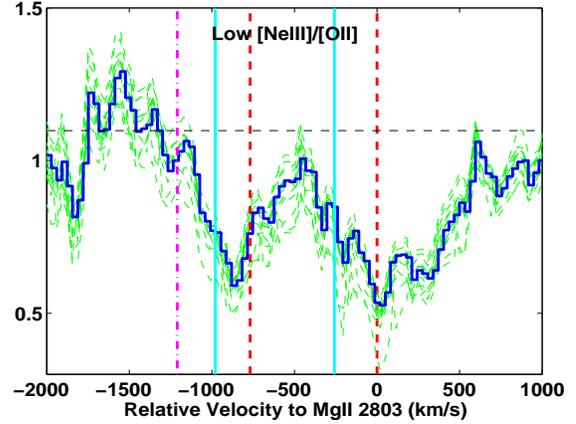} \\
\includegraphics[width=8.5cm,height=6.0cm,angle=0]{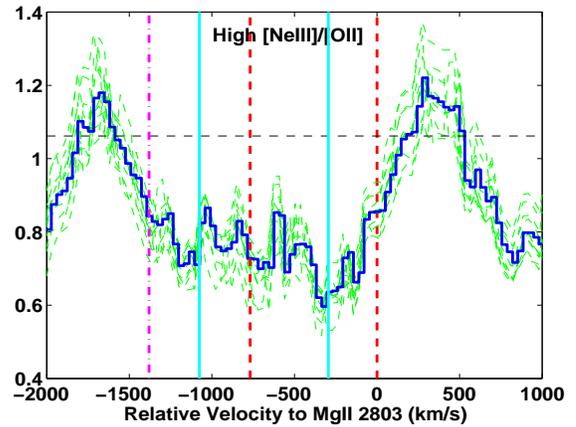} \\
\end{tabular}
\caption{Zoom-in around the MgII absorption feature for both low [NeIII]/[OII] (top) and high [NeIII]/[OII] (bottom) galaxies. The thick dark line is the composite spectrum while the lighter lines show the errors in the absorption line profile from the different jackknife samples. The dashed vertical lines denote the expected position of MgII absorption while the solid vertical lines show the actual measured centroid and the dot-dashed vertical line shows the terminal velocity of MgII 2796. The high [NeIII]/[OII] sample shows a blueshifted tail in MgII absorption extending to velocities as high as $\sim$$-$800 kms$^{-1}$. This is clearly not seen in the low [NeIII]/[OII] sample suggesting that the wind velocity is correlated with the [NeIII]/[OII] line ratio.}
\label{fig:metalstack}
\end{center}
\end{figure} 

We show the MgII doublet in both these subsamples in Figure \ref{fig:metalstack}. Although the line detections aren't highly significant given the limiting S/N of the spectra, the MgII 2800 doublet appears to be separated in the low [NeIII]/[OII] stack while the feature is considerably blended in the high [NeIII]/[OII] stack indicating a range of velocities extending to a tail of $-$800 $\pm$200 kms$^{-1}$. We conclude that galaxies with high [NeIII]/[OII] line ratios power higher velocity winds. There are two possible interpretations for this. Either galaxies with intrinsically lower metallicities drive faster outflows or the [NeIII]/[OII] line ratio allows us to select the most powerful AGN which correspondingly propel stronger winds. 

The first interpretation is supported by the fact that the terminal velocity in the high [NeIII]/[OII] stack is consistent with the $-$700 kms$^{-1}$ terminal velocity measured for the low-metallicity LBG Q2343-BX418 \citep{Erb:10} with a metallicity of 12+log(O/H)=7.9$\pm$0.2. The average metallicity derived from [NeIII]/[OII] for the high [NeIII]/[OII] sample is similar - 12+log(O/H)=7.6$\pm$0.5. However, in the absence of independent metallicity estimates, the current sample does not allow us to discriminate between AGN or metallicity as the main driver for differences in outflow velocities as a function of [NeIII]/[OII]. 

The mean circular velocity of both the high and low [NeIII]/[OII] galaxies in our sample is 80$\pm$30 kms$^{-1}$. The escape velocity is typically $\sim$3 times the circular velocity for reasonable assumptions about the halo mass distribution \citep{Binney:87}. This implies escape velocities of $\sim$240$\pm$90 kms$^{-1}$ for these galaxies. While the FeII and MgII centroid velocities for both low and high [NeIII]/[OII] galaxies are within the 1$\sigma$ error in this escape velocity, the terminal velocity in the high [NeIII]/[OII] stack is almost a factor of 3 larger than the escape velocity. This suggests that at least some of the gas is able to escape these galaxies into the intergalactic medium.

\section{DISCUSSION}

\label{sec:disc}

\subsection{Redshift Evolution of SMG/SFRG Properties}

One of the main aims of our DEIMOS study was to fill in the redshift desert in
spectroscopic samples of SMGs and SFRGs at $z=1.2$--1.7. Previous studies of
these populations have typically used blue-optimised spectrographs focussed on
galaxies at $z\sim 2$ \citep{Chapman:05}. We have seen that the higher redshift 
population has higher bolometric luminosities. While this partly arises due to 
the radio selection of the spectroscopic targets whereby higher redshift sources 
require higher radio luminosities to be detectable, the additional 850$\mu$m selection 
of SMG targets is less sensitive to redshift due to the negative k-correction in the 
submm. Despite their similar 850$\mu$m fluxes, we find that SMGs at z$\sim$1.5 are an order of magnitude less 
luminous than at z$\sim$2.2 suggesting some luminosity evolution between these redshifts. 
The clear dearth of SMGs and SFRGs in our sample with $\gtrsim$1000\,M$_\odot$yr$^{-1}$ star
formation rates suggests that there is considerable evolution in the star-formation
rates of these galaxies between redshift 1--2 with the peak activity in the
population closer to $z\sim2$. However, the star-formation rates at $z\sim1.3$
are still high enough for these galaxies to be classed as ULIRGs. 

In contrast to the strong evolution in the typical star-formation rates of SMGs
and SFRGs, we have seen in $\S$\ref{sec:OII} that the nebular linewidths and
therefore dynamical masses of SMGs and SFRGs at $z\sim1.3$ are virtually
indistinguishable from those at $z\sim2$. We conclude therefore that the masses of submm galaxies
show little evolution between redshifts 1--2.  However, we find that at $z\sim
1.3$ it appears that most of the baryons in the central regions of these
systems are likely to be in the form of stars given the very large stellar masses. This is in contrast to the situation at
somewhat higher redshifts where the cold gas component is more significant. \citet{Tacconi:08}
for example find a cold-gas fraction in z$\sim$2 SMGs of upto 50\%.  

Comparing the star-formation rates as well as dynamical masses of our SMG/SFRG
population to optically-selected star-forming galaxies at similar redshifts
clearly demonstrates that we are probing a much more massive, more
intrinsically luminous population than the optical samples. Many independent
lines of evidence now point to these SMGs and SFRGs being the most likely
progenitors of present-day massive elliptical galaxies at $\sim\, L_{\rm
K}^\ast$--3$\,L_{\rm
K}^\ast$ (M$_{\rm K} \le -24$; \citet{Swinbank:06}). Recent studies
have shown that such massive ellipticals have stellar masses of
$>10^{12}$\,M$_\odot$ already assembled by $z\sim0.7$ (e.g. \citet{Banerji:10}), an order 
of magnitude larger than the typical masses seen in our sample. 

\subsection{Outflow Velocities and Star-Formation Rate}

\label{sec:discSFR}
 
Outflow velocities are generally found to scale with the star-formation rate such that V$_{\rm{wind}} \propto$ SFR$^{0.3}$. This relationship which has been well determined in the local Universe \citep{Martin:05} is also found to hold for populations of normal star-forming galaxies at high redshift \citep{Weiner:09}. Given that the SMGs and SFRGs studied in this work represent some of the most extreme star-forming systems at high redshifts, it is interesting to compare their outflow velocities to those in other systems both in the local Universe and at high redshifts in the context of this V$_{\rm{wind}}$-SFR relation. 

\begin{figure*}
\begin{center}
\begin{minipage}[c]{1.00\textwidth}
\centering
\includegraphics[width=15cm,height=11.5cm,angle=0] {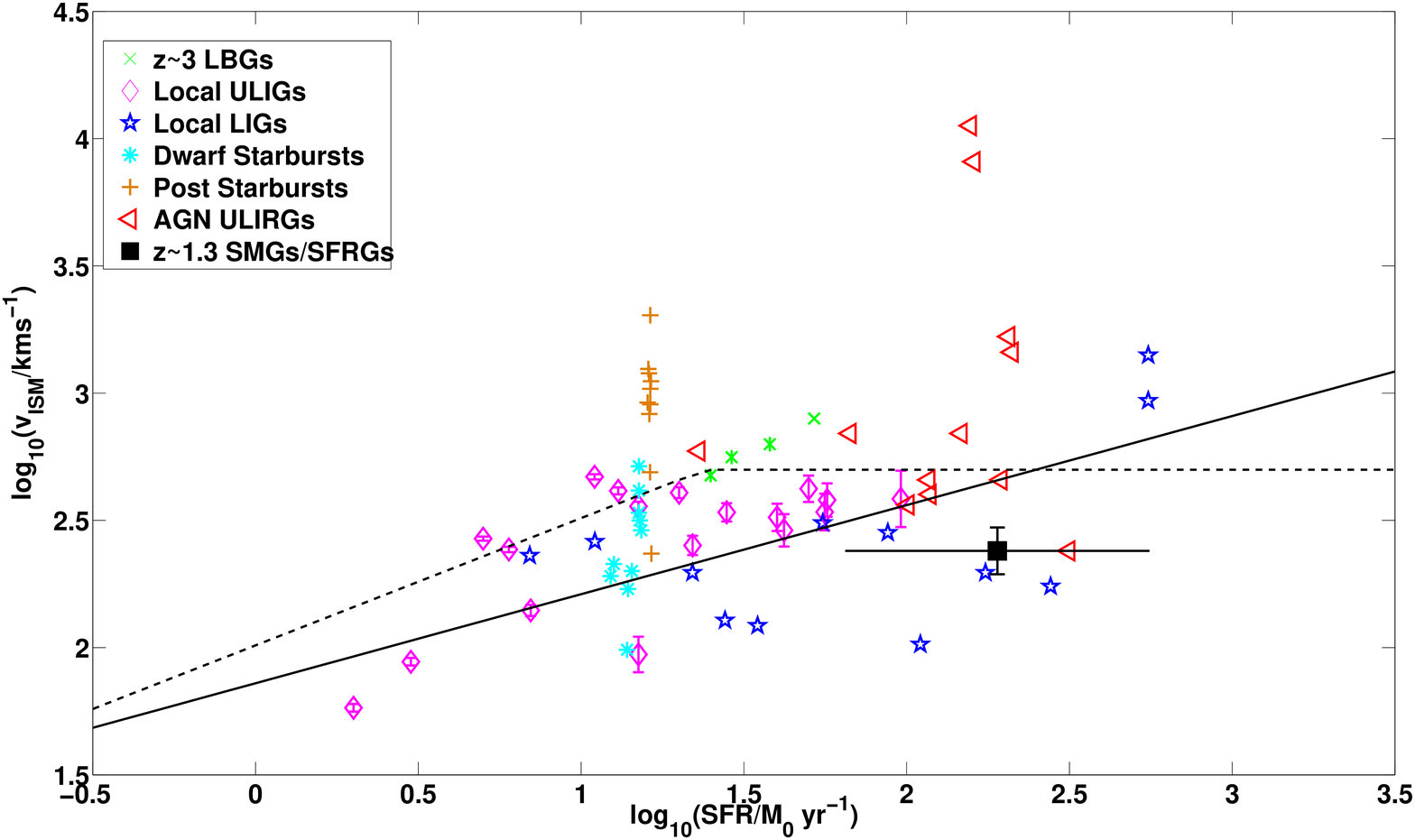}
\end{minipage}
\caption{Variation of outflow velocity as a function of the star-formation rate for several different samples of galaxies. The solid line shows the local relation from \citet{Martin:05} after correcting for projection effects. The dashed line shows the velocities expected if the cold gas is accelerated by momentum-driven winds to the velocity of the hot gas component taken to be 500 kms$^{-1}$. Our sample of SMGs and SFRGs is shown as the dark square. The SMG/SFRG average outflow velocity lies within the local envelope and is consistent with the momentum-driven wind model. The average outflow velocity of our redshift 1.3 SMGs and SFRGs is also indistinguisable from those of local LIRGs and ULIRGs.}
\label{fig:SFRv}
\end{center}
\end{figure*}

In $\S$\ref{sec:outflows} we presented some evidence for galaxies with high-L$_{\rm{IR}}$ and those with high [NeIII]/[OII] powering higher velocity winds. However, as can be seen in Table \ref{tab:velocities}, the differences between these various sub-samples are significant only at $\sim$2$\sigma$ level. For the purposes of this discussion therefore, we only consider the sample average velocity of $-$240$\pm$50 kms$^{-1}$. In Figure \ref{fig:SFRv}, we plot this sample average and compare to various other samples of local as well as high redshift galaxies covering a large dynamic range in star-formation rate of $\sim$3 dex. These include the LBGs from \citet{Shapley:03}, local LIRGs \citep{Heckman:00} and ULIRGs \citep{Martin:05}, local dwarf starbursts \citep{Schwartz:06}, 'post-starbursts' in the SDSS \citep{Tremonti:07} and a sample of AGN-dominated ULIRGs \citep{Rupke:05}. In the case of the dwarf starbursts, the star-formation rate is crudely estimated from the B-band absolute magnitudes quoted in \citet{Schwartz:06}. In the case of the local LIRGs, the star-formation rate is calculated from the total infra-red luminosity from \citet{Heckman:00}. Figure \ref{fig:SFRv} also shows the V$_{\rm{wind}} \propto$ SFR$^{0.3}$ relation as determined by \citep{Martin:05} after correcting for line-of-sight effects i.e. the fact that face-on disks with bipolar flows will appear to have systematically higher velocities. This line should therefore trace the upper envelope of the V$_{\rm{wind}}$-SFR relation assuming there is no evolution in this relation with redshift.

The outflow velocities in our sample are found to be consistent with the local envelope. The median outflow velocities in the local LIRG and ULIRG populations are $-$330$\pm$10 and $-$210$\pm$20 kms$^{-1}$ respectively. Within the errors in our outflow velocity measurements, the redshift 1-2 SMGs and SFRGs are therefore indistinguishable in terms of wind velocity from these local galaxy samples. We also note that the median velocities for local LIRGs and ULIRGs are derived from only those galaxies showing evidence for starburst driven winds. While this is certainly the norm, there are also individual starburst galaxies in the local Universe which do not show blueshifted ISM absorption features. Our outflow velocities, as a consequence of being derived from composite rather than individual spectra, necessarily average over such \textit{no-wind} systems. While these systems would lower the measured equivalent width of the absorption lines, they should not affect the centroid velocities between $-$100 and $-$500 kms$^{-1}$.   

In the local Universe, the velocities of cold interstellar gas are found to scale with the star-formation rate of the galaxy but the hot-wind traced by X-rays is remarkably uniform in temperature \citep{Martin:99,Heckman:00}. One hypothesis is that the cool gas in lower luminosity starbursts is not accelerated to the hot-wind velocities as there is not sufficient momentum for this to happen. In a wind model dominated by momentum injection from supernovae or AGN, the cold gas component traced by the interstellar absorption features is entrained in the hot wind and therefore accelerated to the hot wind velocity. The velocity of the cold gas component, $V_{cold}$ is given by: 

\begin{eqnarray}\nonumber
\frac{V_{\rm{cold}}}{V_{\rm{hot}}}&=&\left(\frac{SFR}{0.24 M_\odot yr^{-1}}\right)^{0.5} \\
& & \left[\left(\frac{10^{20}cm^{-2}}{N_H}\right)\left(\frac{200 pc}{R_0}\right)\left(\frac{500 kms^{-1}}{V_{\rm{hot}}}\right)\right]^{0.5}
\label{eq:SNe}
\end{eqnarray}

\noindent where $R_0$ is the launch radius, $N_H$ is the column density and $V_{\rm{hot}}$ is the velocity of the hot wind component \citep{Martin:05,Murray:05}. We assume the same upper envelope in the $V_{\rm{cold}}-SFR$ relation as fit by \citet{Martin:05} in which $(N_H/10^{22}cm^{-2})(R_0/200 pc) \sim 1$. Figure \ref{fig:SFRv} shows this upper envelope. The hot wind velocity is taken to be a constant 500 kms$^{-1}$ from X-ray measurements \citep{Martin:99, Heckman:00}. Once the cloud reaches this velocity the ram pressure ceases to accelerate the cold gas any further resulting in a plateau in the outflow velocities at high star-formation rates. 

Our composite spectra, do not rule out this plateau. In fact, the outflow velocities in our sample are very similar to those measured in optically-selected galaxies at the same redshifts with considerably lower star-formation rates \citep{Weiner:09}. This suggests perhaps that a plateau has already been reached within the parameter space spanned by our z$\sim$1.3 SMGs and SFRGs. A larger population of high SFR galaxies with robust wind measurements is clearly needed to test this hypothesis fully. 
  
\section{CONCLUSIONS}

\label{sec:conclusion}

In this paper, we have presented a spectroscopic catalogue of SMGs and SFRGs at a median redshift of 1.3 that helps fill in the \textit{redshift desert} in previous studies of these populations. Redshifts are predominantly measured from [OII] 3727 emission seen in the majority of the spectra and corroborated by various other features corresponding to transitions of FeII, MgII, [NeIII], [NeV], [OIII] and the Balmer series. We derive star-formation rates from the radio fluxes, stellar masses from a full SED fit and dynamical masses from the [OII] linewidths for these galaxies. In particular we find that: 

\begin{itemize}

\item{SMGs and SFRGs at redshifts of $\sim$1.3 have lower star-formation rates than those at z$\sim$2 but are still observed to be in the U/LIRG phase.} 

\item{The stellar and dynamical masses of these galaxies show little evolution between redshifts 1--2 but are considerably larger than those of optically-selected star-forming galaxies at similar redshifts. This lends further support to the premise that these dust enshrouded galaxies are the progenitors of modern day ellipticals.}

\item{While the majority of our galaxies have typical spectra for star-forming galaxies, we find significant [NeIII] and [NeV] emission in just under half of the sample indicating the presence of an AGN. In addition, four galaxies show strong H$\delta$ absorption in addition to [OII] emission indicative of a very dusty starburst.}

\end{itemize}

The [OII] line allows us to define the systemic redshift of the galaxies assuming the warm ionized gas from which this emission originates is associated with young stars. We look for shifts in interstellar absorption lines relative to [OII] in order to find evidence for large-scale outflows. While most of the individual spectra do not have sufficient S/N for outflow measurements, an analysis of composite spectra yields the following conclusions: 

\begin{itemize}

\item{Interstellar absorption lines associated with MgII and FeII are found to be at an average blueshift of $-$240$\pm$50 kms$^{-1}$ relative to [OII] in SMGs and SFRGs at z$\sim$1.3.}

\item{SMGs and SFRGs with similar bolometric luminosities are found to drive winds of similar velocities indicating that feedback processes are likely to be comparable in the two populations.}

\item{The outflow velocities in our z$\sim$1.3 sample are consistent with the V $\propto$ SFR$^{0.3}$ relation seen in local ULIRGs and optically-selected star-forming galaxies at redshifts around 1.5.}

\item{The average outflow velocities of our SMGs and SFRGs are well explained by a momentum-driven wind model.}

\end{itemize}

We conclude that SMGs and SFRGs at z$\sim$1.3 allow us to probe a slightly more evolved population than seen at z$\sim$2. At these redshifts, the galaxies show evidence for cool interstellar outflows as traced by rest-UV absorption lines. Their outflow velocities are found to be similar to those seen in optically-selected star-forming galaxies at similar redshifts (which have star-formation rates that are an order of magnitude lower) as well as local LIRGs and ULIRGs (with similar star-formation rates). The observations are, however, limited by the relatively poor S/N of individual spectra in the optical/UV as SMGs and SFRGs are highly obscured at these wavelengths. They are however, exceptionally bright in the millimeter wavelengths that trace molecular gas. Detailed hydrodynamic simulations have suggested that galactic winds can entrain large quantities of molecular gas \citep{Narayanan:08} and this has direct observational consequences for CO morphologies and emission line profiles. There have already been successful attempts at mapping outflows of molecular gas in local ULIRGs \citep{Sakamoto:09} and the advent of ALMA will make similar studies possible at high redshifts. In the ALMA era, rest-UV spectroscopic studies of outflows in SMGs and SFRGs will serve as an important benchmark for the millimeter observations allowing comparison of outflows of the low-ionisation gas with that of the molecular gas. 

\section*{Acknowledgements}

The authors thank the anonymous referee for many useful suggestions that have helped improve the paper. We thank James Allen, Paul Hewett, Richard McMahon, Max Pettini, Anna Quider and Vivienne Wild for useful discussions. MB, IRS and RJI acknowledge support from STFC. AMS acknowledges support from an STFC Advanced Fellowship. JSD acknowledges the support of the Royal Society via a Wolfson Research Merit Award, and the support of the European Research Council via the award of an Advanced Grant.  

\onecolumn
\begin{landscape}
\renewcommand{\thefootnote}{\alph{footnote}}
\renewcommand{\arraystretch}{0.6}
\small
\begin{longtable}{lccccccccccc}
\caption{Summary of DEIMOS Sample of SMGs (top) and SFRGs (bottom)}
\label{tab:deimos1} \\
     Source & RA & DEC & z & S$_{1.4}$ ($\mu$Jy) & S$_{850}$ (mJy) & R$_{AB}$ & $\sigma$([OII]) (kms$^{-1}$)\footnotemark[1] & 12+log(O/H) & Class \\
          \hline
      \hline
      \multicolumn{10}{c}{\textbf{SMGs}} \\ \hline
      sxdf28a & 02:18:06.92 & $-$04:59:12.7 & 1.114 & 94 $\pm$ 16 & 4.8$\pm^{2.2}_{2.7}$ & 22.54  & 190 $\pm$ 30 & $>$8.7  & AGN\footnotemark[2] \\  
      sxdf47a & 02:17:34.36 & $-$04:58:57.2 & 1.180 & 166 $\pm$ 14 & 3.0$\pm^{1.6}_{1.9}$ & 24.85 & 120 $\pm$ 20 & 8.1 $\pm$ 0.2 & SB \\  
      sxdf47b & 02:17:34.40 & $-$04:58:59.8 & 1.178 & 55 $\pm$ 14 & 3.0$\pm^{1.6}_{1.9}$ & 23.02 & 160 $\pm$ 10 &  $>$8.9 & SB \\  
      sxdf55 & 02:17:51.87 & $-$05:04:47.0 & 1.628\footnotemark[3] & 42 $\pm$ 14 & 3.9$\pm^{2.2}_{2.7}$ & --  & -- & -- & SB \\  
      sxdf05 & 02:18:02.86 & $-$05:00:30.9 & 1.087 & 526 $\pm$ 14 & 8.4$\pm^{1.7}_{1.9}$ & 24.50 & 380 $\pm$ 30\footnotemark[4] & $>$8.7 & AGN \\  
      sxdf35 & 02:18:00.87 & $-$04:53:05.7 & 1.254 & 66 $\pm$ 15 & 5.3$\pm^{1.8}_{2.1}$ & 22.93 & 200 $\pm$ 20 & $>$7.9 & AGN \\  
      sxdf28c & 02:18:06.42 & $-$04:59:20.1 & 1.627 & 74 $\pm$ 20 & 4.8$\pm^{2.2}_{2.7}$ & -- & 60 $\pm$ 10 & $>$8.2 & AGN\footnotemark[2] \\  
      sxdf12 & 02:17:59.29 & $-$05:05:04.0 & 1.627 & 39 $\pm$ 16 & 5.7$\pm^{1.7}_{1.8}$ & 24.03 & 90 $\pm$ 40 & $>$8.4 & SB \\  
      sxdf47c & 02:17:33.62 & $-$04:58:58.2 & 1.405 & 64 $\pm$ 13 & 3.0$\pm^{1.6}_{1.9}$ & 23.85 & 140 $\pm$ 30 & -- & AGN \\  
      sxdf88 & 02:18:01.49 & $-$05:04:43.7 & 1.518 & 40 $\pm$ 10 & 4.5$\pm^{2.1}_{2.5}$ & 24.80 & 250 $\pm$ 10\footnotemark[4] & -- & SB \\ 
      & & & & & & & & & & \\
      gn30  & 12:36:52.75 & +62:13:54.6 & 1.359 & 29 $\pm$ 8 & 1.8$\pm$0.5 & 22.7\footnotemark[5] & 120 $\pm$ 5 & 8.2 $\pm$ 0.1 & SB \\  
      gn34 & 12:37:06.22 & +62:21:11.6 & 1.363 & 22 $\pm$ 5 & 5.6$\pm$1.6 & 22.9\footnotemark[5] & 90 $\pm$ 10 & $>$ 9.0 & SB \\
      & & & & & & & & & & \\
      lock76 & 10:51:49.18 & +57:28:40.7 & 0.711 & 44 $\pm$ 9 & 4.7$\pm^{2.5}_{3.1}$ & 21.03 & 100 $\pm$ 20 & 6.7 $\pm$ 0.1 & AGN/e(a) \\  
      lock77 & 10:51:57.14 & +57:22:10.1 & 1.464\footnotemark[3] & 16 $\pm$ 4 & 3.2$\pm^{1.2}_{1.3}$ & 26.50 & -- & -- &  SB \\  
      lock73 & 10:51:42.16 & +57:22:18.1 & 1.423 & 31 $\pm$ 9 & 3.5$\pm^{1.9}_{2.3}$ & 23.63 & 80 $\pm$ 10 & $>$8.7 &  SB \\  
      lock24 & 10:52:00.53 & +57:20:40.5 & 1.415 & 27 $\pm$ 7 & 2.7$\pm$ 1.2 & 23.74 & -- & $>$8.1 &  SB \\  
      lock40 & 10:52:01.87 & +57:19:18.4 & 1.761 & 33 $\pm$ 13 & 3.0$\pm^{1.1}_{1.2}$ & 25.95 & -- & -- & SB \\  
      lock87 & 10:51:53.45 & +57:17:30.4 & 1.660\footnotemark[3] & 85 $\pm$ 5 & 3.4$\pm^{1.5}_{1.7}$ & 25.58 & -- & -- &  SB \\  
      lock38 & 10:53:07.19 & +57:24:31.4 & 1.523 & 56 $\pm$ 7 & 6.5$\pm$1.9 & 23.63 & 120$\pm$10 & -- & SB \\ 
      \hline
      \multicolumn{10}{c}{\textbf{SFRGs}} \\ \hline
      sxdf1293 & 02:18:26.74 & $-$04:57:39.9 & 1.356 & 31 $\pm$ 8 & $-$0.9$\pm$2.0 & 24.74 & 210 $\pm$ 20 & $>$8.8 & SB \\  
      sxdf490 & 02:17:31.26 & $-$04:59:23.3 & 1.583 & 30 $\pm$ 15 & $-$1.3$\pm$1.7 & 24.26 & 110 $\pm$ 10 & -- & SB \\  
      sxdf643 & 02:17:41.56 & $-$04:54:11.5 & 1.274 & 137 $\pm$14 & 0.4$\pm$1.9 & 23.56 & 90 $\pm$ 20 & 7.6 $\pm$ 0.2 & AGN \\  
      sxdf755 & 02:17:49.96 & $-$04:53:47.6 & 1.279 & 108 $\pm$ 15 & $-$0.1$\pm$1.8 & 21.88 & 120 $\pm$ 20 & 7.63 $\pm$ 0.08 &  AGN \\  
      sxdf639 & 02:17:41.35 & $-$04:56:34.6 & 1.341 & 67 $\pm$ 26 & 1.3$\pm$1.8 & 23.75 & 160 $\pm$ 20 & $>$8.6 & SB \\  
      sxdf1204 & 02:18:19.02 & $-$05:07:02.1 & 0.989 & 68 $\pm$ 20 & 0.4$\pm$1.9 & 23.93 & 130 $\pm$ 20 & $>$8.7 & SB \\  
      sxdf844 & 02:17:55.27 & $-$05:04:33.8 & 0.929 & 123 $\pm$ 15 & $-$1.9$\pm$2.0 & 23.42 & 100 $\pm$ 10 & 6.8 $\pm$ 0.2 & AGN \\  
      sxdf965 & 02:18:03.67 & $-$04:59:30.5 & 1.264 & 128 $\pm$ 16 & 1.3$\pm$2.1 & 23.63 & 190 $\pm$ 20 & 7.66 $\pm$ 0.07 & AGN \\  
      sxdf704 & 02:17:46.29 & $-$04:54:39.5 & 1.456\footnotemark[3] & 63 $\pm$ 14 & 2.0$\pm$1.9 & 24.86 & -- & -- & SB \\  
      sxdf1219 & 02:18:20.76 & $-$05:01:17.6 & 1.095 & 115 $\pm$ 25 & 1.1$\pm$1.9 & 22.45 & 140 $\pm$ 20 & 7.3 $\pm$ 0.1 & AGN/e(a) \\  
      sxdf1300 & 02:18:27.35 & $-$05:00:55.9 & 1.356 & 148 $\pm$ 21 & 1.3$\pm$1.8 & 21.30 & 130 $\pm$ 20 & $>$8.9 & SB \\  
      sxdf966 & 02:18:03.69 & $-$04:56:04.6 & 1.291 & 35 $\pm$ 14 & 0.2$\pm$1.7 & 25.05 & 190 $\pm$ 60 & $>$8.8 & SB \\
      & & & & & & & & & & \\
      lock332 & 10:51:37.14 & +57:29:41.8 & 1.143 & 720 $\pm$ 12\footnotemark[6] & $-$1.3$\pm$1.4 & 22.46 & 100 $\pm$ 10 & $>$8.6 & AGN \\  
      lock517 & 10:51:47.13 & +57:18:42.0 & 1.104 & 59 $\pm$ 7 & 1.5$\pm$1.9 & 22.79 & 110 $\pm$ 10 & 8.0 $\pm$ 0.2 & AGN/e(a) \\  
      lock218 & 10:51:50.59 & +57:27:51.1 & 1.341 & 25 $\pm$ 7 & $-$0.4$\pm$1.5 & 25.76 & 120 $\pm$ 50 & 7.5 $\pm$ 0.2 & AGN \\  
      lock492 & 10:51:57.86 & +57:19:40.2 & 1.377 & 30 $\pm$ 11 & 0.5$\pm$1.8 & 23.52 &  240 $\pm$ 10\footnotemark[4] & $>$9.0 & AGN \\  
      lock493 & 10:52:00.42 & +57:18:06.3 & 1.139 & 32 $\pm$ 7 & 1.1$\pm$1.7 & 23.14 & 170 $\pm$ 5 & 7.77 $\pm$ 0.03 & AGN \\
      & & & & & & & & & & \\
      c03.1420 & 03:02:36.28 & +00:10:57.2 & 1.318 & 40 $\pm$18 & 1.18$\pm$0.99 & 23.12\footnotemark[7] & 130 $\pm$ 50 & $>$8.7 & SB \\  
      c03.14 & 03:02:29.75 & +00:08:42.7 & 0.857 & 102 $\pm$ 11 & 0.82$\pm$1.05 & 21.37\footnotemark[7] & 130 $\pm$ 10 & $>$9.0 & AGN\footnotemark[2]/e(a) \\  
      c03.56 & 03:02:49.89 & +00:08:27.0 & 1.320 & 70 $\pm$ 24 & $-$0.63$\pm$1.09 & 23.02\footnotemark[7] & 90 $\pm$ 10 & $>$9.0 & SB \\  
      c03.62 & 03:02:45.99 & +00:08:40.9 & 1.497 & 42 $\pm$ 17 & 1.13$\pm$1.22 & 23.22\footnotemark[7] & 330$\pm$30\footnotemark[3] & $>$8.8 & SB \\
 \hline
\tiny
\footnotetext[1]{Best-fit linewidth including the instrumental profile} \footnotetext[2]{AGN Classification based on 8$\mu$m excess only}
\footnotetext[3]{[OII] line detected at $<=$2-$\sigma$}
\footnotetext[4]{Broad [OII] due to proximity to sky line residuals (sxdf05, sxdf88 and c03.62) or multiple components in 2D spectrum (lock492)}
\footnotetext[5]{i$_{775W}$ AB magnitude from Pope et al. (2006)}
\footnotetext[6]{One lobe of compact FRII}
\footnotetext[7]{I$_{\rm{AB}}$ magnitude as R-band image in this field suffers from poor background variability. Typical R$-$I is expected to be 0.4$\pm$0.3}
\end{longtable}
\twocolumn
\end{landscape}

\onecolumn
\begin{landscape}
\small
\begin{longtable}{lccccc}
\caption{Derived Properties for DEIMOS Sample of SMGs (top) and SFRGs (bottom)}
\label{tab:deimos2} \\
     Source & SFR/M$_\odot$ yr$^{-1}$ & (L$_{\rm{IR}}$/L$_\odot$)$\times 10^{12}$ & log$_{10}$(M$_{*}$/M$_\odot$) & $A_v$ & log$_{10}$(M$_{\rm{dyn}}$/M$_\odot$) \\
          \hline
      \hline
       \multicolumn{6}{c}{\textbf{SMGs}} \\ \hline
      sxdf28a & 170 $\pm$ 30 & 0.95 $\pm$ 0.16 & 11.15$\pm$0.01 & 4.4 & 10.88 $\pm$ 0.34 \\
      sxdf47a  & 340 $\pm$ 30 & 1.95 $\pm$ 0.16 & 10.89$\pm$0.01 & 0.8 & 10.41 $\pm$ 0.34  \\
      sxdf47b  & 110 $\pm$ 30 & 0.64 $\pm$ 0.16 & 10.37$\pm ^{0.10}_{0.03}$ & 2.4 & 10.70 $\pm$ 0.34  \\
      sxdf55  & 200 $\pm$ 70 & 1.14 $\pm$ 0.38 & --  & -- & -- \\
      sxdf05  & 870 $\pm$ 20 & 4.98 $\pm$ 0.13 & 10.28$\pm ^{0.04}_{0.01}$ & 3.8 & 11.48 $\pm$ 0.34  \\
      sxdf35  & 160 $\pm$ 40 & 0.91 $\pm$ 0.20 & 10.50$\pm$0.01 & 3.2 & 10.92 $\pm$ 0.34  \\
      sxdf28c  & 350 $\pm$ 90 & 2.02 $\pm$ 0.54 & 11.26$\pm ^{0.43}_{0.23}$ & 5.4  & 9.52 $\pm$ 0.34 \\
      sxdf12 & 190 $\pm$ 80 & 1.07 $\pm$ 0.45 & 10.43$\pm ^{0.03}_{0.02}$ & 3.4 & 10.17 $\pm$ 0.37 \\
      sxdf47c  & 210 $\pm$ 40 & 1.19 $\pm$ 0.24 & 10.70$\pm ^{0.03}_{0.35}$ & 1.8 & 10.58 $\pm$ 0.34 \\
      sxdf88 & 160 $\pm$ 40 & 0.93 $\pm$ 0.21 & 11.19$\pm$0.01 & 4.8 & 11.10 $\pm$ 0.34 \\
      gn30  &  80 $\pm$ 20 & 0.49 $\pm$ 0.14 & 9.58$\pm ^{0.22}_{0.02}$ & 1.4 & 10.43 $\pm$ 0.34 \\
      gn34 &   70 $\pm$ 15 & 0.38 $\pm$ 0.09 &  9.72$\pm$0.01 & 1.8 & 10.06 $\pm$ 0.34 \\
      lock76 & 20 $\pm$ 5 & 0.14 $\pm$ 0.03 & 10.76$\pm$0.02 & 0.6 & 10.17 $\pm$ 0.34 \\
      lock77 & 60 $\pm$ 10 & 0.33 $\pm$ 0.08 & 10.05$\pm ^{0.02}_{0.15}$ & 1.2 & -- \\
      lock73 & 105 $\pm$ 30 & 0.60 $\pm$ 0.18 & 9.59$\pm$0.01 & 2.2 & 9.96 $\pm$ 0.34 \\
      lock24 & 90 $\pm$ 20 & 0.51 $\pm$ 0.13 & 10.67$\pm ^{0.06}_{0.12}$ & 2.2 & -- \\
      lock40 & 190 $\pm$ 80 & 1.10 $\pm$ 0.44 & 10.82$\pm ^{0.01}_{0.09}$ & 0.8 & -- \\
      lock87 & 430 $\pm$ 30 & 2.45 $\pm$ 0.14 & 10.25$\pm$0.07 & 3.0 & -- \\
      lock38 & 230 $\pm$ 30 & 1.29 $\pm$ 0.15 & 10.77$\pm ^{0.01}_{0.04}$ & 1.0 & 10.41 $\pm$ 0.34 \\
      \hline
       \multicolumn{6}{c}{\textbf{SFRGs}} \\ \hline
      sxdf1293 &  90 $\pm$ 20 & 0.52 $\pm$0.14 & 10.23$\pm ^{0.12}_{0.02}$ & 4.2 & 11.25 $\pm$ 0.34 \\
      sxdf490 & 130 $\pm$ 70 & 0.76 $\pm$ 0.39  &  10.52$\pm$0.06 & 2.2 & 10.65 $\pm$ 0.34 \\
      sxdf643 & 340 $\pm$ 40 & 1.97 $\pm$ 0.20 &  10.22$\pm$0.01 & 3.2 & 10.36 $\pm$ 0.34 \\
      sxdf755 & 270 $\pm$ 40 & 1.57 $\pm$ 0.22 &  10.13$\pm$0.01 & 3.6 & 10.71 $\pm$ 0.34  \\
      sxdf639 & 190 $\pm$ 80 & 1.10 $\pm$ 0.43 &  10.51$\pm ^{0.02}_{0.01}$ & 2.2 & 11.01 $\pm$ 0.34  \\
      sxdf1204 &  90 $\pm$ 30 & 0.50 $\pm$ 0.15 &  9.99$\pm ^{0.14}_{0.01}$ & 4.0 & 10.78 $\pm$ 0.34  \\
      sxdf844 & 130 $\pm$ 20 & 0.77 $\pm$ 0.09 &  10.24$\pm ^{0.12}_{0.05}$ & 0.4 & 10.43 $\pm$ 0.34  \\
      sxdf965 & 310 $\pm$ 40 & 1.80 $\pm$ 0.23 &  10.84$\pm ^{0.01}_{0.04}$ & 2.4 & 11.15 $\pm$ 0.34  \\
      sxdf704 & 220 $\pm$ 50 & 1.29 $\pm$ 0.29 &   10.37$\pm ^{0.01}_{0.07}$ & 1.6 & --  \\
      sxdf1219 & 190 $\pm$ 40 & 1.11 $\pm$ 0.24 &  10.58$\pm ^{0.01}_{0.13}$ & 3.6 & 10.89 $\pm$ 0.34  \\
      sxdf1300 & 440 $\pm$ 60 & 2.50 $\pm$ 0.36 &  11.48$\pm$0.01 & 3.8 & 10.84 $\pm$ 0.34  \\
      sxdf966 & 90 $\pm$ 40 & 0.52 $\pm$ 0.21 & 9.60$\pm ^{0.33}_{0.01}$ & 3.4 & 11.16 $\pm$ 0.35 \\
      lock332 & 1360 $\pm$ 20 & 7.77 $\pm$ 0.13 & 11.68$\pm ^{0.03}_{0.23}$ & 4.8 & 10.48 $\pm$ 0.34  \\
      lock517 & 100 $\pm$ 10 & 0.58 $\pm$ 0.07 &  11.46$\pm ^{0.07}_{0.01}$ & 6.0 & 10.61 $\pm$ 0.34  \\
      lock218 & 70 $\pm$ 20 & 0.41 $\pm$ 0.12 & 11.45$\pm ^{0.57}_{1.25}$ & 2.6 & 10.75 $\pm$ 0.36  \\
      lock492 & 90 $\pm$ 30 & 0.53 $\pm$ 0.19 &  10.97$\pm ^{0.35}_{1.18}$ & 5.2 & 11.38 $\pm$ 0.34   \\
      lock493 & 60 $\pm$ 10 & 0.34 $\pm$ 0.07 & 11.94$\pm$0.01 & 6.0 & 11.07 $\pm$ 0.34 \\
      c03.1420 & 110 $\pm$ 50 & 0.63 $\pm$ 0.29 & 10.81$\pm ^{0.11}_{1.04}$ & 4.8 & 10.81 $\pm$ 0.36  \\
      c03.14 & 90 $\pm$ 10 & 0.52 $\pm$ 0.06 &  11.11$\pm ^{0.54}_{0.02}$ & 5.0 & 10.79 $\pm$ 0.34 \\
      c03.56 & 190 $\pm$ 70 & 1.10 $\pm$ 0.38 &  10.51$\pm ^{0.92}_{0.51}$ & 6.0 & 10.36 $\pm$ 0.34  \\
      c03.62 & 160 $\pm$ 60 & 0.92 $\pm$ 0.37 &  11.15$\pm ^{0.40}_{1.04}$ & 2.2 & 11.66 $\pm$ 0.34 \\
      \hline
\end{longtable}
\end{landscape}
\twocolumn

\onecolumn
\begin{landscape}
\renewcommand{\thefootnote}{\alph{footnote}}
\small
\begin{longtable}{ccccccc}
    \caption{MgII and FeII velocities and equivalent widths for different samples of SMGs and SFRGs. Numbers in brackets are less reliable due to the small equivalent widths of the detected feature and/or blending of features.}
    \label{tab:velocities} \\	
      Sample & v$_{\rm{MgII}}^{\rm{cen}}$ (kms$^{-1}$) &  v$_{\rm{FeII}}^{\rm{cen}}$ (kms$^{-1}$) & v$^{\rm{term}}_{\rm{MgII}}$ (kms$^{-1}$) & W$_{\rm{MgII} 2796}$ (\AA) & W$_{\rm{FeII} 2600}$ (\AA) & W$_{\rm{MgII} 2796}$/W$_{\rm{MgII} 2803}$ \\
      \hline
      \hline
      All (no e(a)) & $-$255$\pm$49 & $-$216$\pm$36 & $-$540$\pm^{40}_{100}$ & 1.48$\pm^{0.11}_{0.15}$ & 1.37$\pm^{0.16}_{0.15}$ & 1.04$\pm^{0.11}_{0.16}$ \\
      All (incl. e(a)) & $-$258$\pm$52 & $-$212$\pm$16 & $-$540$\pm^{30}_{100}$ & 1.09$\pm^{0.10}_{0.14}$ & 1.09$\pm^{0.13}_{0.11}$ & 0.96$\pm^{0.18}_{0.18}$ \\
      & & & & & & \\
      SMG & $-$270$\pm$52 & $-$225$\pm$35 & $-$700$\pm^{40}_{80}$ & 2.70$\pm^{0.34}_{0.18}$ & 1.06$\pm^{0.20}_{0.21}$ & 1.97$\pm^{0.23}_{0.34}$ \\
      SFRG & $-$281$\pm$20 & $-$222$\pm$39 & $-$420$\pm^{120}_{380}$ & -- & 1.88$\pm^{0.20}_{0.23}$ & -- \\
      & & & & & & \\
      AGN & $-$264$\pm$20 & $-$246$\pm$77 & $-$580$\pm^{140}_{40}$ & 1.17$\pm^{0.33}_{0.27}$ & -- & 0.85$\pm^{0.29}_{0.43}$ \\
      SB & $-$268$\pm$66 & $-$242$\pm$48 & $-$540$\pm^{70}_{150}$ & 2.41$\pm^{0.30}_{0.18}$ & 1.98$\pm^{0.17}_{0.18}$ & 1.43$\pm^{0.17}_{0.19}$ \\
      & & & & & & \\
      L$_{\rm{IR}}>$1.09$\times10^{12}$L$_\odot$ & $-$299$\pm$57 & $-$230$\pm$44 & $-$980$\pm^{230}_{150}$ & 1.95$\pm^{0.30}_{0.26}$ & 2.24$\pm^{0.37}_{0.20}$ & 1.07$\pm^{0.19}_{0.22}$ \\
      L$_{\rm{IR}}<$0.65$\times10^{12}$L$_\odot$ & $-$228$\pm$48 & $-$176$\pm$20 & $-$590$\pm^{160}_{110}$ & 1.24$\pm^{0.45}_{0.40}$ & 1.04$\pm^{0.38}_{0.08}$ & 0.68$\pm^{0.38}_{0.37}$ \\
      & & & & & & \\
      $12+\rm{log}(\rm{O}/H)>8.4$ & $-$237$\pm$61 & $-$216$\pm$47 & $-$330$\pm^{140}_{120}$ & 1.02$\pm^{0.48}_{0.24}$ & 2.10$\pm^{0.28}_{0.23}$ & 0.68$\pm^{0.53}_{0.25}$ \\
      $12+\rm{log}(\rm{O}/H)<8.2$\footnotemark[1] & ($-$303$\pm$19) & $-$272$\pm$88 & $-$810$\pm^{220}_{270}$ & (1.28$\pm^{0.34}_{0.26}$) & 1.30$\pm^{0.32}_{0.26}$ & (0.62$\pm^{0.28}_{0.24}$) \\
      & & & & & & \\
      $\rm{log}_{10}(M_*/M_\odot)>10.80$\footnotemark[2] & ($-$243$\pm$27) & $-$263$\pm$43 & ($-$750$\pm^{220}_{210}$) & (0.81$\pm^{0.41}_{0.25}$) & 2.40$\pm^{0.39}_{0.27}$ & -- \\
      $\rm{log}_{10}(M_*/M_\odot)<10.50$ & $-$264$\pm$71 & $-$191$\pm$32 & $-$400$\pm^{20}_{50}$ & 1.90$\pm^{0.21}_{0.26}$ & 0.73$\pm^{0.20}_{0.10}$ & 1.53$\pm^{0.46}_{0.25}$ \\
      & & & & & & \\
      $\rm{log}_{10}(M_{\rm{dyn}}/M_\odot)>10.50$\footnotemark[3] & -- & $-$234$\pm$30 & $-$550$\pm^{240}_{170}$ & -- & 2.00$\pm^{0.17}_{0.15}$ & -- \\
      $\rm{log}_{10}(M_{\rm{dyn}}/M_\odot)<10.30$ & $-$273$\pm$43 & $-$208$\pm$29 & $-$670$\pm^{80}_{170}$ & 2.52$\pm^{0.19}_{0.20}$ & 1.53$\pm^{0.33}_{0.22}$ & 1.47$\pm^{0.12}_{0.13}$ \\
      \hline
      \footnotetext[1]{MgII 2800 doublet is blended}
      \footnotetext[2]{MgII 2796 only marginally detected}
      \footnotetext[3]{Terminal velocity derived from FeII 2586}
\end{longtable} 
\end{landscape}
\twocolumn


\bibliography{}

\begin{thebibliography}{}

\bibitem[\protect\citeauthoryear{{Alatalo} et~al.,}{{Alatalo}
  et~al.}{2011}]{Alatalo:11}
{Alatalo} K.,  et~al., 2011, \apj, 735, 88

\bibitem[\protect\citeauthoryear{{Alexander}, {Bauer}, {Chapman}, {Smail},
  {Blain}, {Brandt} \& {Ivison}}{{Alexander} et~al.}{2005}]{Alexander:05}
{Alexander} D.~M.,  {Bauer} F.~E.,  {Chapman} S.~C.,  {Smail} I.,  {Blain}
  A.~W.,  {Brandt} W.~N.,    {Ivison} R.~J.,  2005, \apj, 632, 736

\bibitem[\protect\citeauthoryear{{Alexander}, {Swinbank}, {Smail}, {McDermid}
  \& {Nesvadba}}{{Alexander} et~al.}{2010}]{Alexander:10}
{Alexander} D.~M.,  {Swinbank} A.~M.,  {Smail} I.,  {McDermid} R.,
  {Nesvadba} N.~P.~H.,  2010, \mnras, 402, 2211

\bibitem[\protect\citeauthoryear{{Banerji}, {Ferreras}, {Abdalla}, {Hewett} \&
  {Lahav}}{{Banerji} et~al.}{2010}]{Banerji:10}
{Banerji} M.,  {Ferreras} I.,  {Abdalla} F.~B.,  {Hewett} P.,    {Lahav} O.,
  2010, \mnras, 402, 2264

\bibitem[\protect\citeauthoryear{{Biggs} \& {Ivison}}{{Biggs} \&
  {Ivison}}{2006}]{Biggs:06}
{Biggs} A.~D.,  {Ivison} R.~J.,  2006, \mnras, 371, 963

\bibitem[\protect\citeauthoryear{{Binney} \& {Tremaine}}{{Binney} \&
  {Tremaine}}{1987}]{Binney:87}
{Binney} J.,  {Tremaine} S.,  1987, {Galactic dynamics}

\bibitem[\protect\citeauthoryear{{Blain}, {Smail}, {Ivison} \& {Kneib}}{{Blain}
  et~al.}{1999}]{Blain:99}
{Blain} A.~W.,  {Smail} I.,  {Ivison} R.~J.,    {Kneib} J.,  1999, \mnras, 302,
  632

\bibitem[\protect\citeauthoryear{{Bolzonella}, {Miralles} \&
  {Pell{\'o}}}{{Bolzonella} et~al.}{2000}]{Bolzonella:Hyperz}
{Bolzonella} M.,  {Miralles} J.,    {Pell{\'o}} R.,  2000, \aap, 363, 476

\bibitem[\protect\citeauthoryear{{Borys}, {Chapman}, {Halpern} \&
  {Scott}}{{Borys} et~al.}{2002}]{Borys:02}
{Borys} C.,  {Chapman} S.~C.,  {Halpern} M.,    {Scott} D.,  2002, \mnras, 330,
  L63

\bibitem[\protect\citeauthoryear{{Bruzual} \& {Charlot}}{{Bruzual} \&
  {Charlot}}{2003}]{BC:03}
{Bruzual} G.,  {Charlot} S.,  2003, \mnras, 344, 1000

\bibitem[\protect\citeauthoryear{{Carilli} \& {Wang}}{{Carilli} \&
  {Wang}}{2006}]{Carilli:06}
{Carilli} C.~L.,  {Wang} R.,  2006, \aj, 132, 2231

\bibitem[\protect\citeauthoryear{{Casey}, {Chapman}, {Beswick}, {Biggs},
  {Blain}, {Hainline}, {Ivison}, {Muxlow} \& {Smail}}{{Casey}
  et~al.}{2009}]{Casey:09a}
{Casey} C.~M.,  {Chapman} S.~C.,  {Beswick} R.~J.,  {Biggs} A.~D.,  {Blain}
  A.~W.,  {Hainline} L.~J.,  {Ivison} R.~J.,  {Muxlow} T.~W.~B.,    {Smail} I.,
   2009, \mnras, 399, 121

\bibitem[\protect\citeauthoryear{{Casey}, {Chapman}, {Neri}, {Bertoldi},
  {Smail}, {Greve}, {Beswick}, {Blain}, {Coppin}, {Cox}, {Genzel}, {Ivison},
  {Muxlow}, {Omont} \& {Swinbank}}{{Casey} et~al.}{2009}]{Casey:09b}
{Casey} C.~M.,  {Chapman} S.~C.,  {Neri} R.,  {Bertoldi} F.,  {Smail} I.,
  {Greve} T.~R.,  {Beswick} R.~J.,  {Blain} A.~W.,  {Coppin} K.,  {Cox} P.,
  {Genzel} R.,  {Ivison} R.~J.,  {Muxlow} T.~W.~B.,  {Omont} A.,    {Swinbank}
  A.~M.,  2009, arXiv:0910.5756

\bibitem[\protect\citeauthoryear{{Chabrier}}{{Chabrier}}{2003}]{Chabrier:03}
{Chabrier} G.,  2003, \pasp, 115, 763

\bibitem[\protect\citeauthoryear{{Chapman}, {Blain}, {Smail} \&
  {Ivison}}{{Chapman} et~al.}{2005}]{Chapman:05}
{Chapman} S.~C.,  {Blain} A.~W.,  {Smail} I.,    {Ivison} R.~J.,  2005, \apj,
  622, 772

\bibitem[\protect\citeauthoryear{{Chapman} et~al.,}{{Chapman}
  et~al.}{2010}]{Chapman:10}
{Chapman} S.~C.,  et~al., 2010, \mnras, 409, L13

\bibitem[\protect\citeauthoryear{{Chapman}, {Smail}, {Blain} \&
  {Ivison}}{{Chapman} et~al.}{2004}]{Chapman:04}
{Chapman} S.~C.,  {Smail} I.,  {Blain} A.~W.,    {Ivison} R.~J.,  2004, \apj,
  614, 671

\bibitem[\protect\citeauthoryear{{Cirasuolo}, {McLure}, {Dunlop}, {Almaini},
  {Foucaud} \& {Simpson}}{{Cirasuolo} et~al.}{2010}]{Cirasuolo:10}
{Cirasuolo} M.,  {McLure} R.~J.,  {Dunlop} J.~S.,  {Almaini} O.,  {Foucaud} S.,
     {Simpson} C.,  2010, \mnras, 401, 1166

\bibitem[\protect\citeauthoryear{{Clements} et~al.,}{{Clements}
  et~al.}{2008}]{Clements:08}
{Clements} D.~L.,  et~al., 2008, \mnras, 387, 247

\bibitem[\protect\citeauthoryear{{Coppin} et~al.,}{{Coppin}
  et~al.}{2006}]{Coppin:06}
{Coppin} K.,  et~al., 2006, \mnras, 372, 1621

\bibitem[\protect\citeauthoryear{{Dye} et~al.,}{{Dye}  et~al.}{2008}]{Dye:08}
{Dye} S.,  et~al., 2008, \mnras, 386, 1107

\bibitem[\protect\citeauthoryear{{Erb}, {Pettini}, {Shapley}, {Steidel}, {Law}
  \& {Reddy}}{{Erb} et~al.}{2010}]{Erb:10}
{Erb} D.~K.,  {Pettini} M.,  {Shapley} A.~E.,  {Steidel} C.~C.,  {Law} D.~R.,
   {Reddy} N.~A.,  2010, \apj, 719, 1168

\bibitem[\protect\citeauthoryear{{Erb}, {Steidel}, {Shapley}, {Pettini},
  {Reddy} \& {Adelberger}}{{Erb} et~al.}{2006}]{Erb:06}
{Erb} D.~K.,  {Steidel} C.~C.,  {Shapley} A.~E.,  {Pettini} M.,  {Reddy} N.~A.,
     {Adelberger} K.~L.,  2006, \apj, 646, 107

\bibitem[\protect\citeauthoryear{{Faber} et~al.,}{{Faber}
  et~al.}{2003}]{Faber:03}
{Faber} S.~M.,  et~al., 2003, in {M.~Iye \& A.~F.~M.~Moorwood} ed., Society of
  Photo-Optical Instrumentation Engineers (SPIE) Conference Series Vol.~4841 of
  Presented at the Society of Photo-Optical Instrumentation Engineers (SPIE)
  Conference, {The DEIMOS spectrograph for the Keck II Telescope: integration
  and testing}.
pp 1657--1669

\bibitem[\protect\citeauthoryear{{Frye}, {Broadhurst} \&
  {Ben{\'{\i}}tez}}{{Frye} et~al.}{2002}]{Frye:02}
{Frye} B.,  {Broadhurst} T.,    {Ben{\'{\i}}tez} N.,  2002, \apj, 568, 558

\bibitem[\protect\citeauthoryear{{Genzel}, {Baker}, {Tacconi}, {Lutz}, {Cox},
  {Guilloteau} \& {Omont}}{{Genzel} et~al.}{2003}]{Genzel:03}
{Genzel} R.,  {Baker} A.~J.,  {Tacconi} L.~J.,  {Lutz} D.,  {Cox} P.,
  {Guilloteau} S.,    {Omont} A.,  2003, \apj, 584, 633

\bibitem[\protect\citeauthoryear{{Giavalisco} et~al.,}{{Giavalisco}
  et~al.}{2004}]{Giavalisco:04}
{Giavalisco} M.,  et~al., 2004, \apjl, 600, L93

\bibitem[\protect\citeauthoryear{{Gonzalez Delgado}, {Leitherer}, {Heckman},
  {Lowenthal}, {Ferguson} \& {Robert}}{{Gonzalez Delgado}
  et~al.}{1998}]{Gonzalez:98}
{Gonzalez Delgado} R.~M.,  {Leitherer} C.,  {Heckman} T.,  {Lowenthal} J.~D.,
  {Ferguson} H.~C.,    {Robert} C.,  1998, \apj, 495, 698

\bibitem[\protect\citeauthoryear{{Hainline}, {Blain}, {Smail}, {Alexander},
  {Armus}, {Chapman} \& {Ivison}}{{Hainline} et~al.}{2011}]{Hainline:10}
{Hainline} L.~J.,  {Blain} A.~W.,  {Smail} I.,  {Alexander} D.~M.,  {Armus} L.,
   {Chapman} S.~C.,    {Ivison} R.~J.,  2011, arXiv:1006.0238

\bibitem[\protect\citeauthoryear{{Heckman}, {Armus} \& {Miley}}{{Heckman}
  et~al.}{1990}]{Heckman:90}
{Heckman} T.~M.,  {Armus} L.,    {Miley} G.~K.,  1990, \apjs, 74, 833

\bibitem[\protect\citeauthoryear{{Heckman}, {Lehnert}, {Strickland} \&
  {Armus}}{{Heckman} et~al.}{2000}]{Heckman:00}
{Heckman} T.~M.,  {Lehnert} M.~D.,  {Strickland} D.~K.,    {Armus} L.,  2000,
  \apjs, 129, 493

\bibitem[\protect\citeauthoryear{{Ivison} et~al.,}{{Ivison}
  et~al.}{2002}]{Ivison:02}
{Ivison} R.~J.,  et~al., 2002, \mnras, 337, 1

\bibitem[\protect\citeauthoryear{{Ivison} et~al.,}{{Ivison}
  et~al.}{2007}]{Ivison:07}
{Ivison} R.~J.,  et~al., 2007, \mnras, 380, 199

\bibitem[\protect\citeauthoryear{{Ivison} et~al.,}{{Ivison}
  et~al.}{2010}]{Ivison:10}
{Ivison} R.~J.,  et~al., 2010, \aap, 518, L31+

\bibitem[\protect\citeauthoryear{{Kennicutt}
  Jr.}{{Kennicutt}}{1998}]{Kennicutt:98}
{Kennicutt} Jr. R.~C.,  1998, \apj, 498, 541

\bibitem[\protect\citeauthoryear{{Lehnert}, {Tasse}, {Nesvadba}, {Best} \& {van
  Driel}}{{Lehnert} et~al.}{2011}]{Lehnert:11}
{Lehnert} M.~D.,  {Tasse} C.,  {Nesvadba} N.~P.~H.,  {Best} P.~N.,    {van
  Driel} W.,  2011, ArXiv e-prints

\bibitem[\protect\citeauthoryear{{Leitherer}, {Tremonti}, {Heckman} \&
  {Calzetti}}{{Leitherer} et~al.}{2011}]{Leitherer:11}
{Leitherer} C.,  {Tremonti} C.~A.,  {Heckman} T.~M.,    {Calzetti} D.,  2011,
  \aj, 141, 37

\bibitem[\protect\citeauthoryear{{Magnelli} et~al.,}{{Magnelli}
  et~al.}{2010}]{Magnelli:10}
{Magnelli} B.,  et~al., 2010, \aap, 518, L28+

\bibitem[\protect\citeauthoryear{{Maiolino} et~al.,}{{Maiolino}
  et~al.}{2008}]{Maiolino:08}
{Maiolino} R.,  et~al., 2008, \aap, 488, 463

\bibitem[\protect\citeauthoryear{{Martin}}{{Martin}}{1999}]{Martin:99}
{Martin} C.~L.,  1999, \apj, 513, 156

\bibitem[\protect\citeauthoryear{{Martin}}{{Martin}}{2005}]{Martin:05}
{Martin} C.~L.,  2005, \apj, 621, 227

\bibitem[\protect\citeauthoryear{{Morganti}, {Tadhunter} \&
  {Oosterloo}}{{Morganti} et~al.}{2005}]{Morganti:05}
{Morganti} R.,  {Tadhunter} C.~N.,    {Oosterloo} T.~A.,  2005, \aap, 444, L9

\bibitem[\protect\citeauthoryear{{Morrison}, {Owen}, {Dickinson}, {Ivison} \&
  {Ibar}}{{Morrison} et~al.}{2010}]{Morrison:10}
{Morrison} G.~E.,  {Owen} F.~N.,  {Dickinson} M.,  {Ivison} R.~J.,    {Ibar}
  E.,  2010, \apjs, 188, 178

\bibitem[\protect\citeauthoryear{{Murray}, {Quataert} \& {Thompson}}{{Murray}
  et~al.}{2005}]{Murray:05}
{Murray} N.,  {Quataert} E.,    {Thompson} T.~A.,  2005, \apj, 618, 569

\bibitem[\protect\citeauthoryear{{Nagao}, {Murayama} \& {Taniguchi}}{{Nagao}
  et~al.}{2001}]{Nagao:01}
{Nagao} T.,  {Murayama} T.,    {Taniguchi} Y.,  2001, \pasj, 53, 629

\bibitem[\protect\citeauthoryear{{Narayanan}, {Cox}, {Kelly}, {Dav{\'e}},
  {Hernquist}, {Di Matteo}, {Hopkins}, {Kulesa}, {Robertson} \&
  {Walker}}{{Narayanan} et~al.}{2008}]{Narayanan:08}
{Narayanan} D.,  {Cox} T.~J.,  {Kelly} B.,  {Dav{\'e}} R.,  {Hernquist} L.,
  {Di Matteo} T.,  {Hopkins} P.~F.,  {Kulesa} C.,  {Robertson} B.,    {Walker}
  C.~K.,  2008, \apjs, 176, 331

\bibitem[\protect\citeauthoryear{{Neri}, {Genzel}, {Ivison}, {Bertoldi},
  {Blain}, {Chapman}, {Cox}, {Greve}, {Omont} \& {Frayer}}{{Neri}
  et~al.}{2003}]{Neri:03}
{Neri} R.,  {Genzel} R.,  {Ivison} R.~J.,  {Bertoldi} F.,  {Blain} A.~W.,
  {Chapman} S.~C.,  {Cox} P.,  {Greve} T.~R.,  {Omont} A.,    {Frayer} D.~T.,
  2003, \apjl, 597, L113

\bibitem[\protect\citeauthoryear{{Nesvadba}, {Lehnert}, {Genzel}, {Eisenhauer},
  {Baker}, {Seitz}, {Davies}, {Lutz}, {Tacconi}, {Tecza}, {Bender} \&
  {Abuter}}{{Nesvadba} et~al.}{2007}]{Nesvadba:07}
{Nesvadba} N.~P.~H.,  {Lehnert} M.~D.,  {Genzel} R.,  {Eisenhauer} F.,  {Baker}
  A.~J.,  {Seitz} S.,  {Davies} R.,  {Lutz} D.,  {Tacconi} L.,  {Tecza} M.,
  {Bender} R.,    {Abuter} R.,  2007, \apj, 657, 725

\bibitem[\protect\citeauthoryear{{Pettini}, {Rix}, {Steidel}, {Adelberger},
  {Hunt} \& {Shapley}}{{Pettini} et~al.}{2002}]{Pettini:02}
{Pettini} M.,  {Rix} S.~A.,  {Steidel} C.~C.,  {Adelberger} K.~L.,  {Hunt}
  M.~P.,    {Shapley} A.~E.,  2002, \apj, 569, 742

\bibitem[\protect\citeauthoryear{{Pettini}, {Shapley}, {Steidel}, {Cuby},
  {Dickinson}, {Moorwood}, {Adelberger} \& {Giavalisco}}{{Pettini}
  et~al.}{2001}]{Pettini:01}
{Pettini} M.,  {Shapley} A.~E.,  {Steidel} C.~C.,  {Cuby} J.,  {Dickinson} M.,
  {Moorwood} A.~F.~M.,  {Adelberger} K.~L.,    {Giavalisco} M.,  2001, \apj,
  554, 981

\bibitem[\protect\citeauthoryear{{Poggianti} \& {Wu}}{{Poggianti} \&
  {Wu}}{2000}]{Poggianti:00}
{Poggianti} B.~M.,  {Wu} H.,  2000, \apj, 529, 157

\bibitem[\protect\citeauthoryear{{Pope}, {Scott}, {Dickinson}, {Chary},
  {Morrison}, {Borys}, {Sajina}, {Alexander}, {Daddi}, {Frayer}, {MacDonald} \&
  {Stern}}{{Pope} et~al.}{2006}]{Pope:06}
{Pope} A.,  {Scott} D.,  {Dickinson} M.,  {Chary} R.,  {Morrison} G.,  {Borys}
  C.,  {Sajina} A.,  {Alexander} D.~M.,  {Daddi} E.,  {Frayer} D.,  {MacDonald}
  E.,    {Stern} D.,  2006, \mnras, 370, 1185

\bibitem[\protect\citeauthoryear{{Rubin}, {Weiner}, {Koo}, {Martin},
  {Prochaska}, {Coil} \& {Newman}}{{Rubin} et~al.}{2010}]{Rubin:09}
{Rubin} K.~H.~R.,  {Weiner} B.~J.,  {Koo} D.~C.,  {Martin} C.~L.,  {Prochaska}
  J.~X.,  {Coil} A.~L.,    {Newman} J.~A.,  2010, \apj, 719, 1503

\bibitem[\protect\citeauthoryear{{Rupke}, {Veilleux} \& {Sanders}}{{Rupke}
  et~al.}{2002}]{Rupke:02}
{Rupke} D.~S.,  {Veilleux} S.,    {Sanders} D.~B.,  2002, \apj, 570, 588

\bibitem[\protect\citeauthoryear{{Rupke}, {Veilleux} \& {Sanders}}{{Rupke}
  et~al.}{2005}]{Rupke:05}
{Rupke} D.~S.,  {Veilleux} S.,    {Sanders} D.~B.,  2005, \apj, 632, 751

\bibitem[\protect\citeauthoryear{{Sakamoto}, {Aalto}, {Wilner}, {Black},
  {Conway}, {Costagliola}, {Peck}, {Spaans}, {Wang} \& {Wiedner}}{{Sakamoto}
  et~al.}{2009}]{Sakamoto:09}
{Sakamoto} K.,  {Aalto} S.,  {Wilner} D.~J.,  {Black} J.~H.,  {Conway} J.~E.,
  {Costagliola} F.,  {Peck} A.~B.,  {Spaans} M.,  {Wang} J.,    {Wiedner}
  M.~C.,  2009, \apjl, 700, L104

\bibitem[\protect\citeauthoryear{{Savage} \& {Sembach}}{{Savage} \&
  {Sembach}}{1991}]{Savage:91}
{Savage} B.~D.,  {Sembach} K.~R.,  1991, \apj, 379, 245

\bibitem[\protect\citeauthoryear{{Schwartz}, {Martin}, {Chandar}, {Leitherer},
  {Heckman} \& {Oey}}{{Schwartz} et~al.}{2006}]{Schwartz:06}
{Schwartz} C.~M.,  {Martin} C.~L.,  {Chandar} R.,  {Leitherer} C.,  {Heckman}
  T.~M.,    {Oey} M.~S.,  2006, \apj, 646, 858

\bibitem[\protect\citeauthoryear{{Shapley}, {Steidel}, {Pettini} \&
  {Adelberger}}{{Shapley} et~al.}{2003}]{Shapley:03}
{Shapley} A.~E.,  {Steidel} C.~C.,  {Pettini} M.,    {Adelberger} K.~L.,  2003,
  \apj, 588, 65

\bibitem[\protect\citeauthoryear{{Shi}, {Zhao} \& {Liang}}{{Shi}
  et~al.}{2007}]{Shi:07}
{Shi} F.,  {Zhao} G.,    {Liang} Y.~C.,  2007, \aap, 475, 409

\bibitem[\protect\citeauthoryear{{Smail}, {Chapman}, {Ivison}, {Blain},
  {Takata}, {Heckman}, {Dunlop} \& {Sekiguchi}}{{Smail}
  et~al.}{2003}]{Smail:03}
{Smail} I.,  {Chapman} S.~C.,  {Ivison} R.~J.,  {Blain} A.~W.,  {Takata} T.,
  {Heckman} T.~M.,  {Dunlop} J.~S.,    {Sekiguchi} K.,  2003, \mnras, 342, 1185

\bibitem[\protect\citeauthoryear{{Snow}, {Lamers}, {Lindholm} \&
  {Odell}}{{Snow} et~al.}{1994}]{Snow:94}
{Snow} T.~P.,  {Lamers} H.~J.~G.~L.~M.,  {Lindholm} D.~M.,    {Odell} A.~P.,
  1994, \apjs, 95, 163

\bibitem[\protect\citeauthoryear{{Steidel}, {Erb}, {Shapley}, {Pettini},
  {Reddy}, {Bogosavljevi{\'c}}, {Rudie} \& {Rakic}}{{Steidel}
  et~al.}{2010}]{Steidel:10}
{Steidel} C.~C.,  {Erb} D.~K.,  {Shapley} A.~E.,  {Pettini} M.,  {Reddy} N.,
  {Bogosavljevi{\'c}} M.,  {Rudie} G.~C.,    {Rakic} O.,  2010, \apj, 717, 289

\bibitem[\protect\citeauthoryear{{Swinbank}, {Chapman}, {Smail}, {Lindner},
  {Borys}, {Blain}, {Ivison} \& {Lewis}}{{Swinbank} et~al.}{2006}]{Swinbank:06}
{Swinbank} A.~M.,  {Chapman} S.~C.,  {Smail} I.,  {Lindner} C.,  {Borys} C.,
  {Blain} A.~W.,  {Ivison} R.~J.,    {Lewis} G.~F.,  2006, \mnras, 371, 465

\bibitem[\protect\citeauthoryear{{Swinbank}, {Lacey}, {Smail}, {Baugh},
  {Frenk}, {Blain}, {Chapman}, {Coppin}, {Ivison}, {Gonzalez} \&
  {Hainline}}{{Swinbank} et~al.}{2008}]{Swinbank:08}
{Swinbank} A.~M.,  {Lacey} C.~G.,  {Smail} I.,  {Baugh} C.~M.,  {Frenk} C.~S.,
  {Blain} A.~W.,  {Chapman} S.~C.,  {Coppin} K.~E.~K.,  {Ivison} R.~J.,
  {Gonzalez} J.~E.,    {Hainline} L.~J.,  2008, \mnras, 391, 420

\bibitem[\protect\citeauthoryear{{Swinbank}, {Smail}, {Bower}, {Borys},
  {Chapman}, {Blain}, {Ivison}, {Howat}, {Keel} \& {Bunker}}{{Swinbank}
  et~al.}{2005}]{Swinbank:05}
{Swinbank} A.~M.,  {Smail} I.,  {Bower} R.~G.,  {Borys} C.,  {Chapman} S.~C.,
  {Blain} A.~W.,  {Ivison} R.~J.,  {Howat} S.~R.,  {Keel} W.~C.,    {Bunker}
  A.~J.,  2005, \mnras, 359, 401

\bibitem[\protect\citeauthoryear{{Swinbank}, {Smail}, {Chapman}, {Blain},
  {Ivison} \& {Keel}}{{Swinbank} et~al.}{2004}]{Swinbank:04}
{Swinbank} A.~M.,  {Smail} I.,  {Chapman} S.~C.,  {Blain} A.~W.,  {Ivison}
  R.~J.,    {Keel} W.~C.,  2004, \apj, 617, 64

\bibitem[\protect\citeauthoryear{{Swinbank}, {Smail}, {Chapman}, {Borys},
  {Alexander}, {Blain}, {Conselice}, {Hainline} \& {Ivison}}{{Swinbank}
  et~al.}{2010}]{Swinbank:10}
{Swinbank} A.~M.,  {Smail} I.,  {Chapman} S.~C.,  {Borys} C.,  {Alexander}
  D.~M.,  {Blain} A.~W.,  {Conselice} C.~J.,  {Hainline} L.~J.,    {Ivison}
  R.~J.,  2010, \mnras, 405, 234

\bibitem[\protect\citeauthoryear{{Tacconi} et~al.,}{{Tacconi}
  et~al.}{2008}]{Tacconi:08}
{Tacconi} L.~J.,  et~al., 2008, \apj, 680, 246

\bibitem[\protect\citeauthoryear{{Targett}, {Dunlop}, {McLure}, {Best},
  {Cirasuolo} \& {Almaini}}{{Targett} et~al.}{2011}]{Targett:11}
{Targett} T.~A.,  {Dunlop} J.~S.,  {McLure} R.~J.,  {Best} P.~N.,  {Cirasuolo}
  M.,    {Almaini} O.,  2011, \mnras, 412, 295

\bibitem[\protect\citeauthoryear{{Tremonti}, {Heckman}, {Kauffmann},
  {Brinchmann}, {Charlot}, {White}, {Seibert}, {Peng}, {Schlegel}, {Uomoto},
  {Fukugita} \& {Brinkmann}}{{Tremonti} et~al.}{2004}]{Tremonti:04}
{Tremonti} C.~A.,  {Heckman} T.~M.,  {Kauffmann} G.,  {Brinchmann} J.,
  {Charlot} S.,  {White} S.~D.~M.,  {Seibert} M.,  {Peng} E.~W.,  {Schlegel}
  D.~J.,  {Uomoto} A.,  {Fukugita} M.,    {Brinkmann} J.,  2004, \apj, 613, 898

\bibitem[\protect\citeauthoryear{{Tremonti}, {Moustakas} \&
  {Diamond-Stanic}}{{Tremonti} et~al.}{2007}]{Tremonti:07}
{Tremonti} C.~A.,  {Moustakas} J.,    {Diamond-Stanic} A.~M.,  2007, \apjl,
  663, L77

\bibitem[\protect\citeauthoryear{{Wardlow} et~al.,}{{Wardlow}
  et~al.}{2011}]{Wardlow:10}
{Wardlow} J.~L.,  et~al., 2011, MNRAS, In Press, arXiv:1006.2137

\bibitem[\protect\citeauthoryear{{Webb}, {Eales}, {Lilly}, {Clements}, {Dunne},
  {Gear}, {Ivison}, {Flores} \& {Yun}}{{Webb} et~al.}{2003}]{Webb:03}
{Webb} T.~M.,  {Eales} S.~A.,  {Lilly} S.~J.,  {Clements} D.~L.,  {Dunne} L.,
  {Gear} W.~K.,  {Ivison} R.~J.,  {Flores} H.,    {Yun} M.,  2003, \apj, 587,
  41

\bibitem[\protect\citeauthoryear{{Weiner}, {Coil}, {Prochaska}, {Newman},
  {Cooper}, {Bundy}, {Conselice}, {Dutton}, {Faber}, {Koo}, {Lotz}, {Rieke} \&
  {Rubin}}{{Weiner} et~al.}{2009}]{Weiner:09}
{Weiner} B.~J.,  {Coil} A.~L.,  {Prochaska} J.~X.,  {Newman} J.~A.,  {Cooper}
  M.~C.,  {Bundy} K.,  {Conselice} C.~J.,  {Dutton} A.~A.,  {Faber} S.~M.,
  {Koo} D.~C.,  {Lotz} J.~M.,  {Rieke} G.~H.,    {Rubin} K.~H.~R.,  2009, \apj,
  692, 187

\bibitem[\protect\citeauthoryear{{Wei{\ss}} et~al.,}{{Wei{\ss}}
  et~al.}{2009}]{Weiss:09}
{Wei{\ss}} A.,  et~al., 2009, \apj, 707, 1201

\end{thebibliography}

\begin{appendix}

\section{Composite Spectra}

\begin{figure*}
\begin{center}
\centering
\caption{Composite Spectra of SMGs, SFRGs and galaxies split by total infra-red luminosity in the ISM absorption line region (left) and the [OII] emission line region (right). The error spectrum obtained from jackknife sampling the individual spectra is shown at the bottom of each panel on the same scale. The vertical lines marked are (in order of increasing wavelength) FeII 2344,2374,2383,2587, MnII 2594, FeII 2600, MnII 2604, MgII 2796,2803, MgI 2853 (left panels) and [NeV] 3452, [OII] 3727, [NeIII] 3869 and the Ca K \& H features at 3933 and 3969\AA (right panels).}
\begin{tabular}{cc}
\includegraphics[width=7.5cm,height=5.0cm,angle=0]{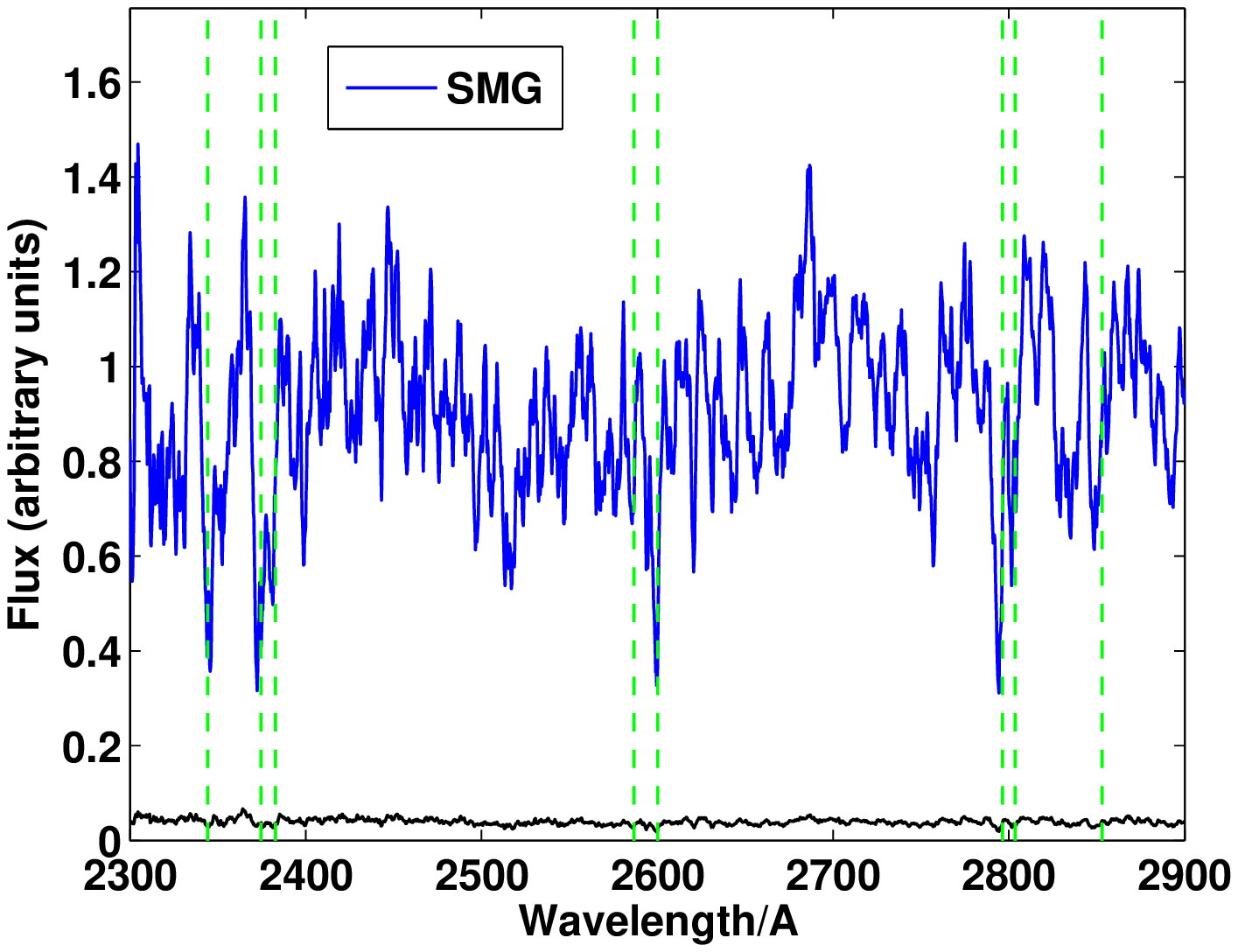} & \includegraphics[width=7.5cm,height=5.0cm,angle=0]{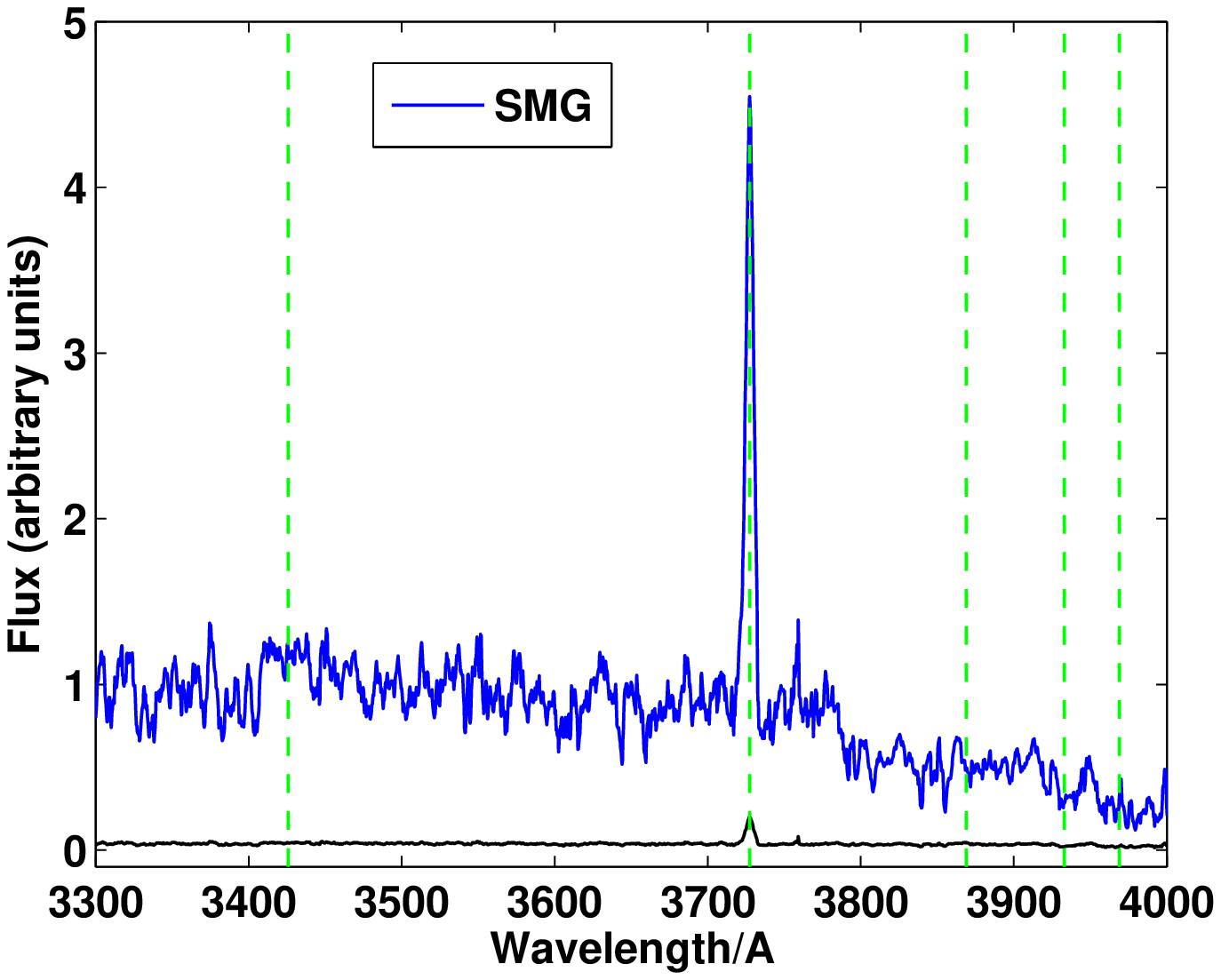} \\
\includegraphics[width=7.5cm,height=5.0cm,angle=0]{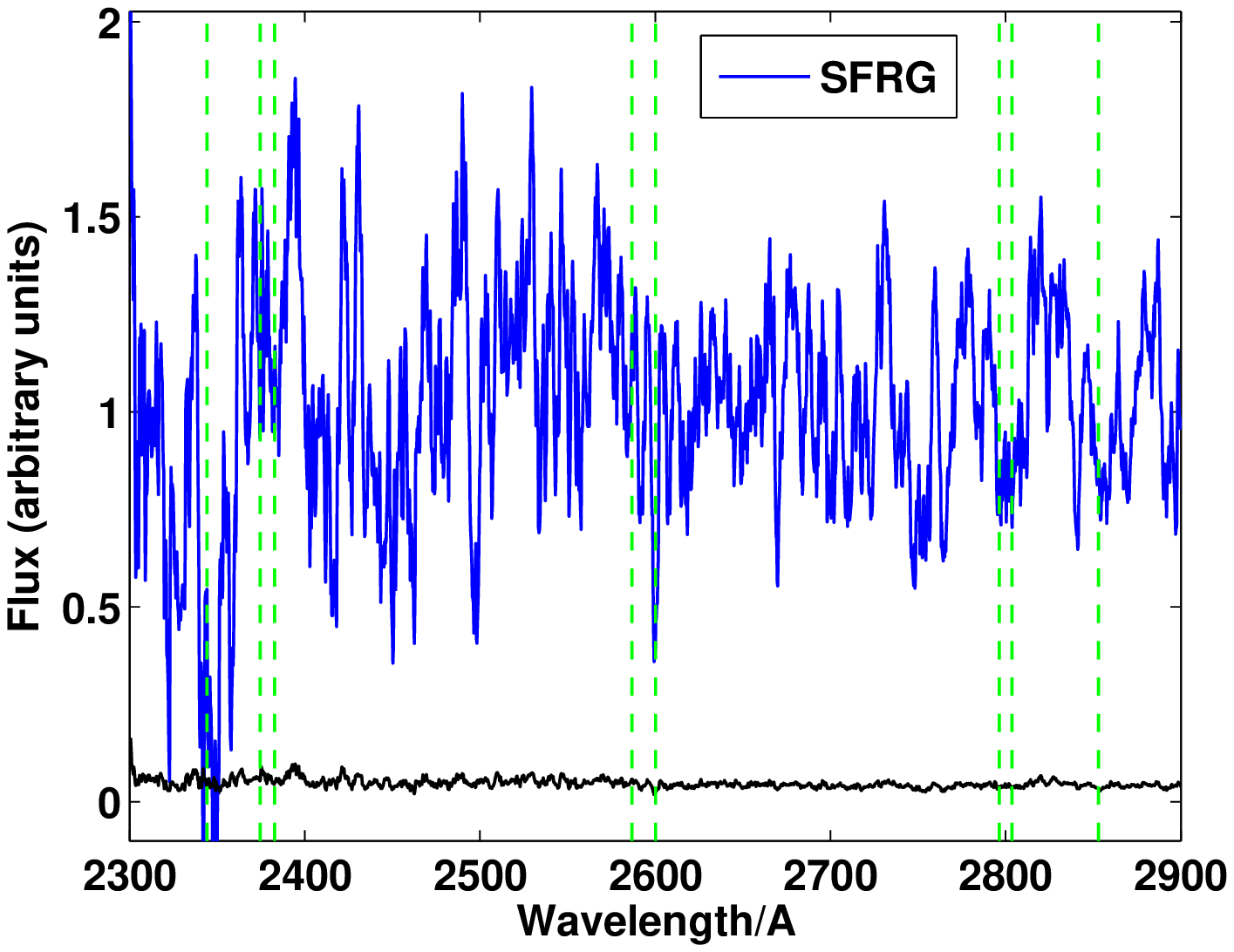} & \includegraphics[width=7.5cm,height=5.0cm,angle=0]{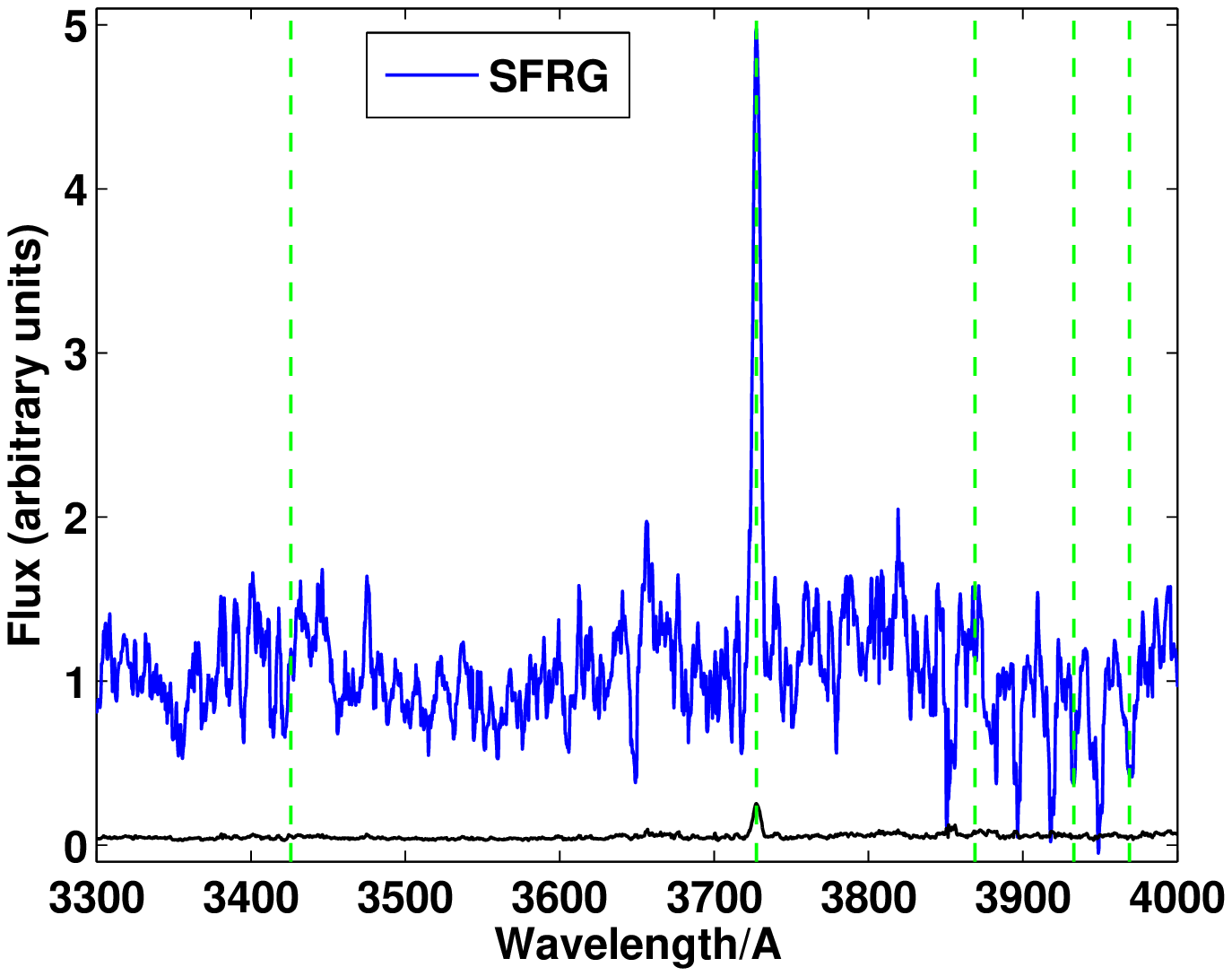} \\
\includegraphics[width=7.5cm,height=5.0cm,angle=0]{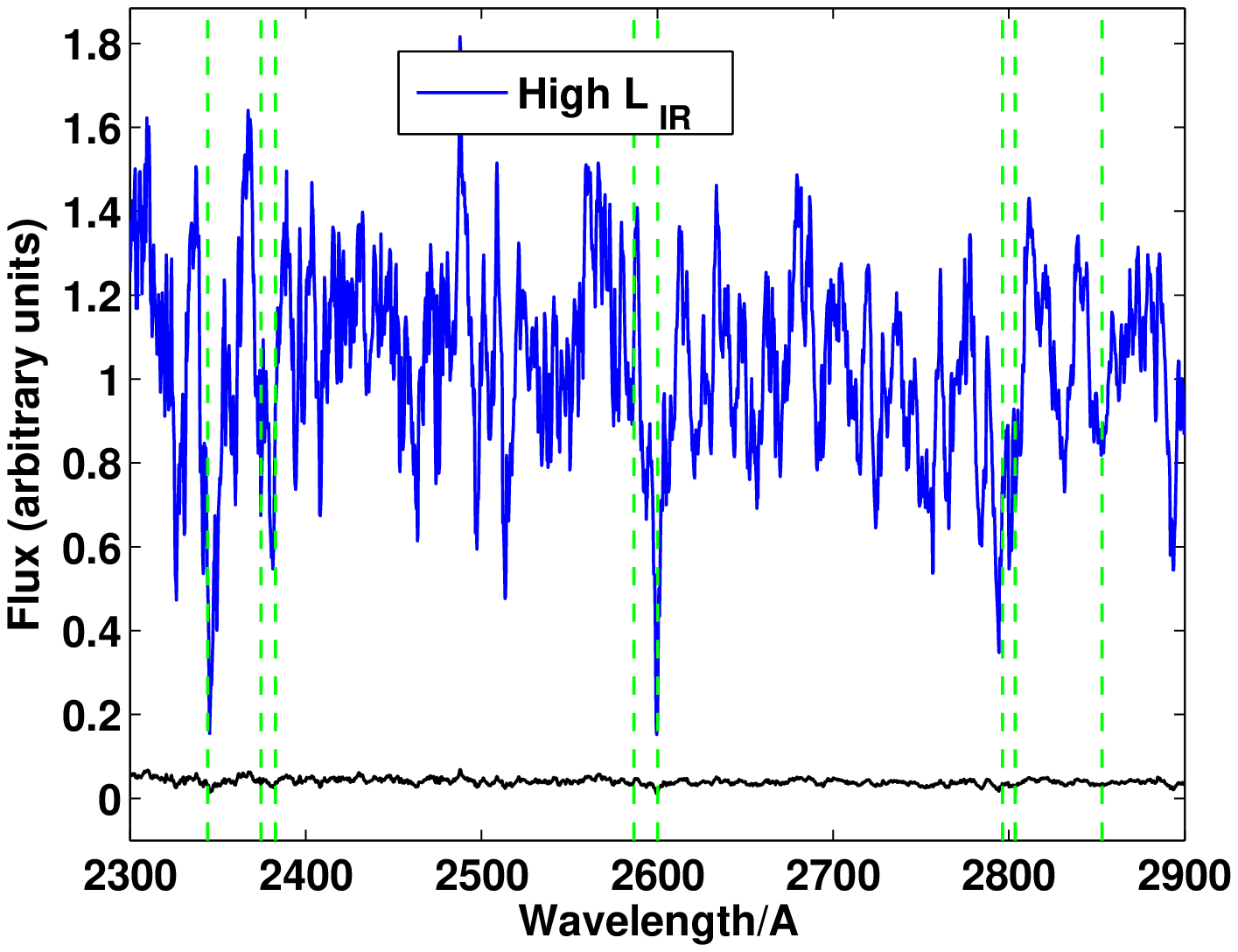} & \includegraphics[width=7.5cm,height=5.0cm,angle=0]{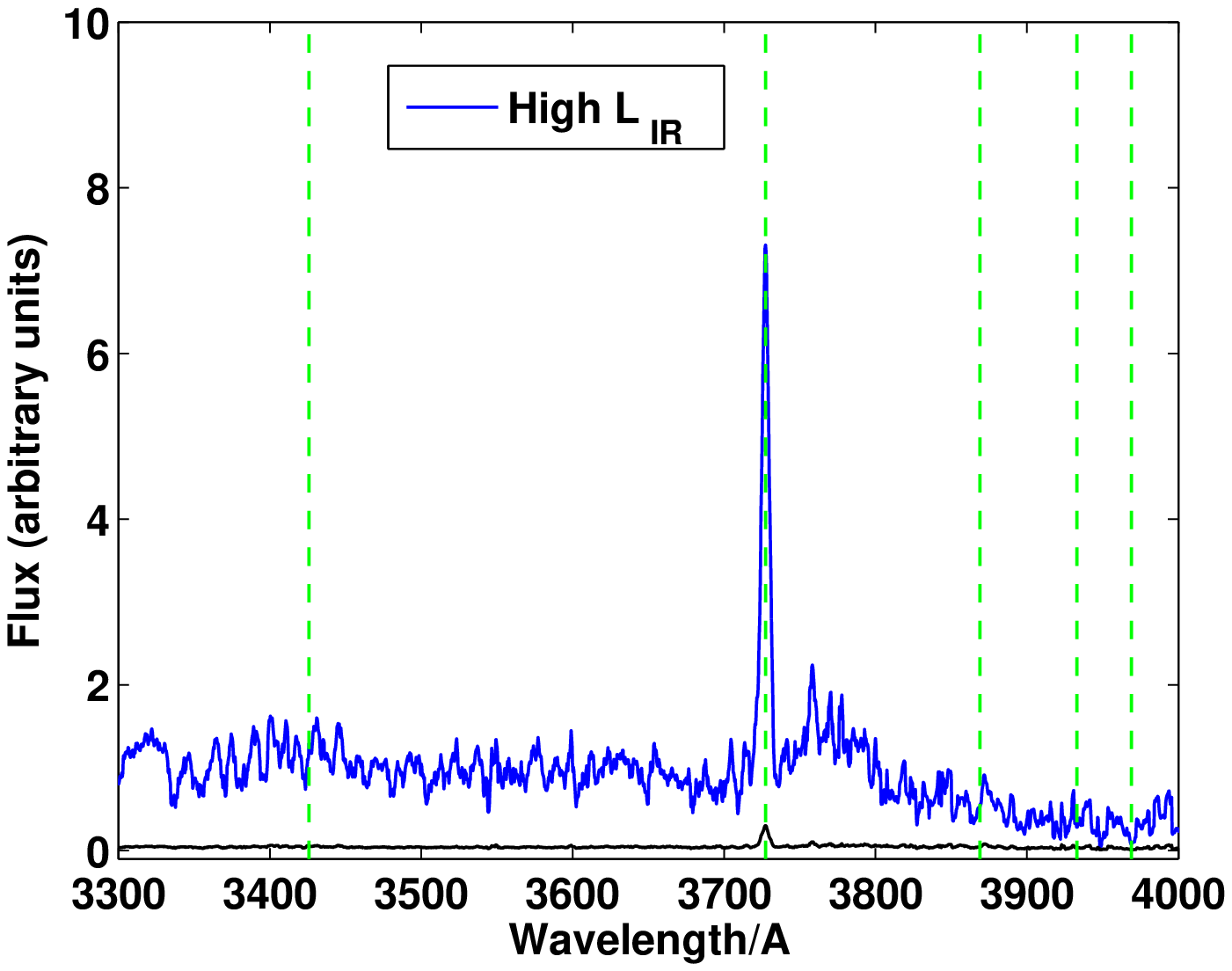} \\
\includegraphics[width=7.5cm,height=5.0cm,angle=0]{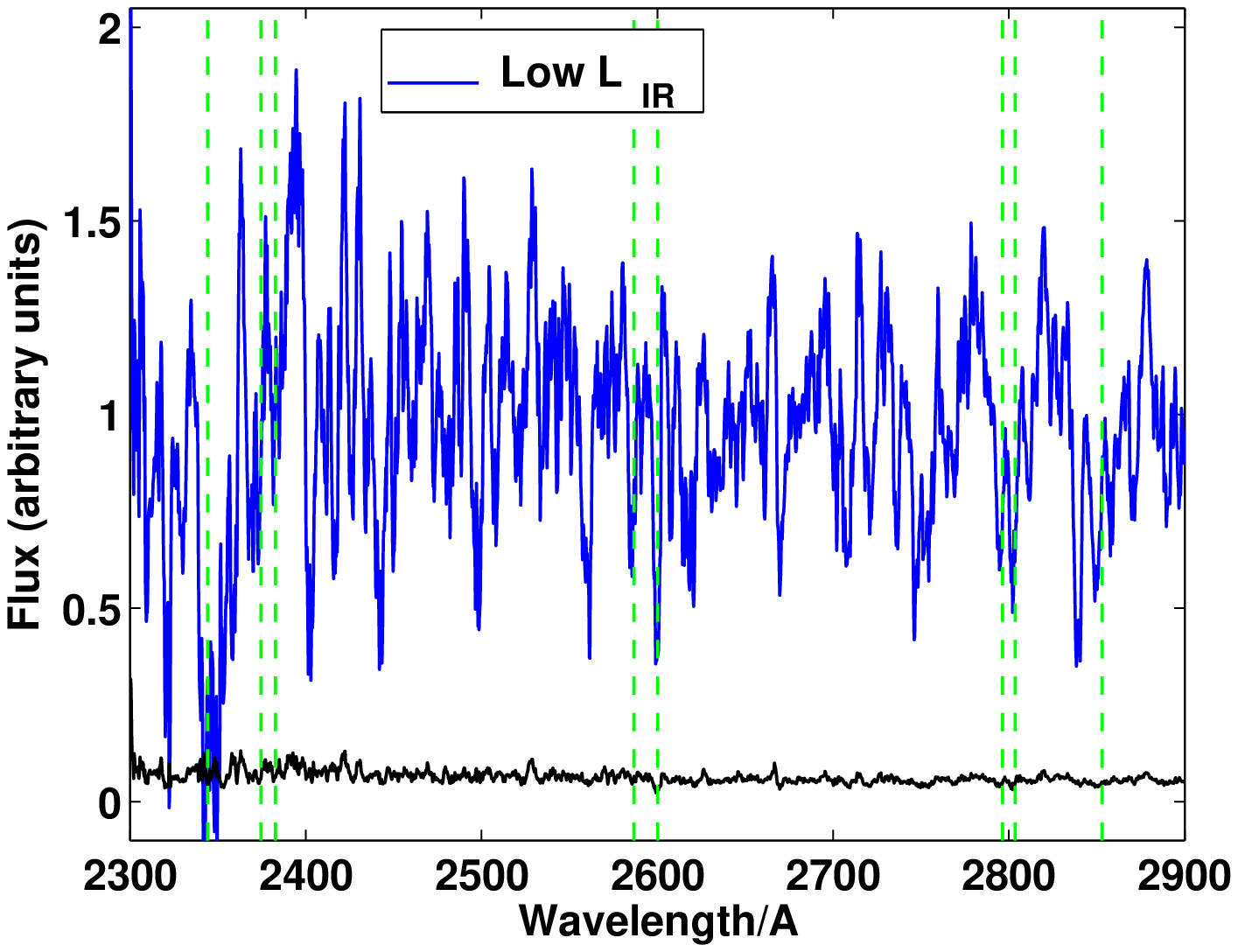} & \includegraphics[width=7.5cm,height=5.0cm,angle=0]{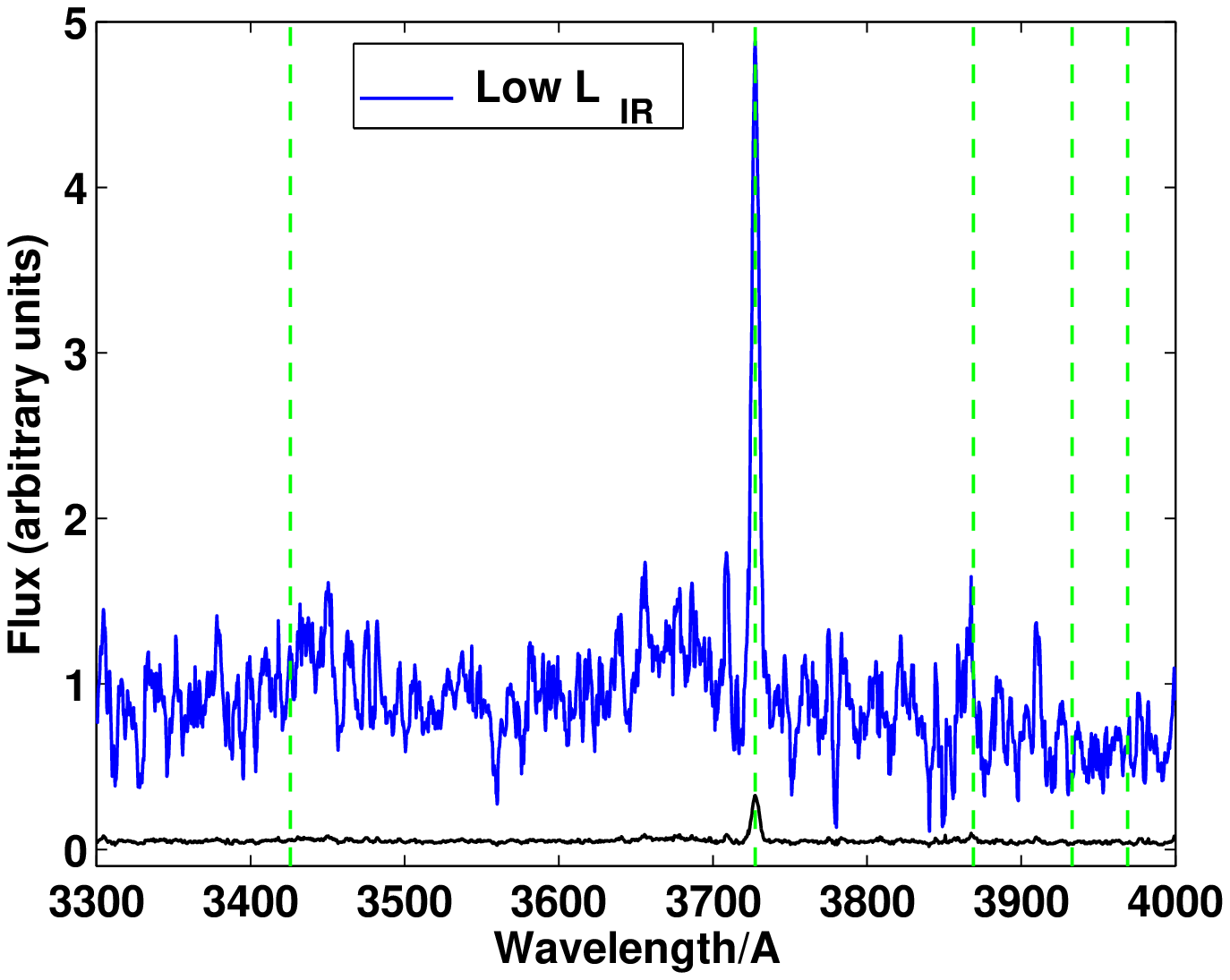} \\
\label{fig:a1}
\end{tabular}
\end{center}
\end{figure*}

\begin{figure*}
\begin{center}
\centering
\caption{Composite Spectra of galaxies split by their [NeIII]/[OII] line ratio as well as dynamical mass in the ISM absorption line region (left) and the [OII] emission line region (right). The error spectrum obtained from jackknife sampling the individual spectra is shown at the bottom of each panel on the same scale. The vertical lines marked are (in order of increasing wavelength) FeII 2344,2374,2383,2587, MnII 2594, FeII 2600, MnII 2604, MgII 2796,2803, MgI 2853 (left panels) and [NeV] 3452, [OII] 3727, [NeIII] 3869 and the Ca K \& H features at 3933 and 3969\AA (right panels).}
\begin{tabular}{cc}
\includegraphics[width=7.5cm,height=5.0cm,angle=0]{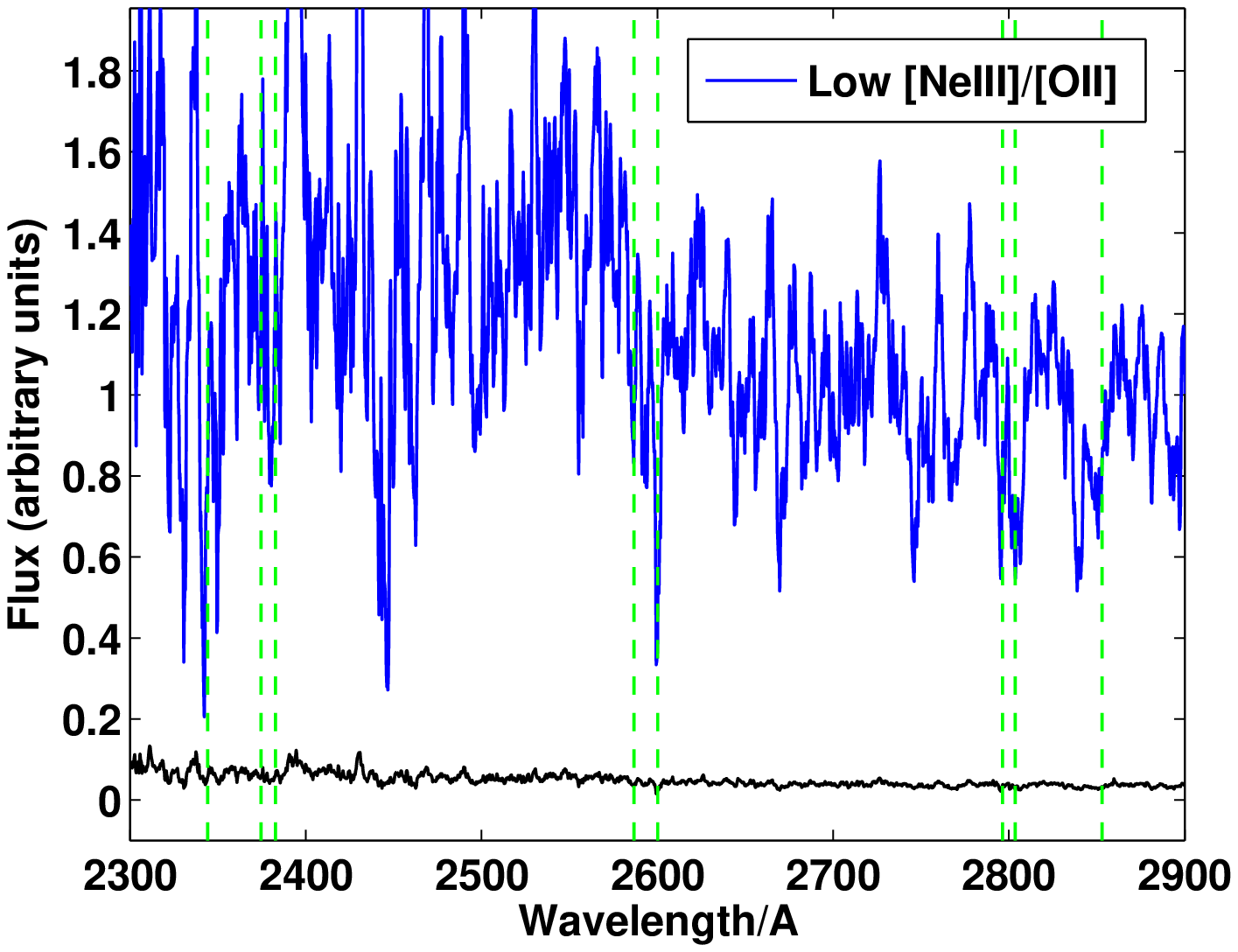} & \includegraphics[width=7.5cm,height=5.0cm,angle=0]{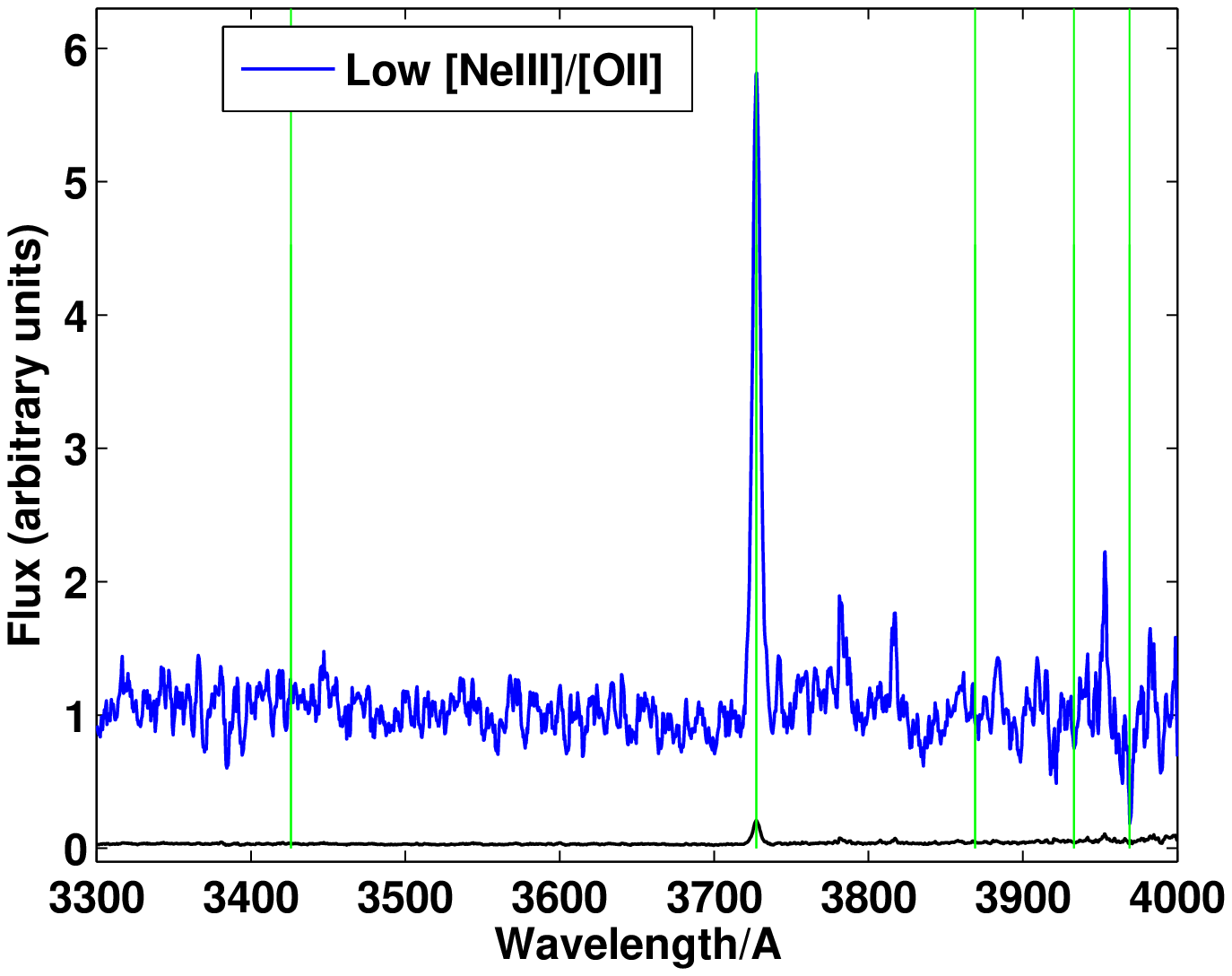} \\
\includegraphics[width=7.5cm,height=5.0cm,angle=0]{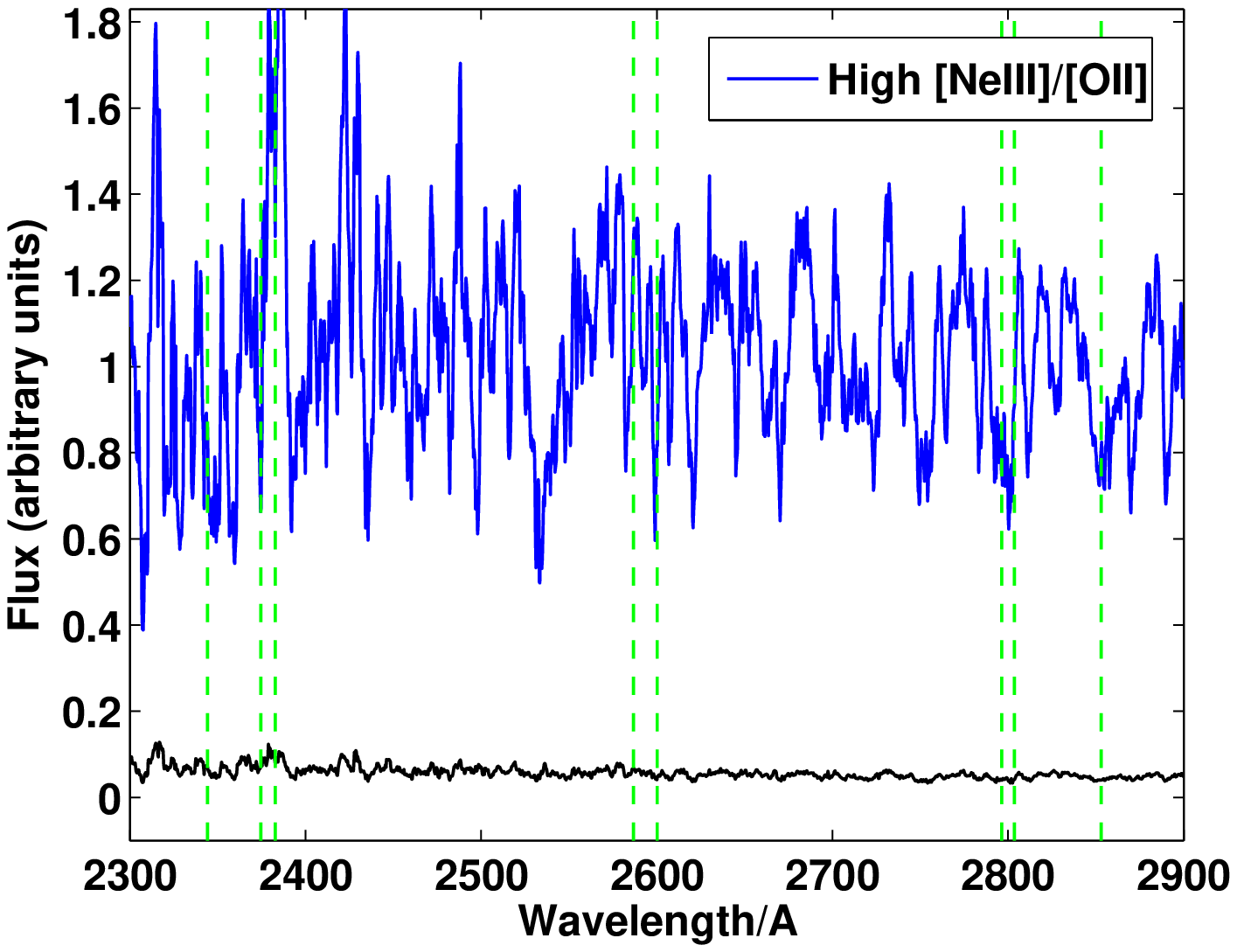} & \includegraphics[width=7.5cm,height=5.0cm,angle=0]{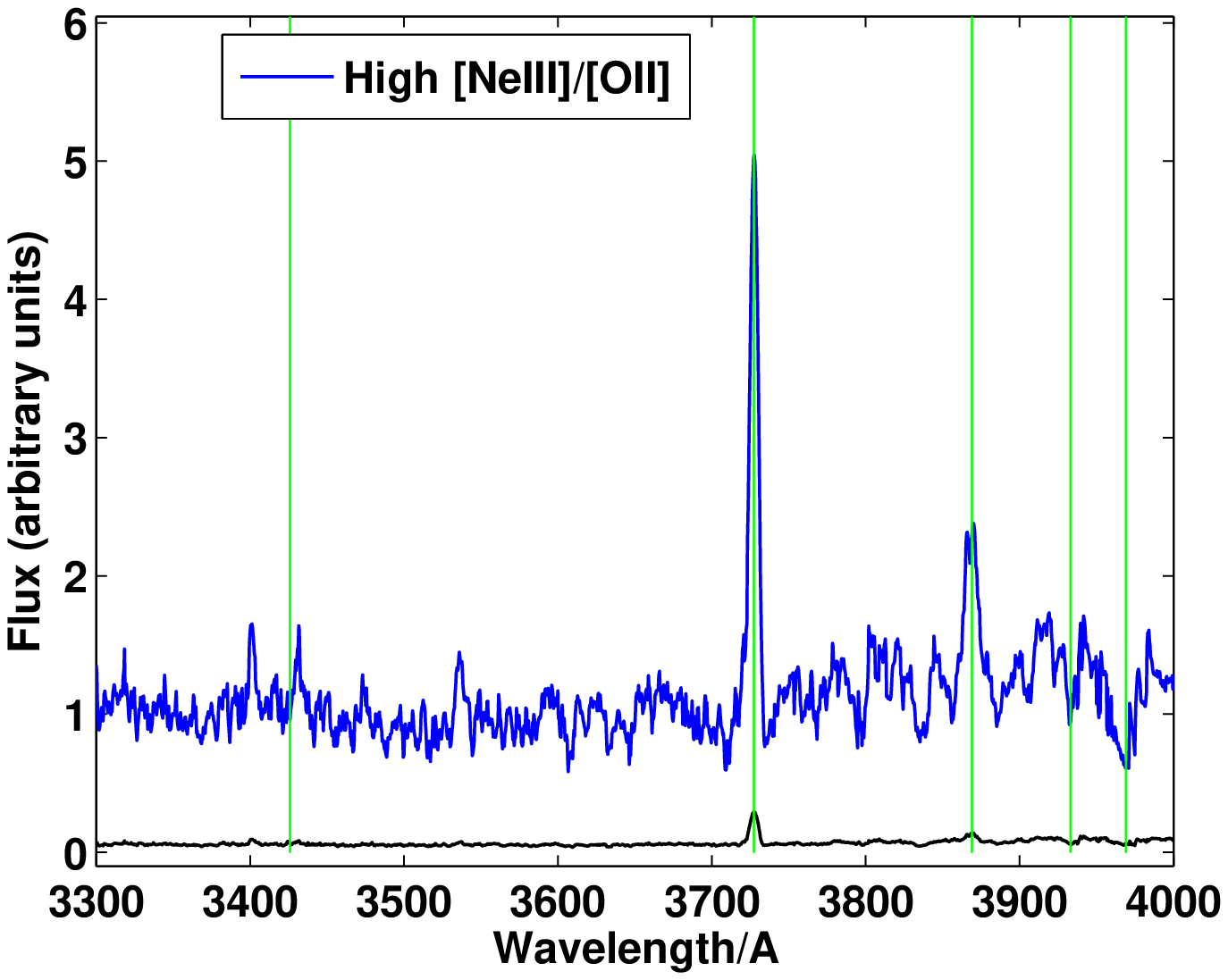} \\
\includegraphics[width=7.5cm,height=5.0cm,angle=0]{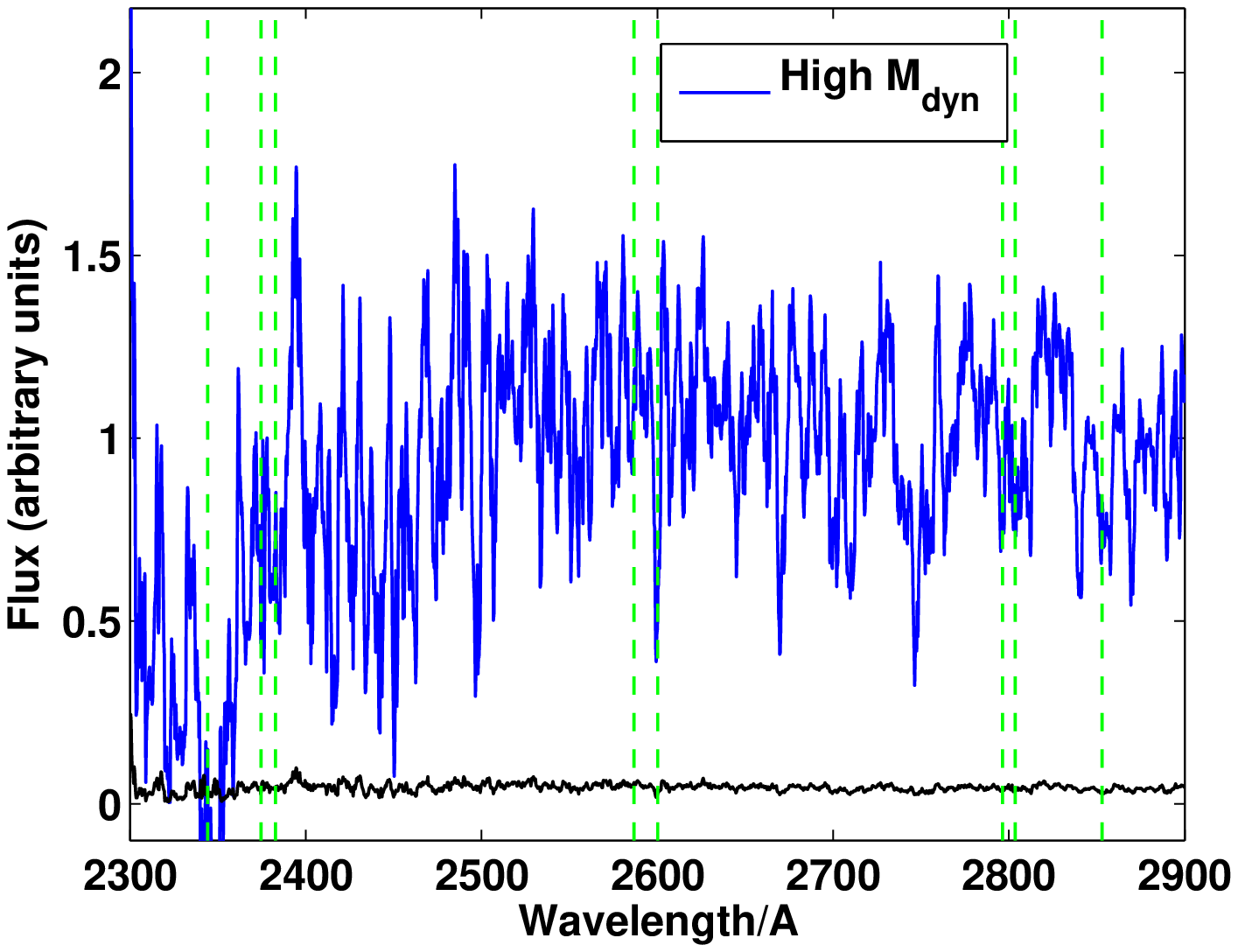} & \includegraphics[width=7.5cm,height=5.0cm,angle=0]{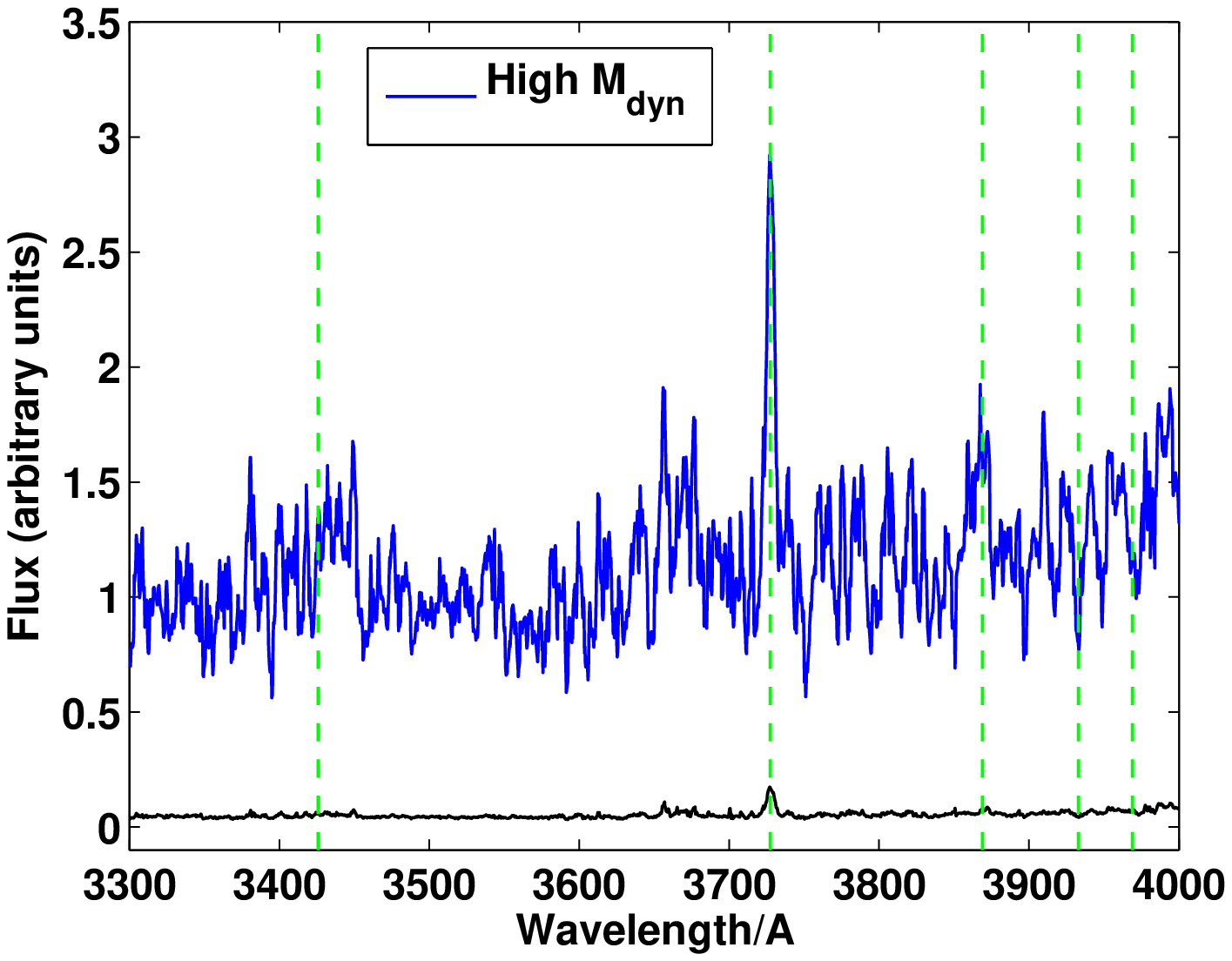} \\
\includegraphics[width=7.5cm,height=5.0cm,angle=0]{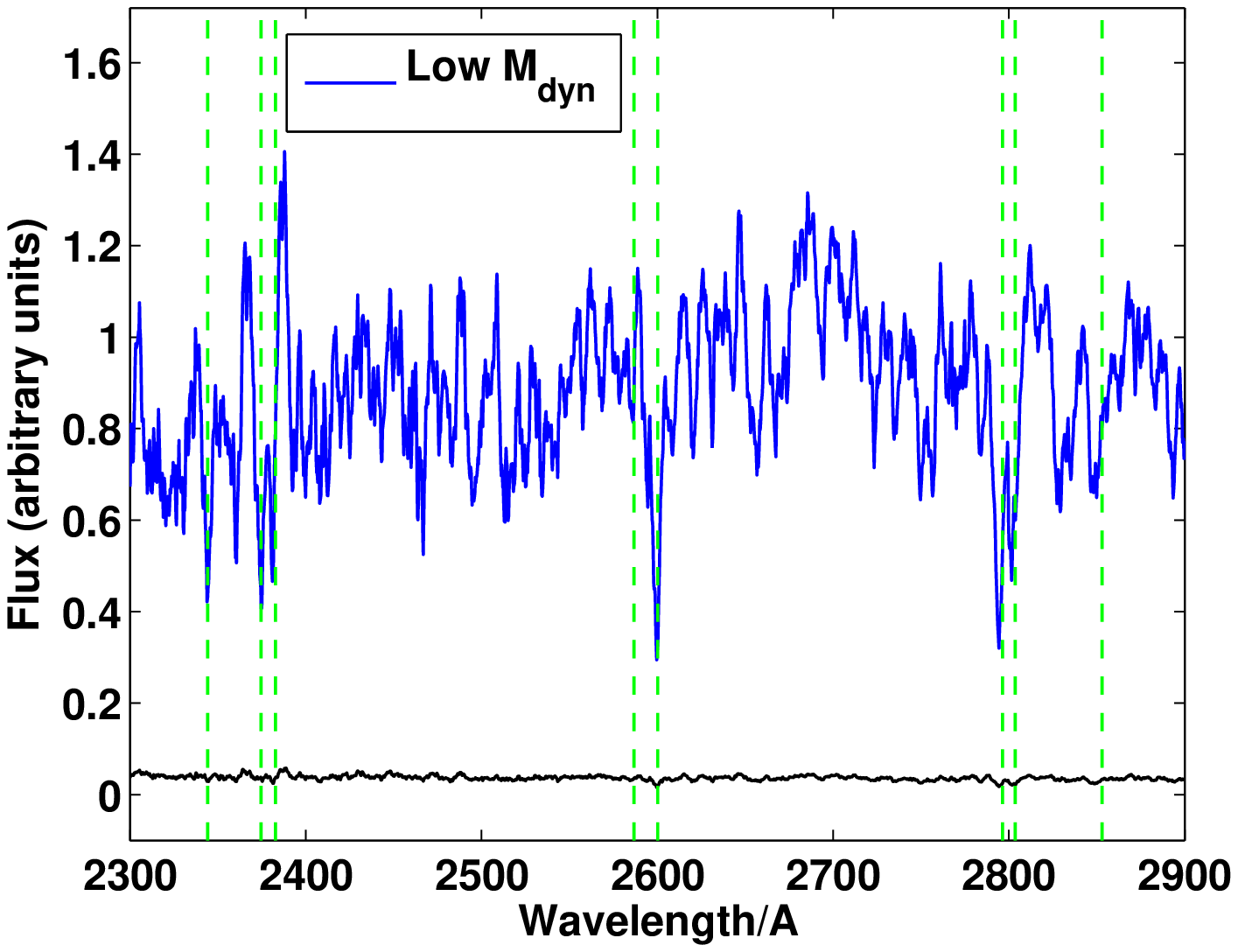} & \includegraphics[width=7.5cm,height=5.0cm,angle=0]{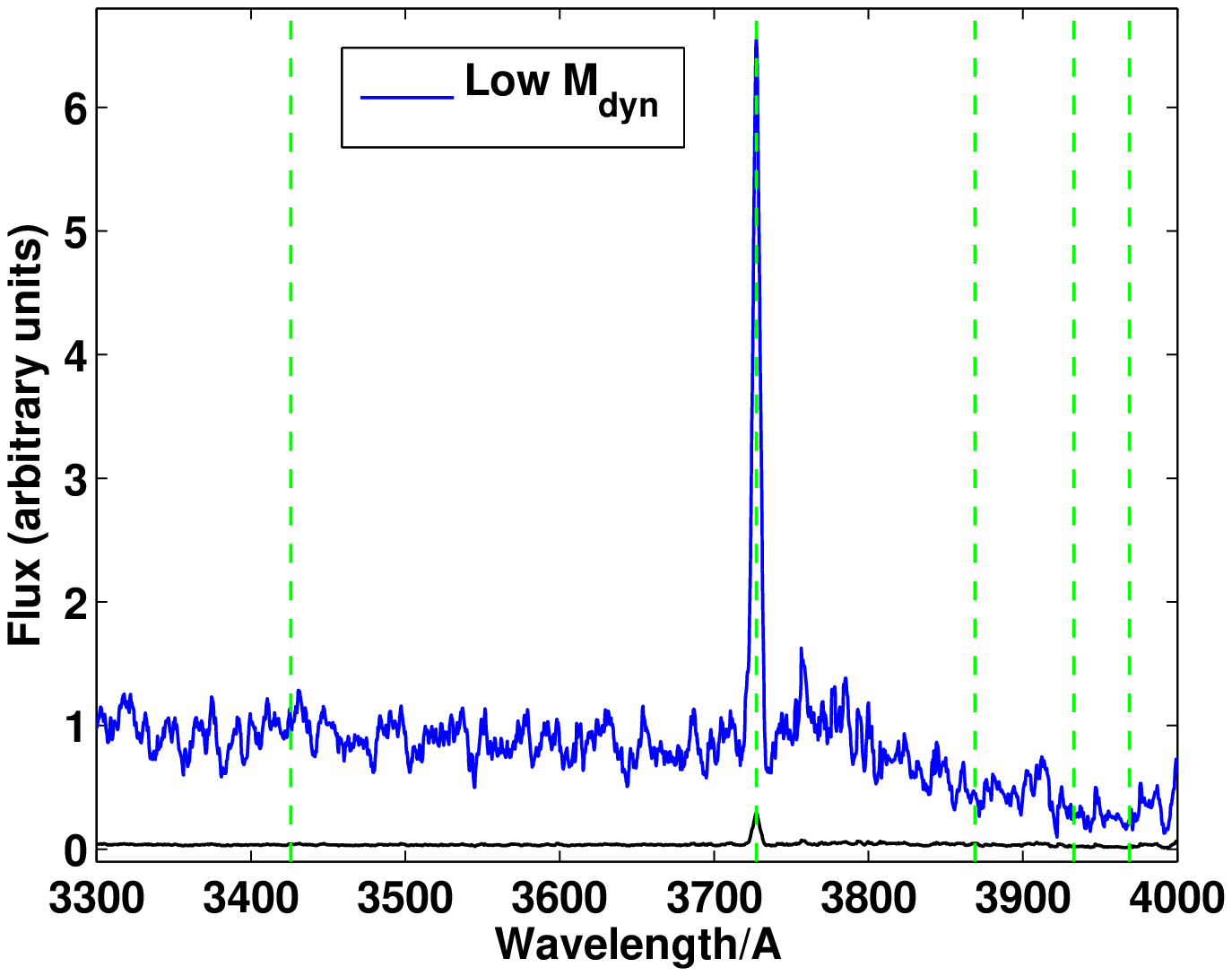} \\
\label{fig:a2}
\end{tabular}
\end{center}
\end{figure*}

In Figure \ref{fig:a1} we present composite spectra of SMGs, SFRGs as well as galaxies split by their infra-red luminosity. In Figure \ref{fig:a2}, we show the composite spectra of galaxies split by their [NeIII]/[OII] line ratio as well as the [OII] linewidth which is used to calculate the dynamical mass of the galaxy. 

\end{appendix}

\end{document}